\documentclass{ws-ijmpd}

\usepackage{color}
\usepackage{graphicx}
\begin{document}

\markboth{Paolo Ciarcelluti}
{Cosmology with Mirror Dark Matter}

%
\catchline{}{}{}{}{}
%

\title{COSMOLOGY WITH MIRROR DARK MATTER}

\author{PAOLO CIARCELLUTI}

\address{paolo.ciarcelluti@gmail.com}

\maketitle

\begin{history}
\received{Day Month Year}
\revised{Day Month Year}
\comby{Managing Editor}
\end{history}

\begin{abstract}

Mirror matter is a stable self-collisional dark matter candidate. 
If parity is a conserved unbroken symmetry of nature, there could exist a parallel hidden (mirror) sector of the Universe composed of particles with the same masses and obeying the same physical laws as our (visible) sector, except for the opposite handedness of weak interactions. 
The two sectors interact predominantly via gravity, therefore mirror matter is naturally ``dark''. 
Here I review the cosmological signatures of mirror dark matter, concerning thermodynamics of the early Universe, Big Bang nucleosynthesis, primordial structure formation and evolution, cosmic microwave background and large scale structure power spectra.
Besides gravity, the effects on primordial nucleosynthesis of the kinetic mixing between photons and mirror photons are considered.
Summarizing the present status of research and comparing theoretical results with observations/experiments, it emerges that mirror matter is not just a viable, but a promising dark matter candidate.

\end{abstract}



\section{Mirror particles}

In 1956, Lee and Yang proposed the non-parity of weak interactions in their famous paper entitled {\it Question of Parity Conservation in Weak Interactions}\cite{Lee:1956qn}.
In the same article they mentioned also a possible way-out to restore the parity, writing the following: {\it ``If such asymmetry is indeed found, the question could still be raised whether there could not exist corresponding elementary particles exhibiting opposite asymmetry such that in the broader sense there will still be overall right--left symmetry. 
If this is the case, it should be pointed out, there must exist two kinds of protons $p_R$ and $p_L$, the right-ended one and the left-ended one.''}
These sentences can be considered the start of the mirror matter adventure.

Subsequently experiments confirmed that the weak interactions indeed violate the parity symmetry, and the interactions of the fundamental particles are non-invariant under mirror reflection of the coordinate system.
The weak nuclear interaction is the culprit, with the asymmetry being particularly striking for the weakly interacting neutrinos. For example, today we know that neutrinos only spin with one orientation. Nobody has ever seen a right-handed neutrino.

Nevertheless, ten years later Kobzarev, Okun and Pomeranchuk\cite{Kobzarev:1966} discussed various phenomenological aspects of the idea suggested by Lee and Yang, a parallel hidden (mirror) sector of particles which is an exact duplicate of the observable particle sector.

In 1974 Pavsic\cite{Pavsic:1974rq} showed that non-conservation of parity in $\beta$-decay does not necessarily imply the mirror asymmetry of basic laws of nature, since by including the internal structure of particles involved in $\beta$-decay we can in principle restore reflection invariance.
What is asymmetric at the $\beta$-decay are initial conditions, while the weak interaction itself is symmetric.

Many years later, in 1991, Foot, Lew and Volkas\cite{Foot:1991bp} presented the old original idea of mirror matter in the modern context of gauge theories.
They developed the model with exact parity (mirror) symmetry between two particle sectors, that are described by the same Lagrangian and coupling constants, and consequently have the same microphysics.\footnote{
In this review I do not consider the so-called ``broken'' or ``asymmetric'' mirror models, where the parity symmetry is spontaneously broken and the two particle sectors have different weak interaction scales and then different microphysics. 
For an example, see Ref.~\refcite{Foot:2000tp}.}
From the modern point of view, the physics of ordinary world is described by the Standard Model $SU(3)\times SU(2)\times U(1)$ with gauge fields (gluons, photons, $W$, $Z$ and Higgs bosons) coupled to ordinary quarks $q$ and leptons $l$.
In the minimal parity-symmetric extension of the Standard Model,\cite{Foot:1991bp} the group structure is $G\otimes G'$, where $G=SU(3) \otimes SU(2) \otimes U(1)$ and the prime $(')$ denotes, as usual, the mirror sector.
The physics of mirror sector is described by the analogous gauge symmetry $SU(3)' \times SU(2)' \times U(1)'$ with the corresponding gauge bosons (mirror gluons, mirror photons, $W'$ and $Z'$ bosons) coupled to mirror quarks $q'$ and leptons $l'$. 
In this model the two sectors are described by the same Lagrangians, but where ordinary particles have left-handed interactions, mirror particles have right-handed interactions. 

Today we know mirror matter as a stable self-interacting\footnote{
Astrophysical constraints on self interactions of dark matter are present in literature, but the reader should keep in mind that they are valid only for homogeneous distributions of dark matter particles, and are therefore not directly applicable to the mirror matter case, as well as to all non-homogeneous dark matter distributions.} 
dark matter candidate that emerges if one, instead of (or in addition to) assuming a symmetry between bosons and fermions (supersymmetry), assumes that nature is parity symmetric.
The main theoretical motivation for the mirror matter hypothesis is that it constitutes the simplest way to restore parity symmetry in the physical laws of nature, since it is a matter of fact that the weak nuclear force does not exhibit parity symmetry.
Thereby the Universe is divided into two sectors with the same particles and interactions, but opposite handedness, that are connected by universal gravity. 

In other words, for each type of particle, such as electron, proton and photon, there is a mirror twin with the same mass, so that the ordinary particles favour the left hand, the mirror particles favour the right hand. 
If such particles exist in nature, then mirror symmetry would be exactly conserved.
In fact one could have the parity symmetry as an exact symmetry of exchange of the ordinary and mirror particles.
Thus, mirror particles are stable exactly as their ordinary counterparts. 

According to this theory the ordinary and mirror particles have the same masses, and the three non-gravitational forces act on ordinary and mirror sectors completely separately (and with opposite handedness: where the ordinary particles are left-handed, the mirror particles are right-handed). 
For example, while ordinary photons interact with ordinary matter, they {\it do not} interact with mirror matter.
Similarly, the ``mirror image'' of this statement must also hold, that is, the mirror photon interacts with mirror matter but does not interact with ordinary matter.  
The upshot is that we cannot see mirror photons because we are made of ordinary matter. 
The mirror  photons would simply pass right through us without interacting at all! 

The mirror symmetry requires that the mirror photons interact with mirror electrons and mirror protons in exactly the same way in which ordinary photons interact with ordinary electrons and ordinary protons. 
A direct consequence of this is that a mirror atom made of mirror electrons and a mirror nucleus, composed of mirror protons and mirror neutrons, can exist. 
In fact, mirror matter made of mirror atoms would also exist with exactly the same internal properties as ordinary matter, but would be completely invisible to us! 

Besides gravity, mirror matter could interact with ordinary matter via some other messengers, as the so-called kinetic mixing of gauge bosons,\cite{Holdom:1985ag}\cdash\cite{Foot:2000vy}
unknown fields that carry both ordinary and mirror charges, neutrino--mirror neutrino mass mixing,\cite{Akhmedov:1992hh}\cdash\cite{Berezinsky:2002fa} and Higgs boson--mirror Higgs boson mixing.\cite{Ignatiev:2000yy}$^,$\footnote{The presence of the mirror sector would cause an invisible decay of the Higgs boson, that could be detected in next measurements at the Large Hadron Collider (LHC).\cite{Ignatiev:2000yy,Li:2007bz}}
If such interactions exist they must be weak and are, therefore, negligible for many cosmological processes, but could be determinant for the detection of mirror dark matter in some contexts.
Since photons do not interact with mirror baryons, or interact only very weakly, the presence of mirror matter is felt mainly by its gravitational effects, which is exactly the definition of ``dark matter''! 
This is a key aspect of the mirror scenario, since it predicts the existence of dark matter in the Universe in a natural way. 

After the first pioneering works on mirror matter during 80's\cite{Holdom:1985ag}\cdash\cite{Carlson:1987si,Blinnikov:1980itep126}\cdash\cite{Khlopov:1989fj}, its many consequences for particle physics and astrophysics have been studied during the last decades.
The reader can refer to Ref.~\refcite{Okun:2006eb} for a history of the development of mirror matter research before 2006.

Like their ordinary counterparts, mirror baryons can form atoms, molecules and astrophysical objects such as planets, stars and globular clusters.
Invisible stars made of mirror baryons\cite{Blinnikov:1996fm}\cdash\cite{Berezhiani:2005vv} are candidates for Massive Astrophysical Compact Halo Objects (MACHOs), which have been observed as microlensing events.
The accretion of mirror matter onto celestial objects\cite{Blinnikov:1980itep126,Blinnikov:1982eh,Blinnikov:1983gh,Khlopov:1989fj} and the presence of mirror matter inside compact stars, for example neutron stars,\cite{Sandin:2008db}\cdash\cite{Iorio:2010hb} could have interesting observable effects.
If the dark matter is composed of mirror matter then it is also possible that some mirror matter exists in our solar system and in other solar systems.
In particular, it can have implications for the study of anomalous events within our solar system,\cite{Ignatiev:2000yw}\cdash\cite{Silagadze:2003xg} close-in extrasolar planets,\cite{Foot:1999ex}\cdash\cite{Foot:2004dh} Pioneer spacecraft anomalies.\cite{Foot:2001pv}
Furthermore, the mirror matter hypothesis has been invoked in various physical and astrophysical questions.\cite{Bell:1998jk}\cdash\cite{Silagadze:2008fa}
The consequences of mirror matter on Big Bang nucleosynthesis (BBN),\cite{Berezhiani:2000gw}\cdash\cite{Ciarcelluti:2009da} primordial structure formation, cosmic microwave background (CMB) and large scale structure (LSS) of the Universe,\cite{Ignatiev:2003js}\cdash\cite{Ciarcelluti:2004ip} have been carefully investigated.
For reviews on mirror matter theory and consequences, see Refs.~\refcite{Ciarcelluti:2009da,Ciarcelluti:2003wm,Foot:2001hc}--\refcite{Berezhiani:2005ek}. 

All these studies provide stringent bounds on the mirror sector and prove that it is a viable candidate for dark matter. 
In addition, mirror matter provides one of the few potential explanations for the recent DAMA/LIBRA annual modulation signal,\cite{Foot:2003iv}\cdash\cite{Foot:2009mw} and possibly other experiments,\cite{Foot:2010hu} as suggested for low energy electron recoil data from the CDMS collaboration,\cite{Foot:2009gk} the two events seen in the CDMSII/Ge experiment,\cite{Foot:2010th} the rising low energy spectrum observed by CoGeNT.\cite{Foot:2010rj}

If there exists a small kinetic mixing between ordinary and mirror photons,\cite{Holdom:1985ag,Foot:2000vy} the mirror particles become sort of ``millicharged'' particles for the ordinary observer, which suggests a very appealing possibility of their detection in the experiments for the direct search of dark matter, as mentioned above. 
It is also extremely interesting that such a kinetic mixing can be independently tested in laboratory ``table-top'' experiments for searching the orthopositronium oscillation into its mirror counterpart.\cite{Glashow:1985ud,Gninenko:1994dr}\cdash\cite{Crivelli:2010bk}
The mixing between the ordinary and mirror neutrinos could provide a possible mechanism for the generation of ultra high energy neutrinos\cite{Berezinsky:1999az} and for the gamma ray bursts as a result of explosion of the mirror supernovae.\cite{Blinnikov:1999ky}\cdash\cite{Foot:2004kd}
Explosions of the mirror supernovae and possible energy transfer of ordinary supernovae in the mirror sector could provide a necessary energy budget for heating the gaseous part of the mirror matter in the galaxies, and hence to prevent its collapse to a disk.
In this case the mirror matter could form spheroidal halos in accord to observations.\cite{Foot:2004wz}
In addition, the efficiency of mirror supernovae explosions can be indirectly tested in the future detectors for the gravitational waves. 

The mirror dark matter paradigm implies a spectrum of dark matter particles with known masses, given by the masses of the stable nuclei (H$'$, He$'$, ..., O$'$, ...,  Fe$'$, ...).\footnote{
As usual, mirror quantities are denoted with a prime ($'$).}$^,$\footnote{
The distribution of these elements in the dark component of the galaxy could be determined, in principle, once we know the chemical evolution of mirror nuclei, starting from their genesis in the mirror primordial nucleosynthesis and going on through successive stages of stellar formation and evolution.
At present we still lack the stellar formation, that constitutes a key ingredient in this study.}
According to this theory, mirror gas and dust should constitute the dark spheroidal galactic halos of galaxies.\cite{Foot:2004wz}
Inside the galaxy, mirror baryons should form astronomical objects and distribute in an complex way, eventually mixed with ordinary baryons,\cite{Blinnikov:1982eh,Blinnikov:1983gh,Khlopov:1989fj} and even accumulate into visible compact stars.\cite{Sandin:2008db}\cdash\cite{Ciarcelluti:2010ji}

It is worthwhile to note that the presence of the mirror sector does not introduce any new parameters in particle physics (if we neglect the possible weak non-gravitational interactions between visible and hidden sectors). 
Even more important, {\em the mirror particles can exist without violating any known experiment or observation}.
Thus, the correct statement is that the experiments have only shown that the interactions of the {\it known} particles are not mirror symmetric, they have not demonstrated that mirror symmetry is broken in nature.

Given its consistency with experiments and observations, and the unfruitful attempts to prove the existence of the other dark matter candidates, scientific community is facing an emergent question: ``is mirror matter the dark matter of the Universe (or at least a significant part of it)?''
One possibility to answer this question is to look at the cosmological signatures of mirror particles, that I try to summarize in this review.
Here I consider the cosmological effects of the gravitational interaction between ordinary and mirror matter, and I add the effects of photon--mirror photon kinetic mixing only when describing the primordial nucleosynthesis.


\paragraph{Photon--mirror photon kinetic mixing.}

Among the possible non-gravitational interactions between ordinary and mirror particles, photon--mirror photon kinetic mixing plays an important role, since it is related to the interpretation of dark matter direct detection experiments.

In fact, it has been shown in Ref.~\refcite{Foot:2008nw}, up-dating earlier studies,\cite{Foot:2003iv,Foot:2004gh,Foot:2005ic} that the mirror dark matter candidate is capable of explaining the positive dark matter signal obtained in the DAMA/LIBRA experiment,\cite{Bernabei:2008yi,Bernabei:2010mq} while also being consistent with the results of the other direct detection experiments.\cite{Foot:2010hu}\cdash\cite{Foot:2010rj}
The simplest possibility sees the mirror particles coupling to the ordinary particles via renormalizable photon--mirror photon kinetic mixing\cite{Foot:1991kb} (such mixing can also be induced radiatively if heavy particles exist charged under both ordinary and mirror $U(1)_{em}$\cite{Holdom:1985ag}) with Lagrangian
\begin{eqnarray}
{\cal L}_{mix} = {\frac{\epsilon}{2}}F^{\mu \nu} F'_{\mu \nu} \;,
\label{Lmix}
\end{eqnarray}
where $F^{\mu \nu} = \partial^{\mu} A^{\nu} - \partial^{\nu}A^{\mu}$ and $F'^{\mu \nu} = \partial^{\mu} A'^{\nu} - \partial^{\nu}A'^{\mu}$ are the field strength tensors for ordinary and mirror electromagnetisms, and $\epsilon$ is a free parameter.
This mixing enables mirror (ordinary) charged particles to couple to ordinary (mirror) photons with charge $\epsilon qe$, where $q=-1$ for $e'$, $q=+1$ for $p'$, {\it etc.}

Thus, the effect of this mixing is equivalent to that of very tiny electric charges, the so-called ``millicharges''.\footnote{
Given the current upper bounds, the correct name should be ``nanocharges'', but the term ``millicharges'' is still commonly used for historical reasons.}
In practice, if mirror and ordinary photons can kinetically mix, the mirror (ordinary) charged particles would behave as millicharges in the ordinary (mirror) sector.
The debate on the existence of fractional electric charges in the Universe is still open, with upper bounds that becomes more and more stringent.
These bounds, coming from laboratory or astrophysics, are computed for generic millicharges. Therefore, one should be careful to apply them to mirror particles, since they can depend on the details of the model, for example the spatial distribution of particles or other interactions concomitant with the kinetic mixing of photons.

An important experimental consequence of Eq.~(\ref{Lmix}) is the mixing of orthopositronium with mirror orthopositronium, leading to oscillations between these states in a vacuum experiment.
The subsequent decays of the mirror state are invisible, resulting in an effective increase of the decay rate.\cite{Glashow:1985ud}
An eventual discrepancy between the theoretical predictions and the experimental measurements may in fact be resolved by this mirror world mechanism.\cite{Gninenko:1994dr,Foot:2000aj}
A new measurement\cite{Vallery:2003iz} performed in 2003 solved the previous longstanding discrepancy, so that currently there is no evidence of the mirror world from orthopositronium studies, placing an upper limit\cite{Foot:2003eq} $\epsilon < 5 \times 10^{-7}$.
More recently, improved measurements of the invisible decay of orthopositronium\cite{Badertscher:2006fm} lowered the upper limit to $\epsilon \leq 1.55 \times 10^{-7}$.
A new proposed experiment\cite{Crivelli:2010bk} promises the expected sensitivity in mixing strength to be $ \epsilon \sim 10^{-9}$, which is in a region of parameter space of great theoretical and phenomenological interest.
In fact, the mirror matter interpretation of the recent DAMA/LIBRA observations of the annual modulation signal\cite{Foot:2008nw} requires just $\epsilon \sim 10^{-9}$, which is currently consistent with laboratory and astrophysical constraints.\cite{Foot:2003eq,Ciarcelluti:2008qk,Ciarcelluti:2010dm}


\section{Mirror baryons as cosmological dark matter}
\label{mir-dm}

In the previous section we have seen that, if we neglect possible but very small non-gravitational interactions, the presence of the mirror sector does not introduce any new parameters in particle physics.

However, the fact that microphysics is the same does not mean that also macroscopic realizations of the two particle sectors should be the same.
The different macrophysics is usually parametrized in terms of two ``cosmological'' free parameters: the ratio $ x $ of temperatures of two sectors, in terms of temperatures of the ordinary and mirror photons in the cosmic background radiation; the relative amount $ \beta $ of mirror baryons compared to the ordinary ones.
Indeed these two parameters are not completely free, since there are astrophysical bounds, that become more and more stringent with the progress of research, as we will see in the following.

All the differences with respect to the ordinary world can be described in terms of only two free parameters
\begin{eqnarray}\label{mir-param}
x \equiv \left( \frac{S'}{S} \right)^{1/3} \simeq \frac{T'}{T}
~~~~~~ {\rm and} ~~~~~~
\beta \equiv \frac{\Omega'_{b}}{\Omega_{b}} ~~,
\end{eqnarray}
where $T$ ($T'$), $\Omega_{b}$ ($\Omega'_{b}$), and $S$ ($S'$) are respectively the ordinary (mirror) photon temperature, cosmological baryon density (normalized, as usual, to the critical density of the Universe), and entropy per comoving volume.
These parameters are defined in order to be invariant along the evolution of the Universe, if entropies per comoving volume and baryon numbers are separately conserved in two sectors.
As will be more clear in next section, the parameter $x$ is not exactly the ratio of temperatures during all cosmic history, but it is a good approximation during most of the time for the usually considered values of $x$.

Since it interacts only gravitationally with our ordinary sector, mirror matter is a {\it natural dark} matter candidate. 
At present there are observational evidences that dark matter exists and it's density is about 5 times that of ordinary baryons.
This is not a problem for mirror scenario, because the same microphysics does not imply the same initial conditions in both ordinary and mirror sectors.
In particular, various baryogenesis scenarios with a hidden sector\cite{Berezhiani:2000gw,Bento:2001rc}\cdash\cite{Foot:2004pq} support a mirror baryonic density at least equal to the ordinary one.\footnote{
As far as the two sectors have the same particle physics, it is natural to think that the mirror baryon number density $n'_b$ is determined by the baryogenesis mechanism which is similar to the one which fixes the ordinary baryon density $n_b$. 
Thus, since the mass of mirror baryons is the same as the ordinary ones, one could question whether the ratio $ \beta $ could be naturally of order 1 or somewhat bigger.     
There are several baryogenesis mechanisms as are GUT baryogenesis, leptogenesis, electroweak baryogenesis, {\it etc.} 
At present it is not possible to say definitely which of the known mechanisms is responsible for the observed baryon asymmetry in the ordinary world. 
However, it is most likely that the baryon asymmetry in the mirror world is produced by the same mechanism and moreover, the properties of the $B$ and CP violation processes are parametrically the same in both cases.

But the mirror sector has a lower temperature than the ordinary one, and so at epochs relevant for baryogenesis the out-of-equilibrium conditions should be easier fulfilled for the mirror sector.
In particular, we know that in certain baryogenesis scenarios the mirror world gets a larger baryon asymmetry than the ordinary sector, and it is pretty plausible that $\beta \gtrsim 1$.\cite{Berezhiani:2000gw}
This situation emerges in a particularly natural way in the leptogenesis scenario due to the lepton number exchange from one sector to the other, which leads to  $n'_b \geq n_b$, and can thus explain the near coincidence of visible and dark components in a rather natural way.\cite{Bento:2001rc}\cdash\cite{Foot:2004pq}   
}
The bounds on the parameter $\beta$ come from the requirements that mirror baryon density cannot be more than that inferred for dark matter, and should be at least of the same magnitude of the ordinary baryon density, in order to give a significant contribution to the total matter density.
This translates into
\begin{eqnarray}\label{beta-bounds}
1 \lesssim \beta \lesssim \frac{\Omega_{DM}}{\Omega_{b}} \approx 5 .
\end{eqnarray}

If the mirror sector exists, then the Universe along with the ordinary electrons, nucleons, neutrinos and photons, should also contain their mirror partners.  
One could naively think that due to mirror parity the ordinary and mirror particles should have the same cosmological abundances and hence the ordinary and mirror sectors should have the same cosmological evolution. 
In this case the mirror photons $\gamma'$, electrons ${e^{\pm}}'$ and neutrinos $\nu'_{e,\mu,\tau}$ would give a contribution to the Hubble expansion rate equivalent to that of the ordinary ones, {\it i.e.} in terms of energetic degrees of freedom $g'=g=10.75$.
Since each neutrino family contributes $\Delta g = (7/8) \cdot 2$, the mirror contribution is equivalent to an effective number of extra neutrino families\cite{Berezhiani:1995am} $\Delta N_\nu = N_\nu -3 = (10.75/1.75) \simeq 6.14$.
However, this is forbidden by current estimates\cite{Izotov:2010ca} $N_\nu = 3.68^{+0.80}_{-0.70}$ or $N_\nu = 3.80^{+0.80}_{-0.70}$, depending on the value used for the neutron lifetime $\tau_n$.
The only known possibility is then to have a lower temperature for the mirror sector, so that the contribution of relativistic degrees of freedom at BBN epoch is suppressed by a $(T'/T)^4$ factor.
In this case the contribution of the mirror sector translates into\cite{Berezhiani:2000gw,Berezhiani:1995am} $\Delta N_\nu \approx 6.14 x^4$, where $x \simeq T'/T$ (for more details see next section). 
Thus, the conservative upper bound $\Delta N_\nu \lesssim 1$ implies the upper limit
\begin{equation} \label{x-bound}
x \lesssim 0.64 (\Delta N_\nu)^{1/4} \simeq 0.64 \:,
\end{equation}
with rather mild dependence on $\Delta N_\nu$, {\it e.g.} $\Delta N_\nu <0.5$ implies roughly $x < 0.5$ while $\Delta N_\nu <0.2$ implies $x < 0.4$.  
Considering the less conservative upper bound of Ref.~\refcite{Izotov:2010ca} even values $ x \leq 0.7 $ are acceptable.
It is remarkable that the aforementioned indirectly measured values of $N_\nu$ are only marginally consistent with the Standard Model expectation,\cite{Izotov:2010ca} and the presence of mirror particles with a lower temperature than ordinary ones would improve the consistency.
 
Hence, in the early Universe the mirror system should have a somewhat lower temperature than ordinary particles. 
Therefore, even if their microphysics is the same, initial conditions are not the same in both sectors, and in the early Universe the mirror particles are colder than the ordinary ones. 
This situation is plausible if the following conditions are satisfied.
\begin{enumerate}
\item[(A)] At the Big Bang the two systems are born with different temperatures.
According to the inflationary paradigm, at the post-inflationary epoch the mirror sector is (re)heated at lower temperature than in the ordinary one, which can be naturally achieved in certain models.\cite{Kolb:1985bf,Berezhiani:2000gw,Berezinsky:1999az,Berezhiani:1995am,Hodges:1993yb}
\item[(B)] At temperatures below the reheating temperature the two systems interact very weakly, so that they do not come into thermal equilibrium with each other after reheating.
This condition is automatically fulfilled if the two worlds communicate only via gravity.
If there are some other effective couplings between the ordinary and mirror particles, they have to be properly suppressed. 
\item[(C)] Both systems expand adiabatically, without significant entropy production.
If the two sectors have different reheating temperatures, during the expansion of the Universe they evolve independently, their temperatures remain different at later stages, $T' < T$, and the presence of the mirror sector would not affect primordial nucleosynthesis in the ordinary world.  
\end{enumerate}

Therefore, the BBN bounds require that at the nucleosynthesis epoch the temperature of the mirror sector be smaller than that of the ordinary one, $T' < T$.  
As far as the mirror world is cooler than the ordinary one, $x < 1$, in the mirror world all key epochs (as are baryogenesis, nucleosynthesis, recombination, {\it etc.}) proceed in somewhat different conditions than in ordinary world.
Namely, in the sector world the relevant processes go out of equilibrium earlier than in the ordinary sector, which has many far going implications, in particular on the key cosmological epochs. 
In next sections we will see the consequences for the thermodynamics of the early Universe, BBN, structure formation and evolution.

In the most general context, the present energy density contains relativistic (radiation) component $ \Omega_r $, non-relativistic (matter) component $ \Omega_m $ and the vacuum energy (cosmological term or dark energy) density $ \Omega_\Lambda $. 
According to the inflationary paradigm the Universe should be almost flat, $\Omega_0=\Omega_m + \Omega_r + \Omega_\Lambda \approx 1$, which agrees well with the results on the CMB anisotropy.

Then, in a so-called Mirror Universe\footnote{
The expression ``Mirror Universe'' is perhaps misleading, since one could think that there is another Universe, while it is just one, but made of two particle sectors, ordinary and mirror. 
Nevertheless, this expression is sometimes used to shortly refer to this scenario.} 
$\Omega_{r}$ and $\Omega_{m}$ represent the total amount of radiation and matter of both ordinary and mirror sectors: 
$ \Omega_{r} = (\Omega_{r})_{\rm O} + (\Omega_{r})_{\rm M} $ and 
$ \Omega_{m} = (\Omega_{m})_{\rm O} + (\Omega_{m})_{\rm M} $. 

In the context of our model, as explained in next section, the relativistic fraction is represented by the ordinary and mirror photons and neutrinos.
Using the expression for the ordinary energetic degrees of freedom in a Mirror Universe, $ \bar g(T) = g(T) (1 + x^4) $ (see Sec.~\ref{therm_approx} for details), and the value of the observable radiation energy density 
$ (\Omega_r)_{\rm O} \, h^2 \simeq 4.2 {\times} 10^{-5} $ (where $h$ is, as usual, the Hubble parameter), it is given by 
\begin{equation}
\Omega_r = 4.2 \times 10^{-5}\,h^{-2}\,(1+x^4) \simeq 4.2 \times 10^{-5}\,h^{-2} \;,
\end{equation}
where the contribution of the mirror species, expressed by the additional term $ x^4 $, is very low in view of the BBN constraint $ x < 0.64 $. 
As for the non-relativistic component, it contains the ordinary baryon fraction $\Omega_b$ and the mirror baryon fraction $\Omega'_b = \beta\Omega_b$, while the other types of dark matter could also be present.
Obviously, since mirror parity doubles {\it all} the ordinary particles, even if they are ``dark'' (i.e., we are not able to detect them now), whatever the form of dark matter made by some exotic ordinary particles, there will exist a mirror partner made by the mirror counterpart of these particles.\footnote{
In the context of supersymmetry, the cold dark matter (CDM) component could exist in the form of the lightest supersymmetric particle (LSP).  
It is interesting to remark that the mass fractions of the ordinary and mirror LSP are related as $\Omega'_{LSP} \simeq x\Omega_{LSP}$. 
In addition, a significant HDM component $\Omega_\nu$ could be due to neutrinos with order eV mass. 
The contribution of the mirror massive neutrinos scales as $\Omega'_\nu = x^3 \Omega_\nu$ and thus it is irrelevant.
In any case, considering the only CDM component, which is now the preferred candidate, we can combine both the ordinary and mirror components, since their physical effects are the same.}
Thus we consider a matter composition of the Universe expressed in general by
\begin{equation}
\Omega_m = \Omega_b+\Omega'_b+\Omega_{DM}
         = \Omega_b (1+\beta) + \Omega_{DM}\;,
\end{equation}
where the term $\Omega_{DM}$ includes the contributions of any other possible dark matter particles but mirror baryons.


\paragraph{Cosmological key epochs.}

As shown in Refs.~\refcite{Berezhiani:2000gw,Berezhiani:2003wj}--\refcite{Ciarcelluti:2004ik},
due to the temperature difference between the two sectors, the cosmological key epochs take place at different redshifts, and in particular they happen in the mirror sector before than in the ordinary one. 

\begin{figure}[pb]
\centerline{\psfig{file={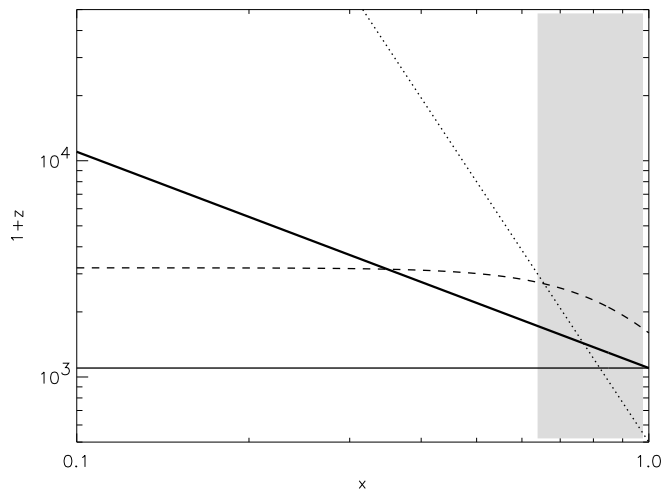},width=9.5cm}}
\vspace*{-144pt}
\caption{The mirror photon decoupling redshift $1+z_{\rm dec}'$ as a function of $x$ (thick solid). The horizontal thin solid line marks the ordinary photon decoupling redshift $1+z_{\rm dec} = 1100$. We also show the matter-radiation equality redshift $1+z_{\rm eq}$ (dash) and the mirror Jeans-horizon mass equality redshift $1+z'_c$ (dot) for the case $\Omega_m h^2 =0.135$ (see Sec.~\ref{mirror_bar_struc}). The shaded area $x> 0.64$ is excluded by the BBN limits.
\label{figzz}}
\end{figure}

The relevant epochs for the cosmic structure formation are related to the matter-radiation equality (MRE) $z_{\rm eq}$, the matter-radiation decouplings (MRD) $z_{\rm dec}$ and $z'_{\rm dec}$ due to the plasma recombinations in both sectors, and the photon-baryon equipartitions $z_{b\gamma}$ and $z'_{b\gamma}$. 
The MRE occurs at the redshift 
\begin{equation} \label{z-eq} 
1+z_{\rm eq}= {\frac{\Omega_m}{\Omega_r}} \approx 
 2.4\cdot 10^4 {\frac{\Omega_{m}h^2}{1+x^4}} \;,
\end{equation}
which is {\it always smaller than the value obtained for an ordinary Universe}, but approximates it for low $x$ (see Fig.~\ref{figzz}). 
If we consider only ordinary and mirror baryons and photons, we find
\begin{equation} \label{z-eq_2} 
1+z_{\rm eq} = 
 {\frac{\rho_b \left( 1+\beta \right)}{\rho_{\gamma } \left( 1+x^4 \right)}} 
 {\left( 1+z \right)} =
 {\frac{\rho_b' \left( 1+\beta^{-1} \right)}{\rho_{\gamma }' \left( 1+x^{-4} \right)}} 
 {\left( 1+z \right)} \;,
\end{equation}
where the baryon and photon densities refer to the redshift $z$.
This implies that, with the addition of a mirror sector, the matter-radiation equality epoch shifts toward earlier times as \footnote{
In the most general case, where there is also some other dark matter component, the parameter $\beta$ could be the sum of two terms, $ \beta + \beta_{DM} $.}
\begin{equation} \label{shifteq} 
1+z_{\rm eq} ~~ \longrightarrow ~~ 
  { \frac{\left( 1+\beta \right)}{\left( 1+x^4 \right)} } ~ (1+z_{\rm eq}) \;.
\end{equation}

The MRD, instead, takes place in every sector only after most electrons and protons recombine into neutral hydrogen and the free electron number density $n_{e}$ diminishes, so that the interaction rate of the photons
$\Gamma_\gamma=n_{e}\sigma_{T}=X_{e}\eta n_{\gamma} \sigma_{T}$ drops below the Hubble expansion rate $H(T)$, where 
$\sigma_T$ is the Thomson cross section, $X_e=n_e/n_b$ is the fractional ionization, and $ \eta = n_b/n_\gamma $ is the baryon to photon ratio.
In the condition of chemical equilibrium, $ X_e $ is given by the Saha equation, which for $ X_e \ll 1$ reads
\begin{equation} \label{Saha} 
X_e \approx (1-Y_4)^{1/2}\; {\frac{0.51}{\eta^{1/2}}} 
\left(\frac{T}{m_e}\right)^{-3/4} e^{-B/2T} \;,
\end{equation}
where $ B=13.6 $ eV is the hydrogen binding energy and $ Y_4 $ is the $^4$He abundance.
Thus we obtain the familiar result that in our Universe the MRD takes place in the matter domination period, at the temperature $T_{\rm dec} \simeq 0.26$ eV, which corresponds to redshift $1+z_{\rm dec}=T_{\rm dec}/T_0 \simeq 1100$.

The MRD temperature in the mirror sector $T'_{\rm dec}$ can be calculated following the same lines as in the ordinary one. 
Due to the fact that in either case the photon decoupling occurs when the exponential factor in Eq.~(\ref{Saha}) becomes very small, we have $T'_{\rm dec} \simeq T_{\rm dec}$, up to small corrections related to $\eta'$, $Y'_4$ which are respectively different from $\eta$, $Y_4$.
Hence, considering that $T' = x \cdot T$, we obtain
\begin{equation} \label{z'_dec}
1+z'_{\rm dec} \simeq x^{-1} (1+z_{\rm dec}) 
\simeq 1.1\cdot 10^3 x^{-1} \;, ~~
\end{equation}
so that {\it the MRD in the mirror sector occurs earlier than in the ordinary one}. 
Moreover, comparing Eqs.~(\ref{z-eq}) and (\ref{z'_dec}), which have different dependences on $x$, we find that, for $x$ smaller than a typical value $x_{\rm eq}$ expressed by
\begin{equation}
x_{\rm eq} \approx 0.046(\Omega_m h^2)^{-1} \;, 
\end{equation}
the mirror photons would decouple yet during the radiation dominated period (see Fig.~\ref{figzz}). 
This quantity has a key role in structure formation with mirror dark matter.
In fact we expect that for values $ x < x_{\rm eq} $ the evolution of primordial perturbations in the linear regime is practically identical to the standard CDM case.
Assuming, e.g., the value $\Omega_m h^2 = 0.135$, we obtain $x_{\rm eq} \approx 0.34$, which indicates that below about this value the mirror decoupling happens in the radiation dominated period, with consequences on structure formation (as we will see in the following sections and as shown in Refs.~\refcite{Berezhiani:2000gw,Ignatiev:2003js}--\refcite{Ciarcelluti:2004ik}).

The redshifts of photon-baryon equipartitions for ordinary and mirror particles are related by
\begin{equation} \label{shiftzbg} 
1+z_{b\gamma}' 
  = {\frac{\Omega_b'}{\Omega_{\gamma}'} } 
  \simeq \frac{ \Omega_b \, \beta}{\Omega_\gamma \, x^4} 
  = (1+z_{b\gamma}) { \frac{\beta}{x^4} } > 1+z_{b\gamma} \;.
\end{equation}
This means that also {\it photon-baryon equipartition happens in the mirror sector earlier than in the ordinary one}. 

In Sec.~\ref{sec-evol-pert} we will study the important consequences of these effects on structure formation.


\section{Thermodynamics of the early Universe}
\label{thermuniv}


\subsection{Thermodynamical equilibrium}
\label{sec_eq_therm}

We extend the standard theory of thermodynamics in the early Universe in order to take the existence of mirror particles into account.

We consider the Universe as a thermodynamical system composed of different species (ordinary and mirror electrons, photons, neutrinos, nucleons, {\it etc.}) which, in the early phases, were to a good approximation in thermodynamical equilibrium, established through rapid interactions, in the two sectors separately.

We assume, as usual, that the Universe is homogeneous, and the chemical potentials of all particle species A are negligible,\cite{Kolb:1990vq,paddybook} i.e. $\mu_A \ll T$.
The latter assumption implies the conservation of the total entropy of the Universe $S$.
Therefore we can use the equilibrium Bose-Einstein or Fermi-Dirac distribution functions and calculate the energy density $\rho$, the pressure $p$, and the entropy density $s$ for every particle species in thermal equilibrium
\begin{eqnarray} \label{energy_density}
  \rho_A(T) = \frac{g_A}{2\pi^2} \int_{m_A}^{\infty}
    \frac{(E^2-m_A^2)^{\frac{1}{2}} \:E^2}
    {\exp \left(E/T\right)\pm 1} \:\mathrm{d}E \;,
\end{eqnarray}
\begin{eqnarray} \label{pressure}
  p_A(T) = \frac{g_A}{6\pi^2} \int_{m_A}^{\infty}
    \frac{(E^2-m_A^2)^{\frac{3}{2}} }
    {\exp \left(E/T\right)\pm 1} \:\mathrm{d}E \;,
\end{eqnarray}
\begin{eqnarray} \label{entropy_generic}
  s_A(T) = \frac{p_A(T) + \rho_A(T)}{T} \;,
\end{eqnarray}
where $g_A$ is the spin-degeneracy factor of the species $A$, the signs + and -- correspond respectively to fermions and bosons, and $E = \sqrt{\mathbf{p}^2 + m^2}$, with $\mathbf{p}$ the momentum.
We can use the usual parametrization of the entropy density
\begin{eqnarray} \label{entropy}
  s(T) = \frac{2\pi^2}{45}\: q(T)\: T^3 \;,
\end{eqnarray}
where
\begin{eqnarray} \label{q_tot__def}
  q(T) \equiv \sum_{\rm bosons} g_b (T) \left(\frac{T_b}{T}\right)^3
       +\frac{7}{8}\sum_{\rm fermions} g_f (T) \left(\frac{T_f}{T}\right)^3
\end{eqnarray}
is the number of {\it effective entropic} degrees of freedom (d.o.f.), that group together the d.o.f. for all bosons ($g_b$) and fermions ($g_f$).
An analogous formalism can be used for the total energy density $\rho$, which can be parametrized as
\begin{eqnarray} \label{energy}
  \rho (T)&=& \frac{\pi^2}{30}\: g(T)\: T^4 \;,
\end{eqnarray}
where
\begin{eqnarray} \label{g_tot__def}
  g(T) \equiv \sum_{\rm bosons} g_b (T) \left(\frac{T_b}{T}\right)^4 
       +\frac{7}{8}\sum_{\rm fermions} g_f (T) \left(\frac{T_f}{T}\right)^4
\end{eqnarray}
is the number of {\it effective energetic} degrees of freedom.


\subsection{A useful simple approximation}
\label{therm_approx}

Once the ordinary and mirror systems are decoupled already after reheating, at later times $t$ they will have different temperatures $T(t)$ and $T'(t)$, and so different energy and entropy densities
\begin{equation} \label{rho}
\rho(t) = \frac{\pi^{2}}{30} g(T) T^{4} \;, ~~~ 
\rho'(t) = \frac{\pi^{2}}{30} g'(T') T'^{4} \;,~  
\end{equation}
\begin{equation} \label{s}
s(t) = \frac{2\pi^{2}}{45} q(T) T^{3} \;, ~~~ 
s'(t) = \frac{2\pi^{2}}{45} q'(T') T'^{3} \;.~
\end{equation}
The factors $g$, $q$ and $g'$, $q'$ account for the effective number of the degrees of freedom in two systems, and in general can be different from each other.

The Hubble expansion rate is determined by the total energy density ${\bar{\rho}}=\rho+\rho'$, $H=\sqrt{(8\pi/3) G {\bar{\rho}}}$. Therefore, at a given time $t$ in a radiation dominated epoch we have 
\begin{equation} \label{Hubble}
H(t) = \frac{1}{2t} = 1.66 \sqrt{\bar{g}(T)} {\frac{T^2}{M_{\rm Pl}}} = 
1.66 \sqrt{\bar{g}'(T')} {\frac{T'^2}{M_{\rm Pl}}} ~  
\end{equation}
in terms of ordinary and mirror temperatures $T(t)$ and $T'(t)$, where 
\begin{eqnarray} \label{g-ast}
\bar{g}(T)   &=& g(T) (1 + ax^4) \simeq g(T) (1 + x^4) \;,
\end{eqnarray}
\begin{eqnarray} \label{gm-ast}
\bar{g}'(T') &=& g'(T')\left(1 + {\frac{1}{ax^4}}\right) 
              \simeq g'(T')\left(1 + x^{-4}\right) \;, 
\end{eqnarray}
and $M_{\rm Pl} \simeq 1.22 \times 10^{22}$ MeV is the Planck mass.
Here the factor $ a(T,T') = [g'(T') / g(T)] \cdot [q(T) / q'(T')]^{4/3} $ takes into account that for $T'\neq T$ the relativistic particle contents of the two worlds can be different. However, except for very small values of $x$, during most of the time in history of the Universe a {\it useful simple approximation} is $a \sim 1$. 
In particular, in the modern Universe we have $a(T_0,T'_0) =1$, $q(T_0) = q'(T'_0)=3.91$, and $x= T'_0/T_0$, where $T_0,T'_0$ are the present temperatures of the ordinary and mirror relic photons.

As far as $x^4\ll 1$, the effective degrees of freedom and the Hubble expansion rate are dominated by the density of ordinary particles and the presence of mirror sector has negligible effects on the standard cosmology of the early ordinary Universe. 
The opposite happens instead for the mirror sector, that is largely influenced by the density of ordinary relativistic particles, as evident from the factor $x^{-4}$ in Eq.~(\ref{gm-ast}).
This makes the cosmology of the early mirror world different from the standard one as far as the crucial epochs like baryogenesis, nucleosynthesis, {\it etc.} are concerned. 
Any of these epochs is related to an instant when the rate of the relevant particle process $\Gamma(T)$, which is generically a function of the temperature, becomes equal to the Hubble expansion rate $H(T)$. 
Obviously, in the mirror sector these events take place earlier than in the ordinary sector, and as a rule, the relevant processes in the former freeze out at larger temperatures than in the latter. 

It is useful to note that, due to the difference in the initial temperature conditions in the two sectors, reactions at the same temperature  $T_* = T_*'$ occur at different times $t_*' = x^2 t_*$, which implies different rates of the Hubble expansion (in particular $H(t_*') > H(t_*)$, given $x < 1$), while reactions at the same time $t_* = t_*'$ occur at different temperatures  $T_*' = x T_*$.
In particular, this behaviour has important consequences on Big Bang nucleosynthesis, that are shown in details in Sec.~\ref{sec-bbn}.


\subsection{A more accurate study}

During the expansion of the Universe, the two sectors evolve with separately conserved entropies, which means that the ratio of entropy densities is also conserved, so that in general we can approximate
\begin{equation} \label{xmir}
  x = \left( \frac{s'}{s} \right)^{1/3}
    = \left[\frac{q'(T')}{q(T)} \right]^{1/3}{\frac{T'}{T}} \approx {\frac{T'}{T}} \;.
\end{equation}
This approximation is valid only if the temperatures of the two sectors are not too different, or otherwise if we are far enough from crucial epochs, like $e^+$-$e^-$ annihilation.\cite{Berezhiani:2000gw}
If, instead, we are studying the range of temperatures interested by this phenomenon, the approximation (\ref{xmir}) is no more valid, and a more accurate study is required for different reasons.
First of all, we need a detailed study of thermodynamical evolution of the early Universe in order to obtain a reliable description of the thermal cosmic history in presence of mirror dark matter. 
Second, an accurate study of the trend of number of degrees of freedom is necessary for any future simulation of nucleosynthesis of primordial elements in both sectors, that may be compared with observations.
In particular, it is even more important in view of the recently proposed interpretation of the DAMA/LIBRA experiment in terms of interactions with mirror heavy elements.\cite{Foot:2008nw}
Third, claims for variations of the effective number of extra neutrino families, computed at BBN ($\sim 1$ MeV) and CMB formation ($\lesssim 1$ eV) epochs, require investigations for possible variations of these numbers as consequences of physics beyond the Standard Model (see Ref.~\refcite{Mangano:2006ur} and references therein for previous works).


\subsubsection{Equations}
\label{sec_equations}

We calculate the equations which link the ordinary and mirror sector temperatures and thermodynamical quantities; then we solve these equations numerically.
Once the temperatures are known, it is possible to obtain the exact total number of d.o.f. in both sectors, which, as common in literature, can be expressed in terms of extra-neutrino number.
We report in Sec.~\ref{sec_Num_calc} the results of these calculations.

The presence of the other sector leads in both sectors to the same effects of having more particles.
As already stated, we do not take interactions between the two sectors besides gravity into account.
This implies that the entropies of the two sectors are conserved separately, and the parameter $x$ is constant during the cosmic evolution
\begin{eqnarray} \label{xSs}
  x =\left(\frac{S'}{S}\right)^{1/3}
    =\left(\frac{s' \cdot a^3}{s \cdot a^3}\right)^{1/3}
    =\left(\frac{s'}{s}\right)^{1/3}
    = {\rm const.} \;,
\end{eqnarray}
where $a$ is the scale factor of the Universe.

As already stressed, ordinary and mirror sectors have the same microphysics; therefore we can assume, to a first approximation, that the neutrino decoupling temperature\footnote{
The assumption that the neutrino decoupling is an instantaneous process taking place when photons have the temperature $T=T_{D\nu}$ introduces just a small error ($< 1\%$), that is negligible for the precision required here.
} $T_{D\nu}$ is the same in both of them, that is $T_{D\nu}=T_{D\nu'}'$.
But the temperatures in the ordinary sector when the ordinary and mirror neutrino decouplings take place are different, $T_{D\nu} \ne T_{D\nu'}$, with $T_{D\nu}<T_{D\nu'}$ since $x<1$.
Therefore, at $T \gg T_{D\nu}$ ($T_{D\nu}'$), ordinary (mirror) neutrinos are in thermal equilibrium with the ordinary (mirror) plasma and their temperature is $T_{\nu} = T$ ($T_{\nu}' = T'$), while after decoupling it scales as $a^{-1}$.

In each sector, shortly after the $\nu$ decoupling, the $e^+$-$e^-$ annihilate because the temperature becomes lower than $2 m_e$, which is the threshold for the reaction $\gamma \leftrightarrow e^+e^-$ in both ordinary and mirror worlds.
Thus electrons and positrons transfer their entropy to the corresponding photons, which become hotter than neutrinos.

This fact will be used together with the entropy conservations (total and in each sector separately) to find the equations that govern the evolution of the mirror photon temperature $T'$ as a function of the ordinary one $T$.
In fact, ordinary and mirror neutrino decouplings are key events, together with both $e^+$-$e^-$ annihilation processes, for the thermodynamics in this range of temperatures.
Once we call $T_{D\nu'}$ the ordinary world temperature when the mirror neutrino decoupling takes place, we can split the early Universe evolution for temperatures $T \lesssim 100$ MeV into three phases.

\paragraph{Phase~ $T > T_{D\nu'}$.}

Photons, electrons, positrons and neutrinos are in thermal equilibrium in each sector separately, that is $T_{\nu} = T_e = T \; , \; T_{\nu}' = T_e' = T'$.\footnote{
For simplicity we write $T'_{\nu'} = T'_{\nu}$ , $T'_{e'} = T'_e$ , $T'_{\gamma'} = T'_{\gamma} = T'$ , $T_{\gamma} = T$.}
Using Eqs.~(\ref{energy_density})--(\ref{g_tot__def}) for particles in both sectors we are able to calculate only the d.o.f. number in ordinary or mirror worlds; but to work the {\it total} {\rm d.o.f.} number out, we need the mirror temperature $T'$ as a function of the ordinary one $T$ or vice versa.

As we neglect the entropy exchanges between the sectors (valid since there are only gravitational interactions between them), we can obtain both these functions using Eq.~(\ref{xSs}) and imposing $x=$const. in
\begin{eqnarray} \label{x_cons_1}
  x^3
 =\frac{s'_e + s'_{\gamma} + s'_{\nu}}{s_e + s_{\gamma} + s_{\nu}}
 =\frac{\left[ \dfrac{7}{8} q_e(T')+ q_{\gamma} + \dfrac{7}{8} q_{\nu} \right]T'^3} 
  {\left[ \dfrac{7}{8} q_e(T)+ q_{\gamma} + \dfrac{7}{8} q_{\nu} \right] T^3} \;,
\end{eqnarray}
where $q_{\nu} = 6$ and $q_{\gamma} = 2$, while $q_e (T)$ stands for
\begin{eqnarray} \label{}
  q_e (T) = \frac{8}{7} \; \frac{s_e(T)}{\dfrac{2\pi^2}{45} T^3} \;,
\end{eqnarray}
with $s_e(T)$ defined in (\ref{entropy_generic}), and we have used $T'_e = T'_{\nu} = T'_{\gamma} = T'$,  $T_e = T_{\nu} = T_{\gamma} = T$.
We have used also that, since particles and physics are the same in both sectors, the equilibrium distribution functions, and thus the spin-degeneracy factors, are the same:
$q_e = q_e'$, $q_{\gamma} = q_{\gamma}'$, $q_{\nu} = q_{\nu}'$.

Equation (\ref{x_cons_1}) can be solved numerically in order to obtain the function $T'(T)$ for every $T > T_{D\nu'}$.

\paragraph{Phase~ $T_{D\nu} < T \leq T_{D\nu'}$.}

\noindent At $T \simeq T_{D\nu'}$ mirror neutrinos decouple and right after mirror electrons and positrons annihilate, transferring their entropy only to the mirror photons, and hence raising their temperature.
In the mirror sector the entropies of the system (${e^\pm}'$,$\gamma'$) and of $\nu'$ are conserved separately.
Nevertheless, ordinary photons and neutrinos still have the same temperature $T$.
Therefore we have two equations. The first one comes from the conservation of entropies in the mirror sector, so that their ratio is equal to the asymptotic value computed at high temperatures, when the ${e^+}'$-${e^-}'$ annihilation process had not begun yet
\begin{eqnarray} \label{m_dec}
  \frac{S_e'+S_{\gamma}'}{S_{\nu}'}
 =\frac{s_e'+s_{\gamma}'}{s_{\nu}'}
 =\frac{\dfrac{7}{8}q_{e}(T')+q_{\gamma}}{\dfrac{7}{8}q_{\nu}}
  \left(\frac{T'}{T_{\nu}'}\right)^3
 ={\frac{\dfrac{7}{8} \cdot 4 + 2}{\dfrac{7}{8} \cdot 6}}
 =\frac{22}{21} \;,
\end{eqnarray}
and its solution gives $T_{\nu}'$ as a function of $T'$. The second one is the conservation of ratio of entropies in the two sectors
\begin{eqnarray} \label{x_cons_2}
  x^3
 =\frac{\left[ \dfrac{7}{8}q_e(T')+ q_{\gamma} \right] T'^3 
  + \dfrac{7}{8} q_{\nu} T_{\nu}'\,^3} {\left[  \dfrac{7}{8}q_e(T)+ q_{\gamma} 
  + \dfrac{7}{8} q_{\nu} \right] T^3} \;,
\end{eqnarray}
and its solution, obtained using Eq.~(\ref{m_dec}), gives $T'$ as a function of $T$.
Eq.~(\ref{x_cons_2}) is the same as Eq.~(\ref{x_cons_1}) but with $T_e' = T_{\gamma}' = T' \ne T_{\nu}'$.

\paragraph{Phase~ $T \leq T_{D\nu}$.}

\noindent At $T\simeq T_{D\nu}$ ordinary neutrinos decouple and right after ordinary electrons and positrons annihilate. Now also in the ordinary sector the entropies of the system ($e^\pm$,$\gamma$) and $\nu$ are conserved separately; therefore we need one more equation to calculate the ordinary neutrino temperature $T_{\nu}$ as a function of the ordinary photon one $T$
\begin{eqnarray} \label{}
  \frac{S_e+S_{\gamma}}{S_{\nu}}
 =\frac{\dfrac{7}{8}q_{e}(T)+q_{\gamma}}{\dfrac{7}{8}q_{\nu}}
  \left(\frac{T}{T_{\nu}}\right)^3 
 =\frac{22}{21} \;,
\end{eqnarray}
\begin{eqnarray} \label{}
  \frac{S_e'+S_{\gamma}'}{S_{\nu}'}
 =\frac{\dfrac{7}{8}q_{e}(T')+q_{\gamma}}{\dfrac{7}{8}q_{\nu}}
  \left(\frac{T'}{T_{\nu}'}\right)^3 
 =\frac{22}{21} \;,
\end{eqnarray}
\begin{eqnarray} \label{x_cons_3}
  x^3
 = \frac{\left[ \dfrac{7}{8}q_e(T')+ q_{\gamma} \right] T'^3 
 + \dfrac{7}{8} q_{\nu} T_{\nu}'\,^3}
 {\left[  \dfrac{7}{8}q_e(T)+ q_{\gamma} \right] T^3 
 + \dfrac{7}{8} q_{\nu} T_{\nu}^3} \;.
\end{eqnarray}
Eq.~(\ref{x_cons_3}) is the same as Eq.~(\ref{x_cons_2}) but with $T_e = T_{\gamma} = T \ne T_{\nu}$.

Once both ordinary and mirror photon temperatures are known, it is straightforward to calculate the total energy and entropy densities; then, reversing Eqs.~(\ref{entropy}) and (\ref{energy}), we can obtain the entropic $(q)$ and energetic $(g)$ number of d.o.f.
Calculations have been made for several different values of $x$, as reported in the following subsection.


\subsubsection{Numerical calculations}
\label{sec_Num_calc}

\begin{figure}[h]
\centerline{\psfig{file={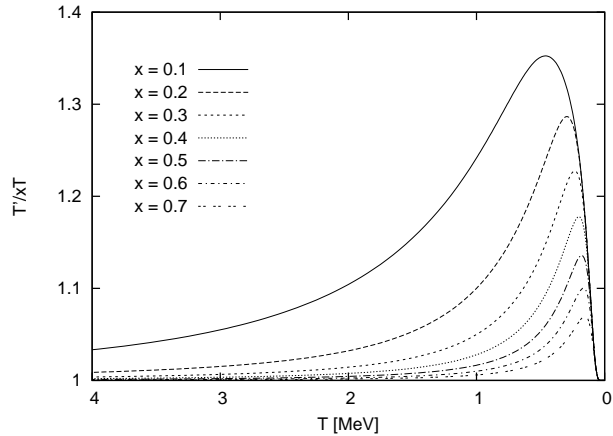},width=8.4cm}}
\caption{The ratio $T'/{xT}$ for several values of $x$.
The asymptotic values of this ratio are $1$ both for high and low temperatures, as expected.
\vspace{5mm} \label{fig_Tm_xT}}
\end{figure}

The equations we introduced above have been solved numerically.
Since $x$ is a free parameter in our theory, several values have been used for it --- from $0.1$ to $0.7$ with step $0.1$ or less.
In the extreme asymptotic cases $T \gg T_{D\nu'} \simeq T_{D\nu}/x$ or $T \ll T_{{\rm ann} \: e^{\pm}}\simeq 1$ MeV, we expect $q(T)\simeq q'[T'(T)]$; therefore, using Eq.~(\ref{xmir}), in these limits the ratio of mirror and ordinary photon temperatures should be $x$, that is $T'/{xT} \simeq 1$.
Instead, when $T_{D\nu'} \gtrsim T \gtrsim T_{{\rm ann} \: e^{\pm}}$ we expect $q'[T'(T)] \leq q(T)$ because the $e^+$-$e^-$ annihilation takes place before in the mirror world, leading to a decrease of $q'(T')$ and a corresponding increase of $T'$ in order to keep the ratio of entropy densities constant; thus we expect $T'/{xT} > 1$.
Later on, when the ordinary electrons and positrons annihilate, they make even $T$ increase and thus the ratio $T'/{xT}$ decreases to the asymptotic value 
$1$.
These remarks have been verified numerically; the ratio $T'/{xT}$ is plotted in Fig.~\ref{fig_Tm_xT} for different values of $x$.

Mirror $e^+$-$e^-$ annihilation happens at higher ordinary temperatures for lower $x$ values, raising before the temperature of mirror photons; this results in a shift of the peaks of Fig.~\ref{fig_Tm_xT} towards higher $T$.
In addition, for lower $x$ the difference in $T$ between the two annihilation processes is higher, so mirror photons have more time to raise their temperatures before the ordinary ones start to do the same.
This leads to the change of the shape of the curves.


\paragraph{Number of degrees of freedom.}
\label{sec_DoF_num}

Using Eqs.~ (\ref{energy_density})--(\ref{entropy_generic}) to reverse (\ref{entropy}) and (\ref{energy}), we can obtain the total number of ordinary entropic ($\bar q$) and energetic ($\bar g$) d.o.f. at any temperatures $T$.
We can apply the same procedure to calculate the {\it standard} values $q_{\rm std}$ and $g_{\rm std}$, as well as the mirror ones $\bar q'$ and $\bar g'$, and the influence of a sector on the other one.

To a first approximation, we expect $\bar q$ ($\bar g$) to have a cubic (quartic) dependence on $x$ when the temperature is not close to the $\nu$ decoupling and the $e^+$-$e^-$ annihilation phases: $\bar q = q_{\rm std} (1+x^3)$,  $\bar q' = q_{\rm std} (1+x^{-3})$, $\bar g = g_{\rm std} (1+x^4)$, $\bar g' = g_{\rm std} (1+x^{-4})$ (see Sec.~\ref{therm_approx}). 

In the continuation we present the results of accurate numerical calculations.
In Table~\ref{tab_DoF} some values are reported for special temperatures and several values of $x$.
As expected, the total d.o.f. numbers are always higher than the standard and increase with $x$.
Moreover, the mirror sector values are higher than the ordinary ones by a 
factor of order $x^{-3}$ (for $\bar q$) or $x^{-4}$ (for $\bar g$).

\begin{table}[h]
\tbl{Standard and non-standard total d.o.f. numbers for several $x$ values in both ordinary and mirror sectors. Temperatures are in MeV.}
{\begin{tabular}{lccccc} \toprule
$T $ & $0.005$ & $0.1$ & $0.5$ & $1$ & $5$ \\ \colrule
$q_{\rm std}$ & 3.91 & 4.78 & 10.0 & 10.6 & 10.75 \\
$g_{\rm std}$ & 3.36 & 4.30 & 10.0 & 10.6 & 10.75 \\
\multicolumn{6}{c}{$\mathbf{x = 0.1}$} \\
$T'$ & 0.0005 & 0.0107 & 0.0675 & 0.124 & 0.511 \\
$\bar q$ & 3.91 & 4.78 & 10.0 & 10.6 & 10.75 \\
$\bar q'$ & 3913 & 3913 & 4072 & 5522 & 10070 \\
$\bar g$ & 3.36 & 4.30 & 10.0 & 10.6 & 10.75 \\
$\bar g'$ & 33629 & 32894 & 30032 & 44430 & 98436 \\
\multicolumn{6}{c}{$\mathbf{x = 0.3}$} \\
$T'$ & 0.0015 & 0.0321 & 0.170 & 0.315 & 1.50 \\
$\bar q$ & 4.015 & 4.91 & 10.3 & 10.8 & 11.0 \\
$\bar q'$ & 149 & 149 & 261 & 347 & 406 \\
$\bar g$ & 3.39 & 4.34 & 10.1 & 10.6 & 10.8 \\
$\bar g'$ & 418.5 & 409 & 750 & 1082 & 1325 \\
\multicolumn{6}{c}{$\mathbf{x = 0.5}$} \\
$T'$ & 0.0025 & 0.0533 & 0.263 & 0.508 & 2.50 \\
$\bar q$ & 4.40 & 5.38 & 11.3 & 11.9 & 12.1 \\
$\bar q'$ & 35.2 & 35.5 & 77.3 & 90.5 & 96.5 \\
$\bar g$ & 3.57 & 4.58 & 10.7 & 11.2 & 11.4 \\
$\bar g'$ & 57.2 & 56.6 & 138 & 168 & 182 \\
\multicolumn{6}{c}{$\mathbf{x = 0.7}$} \\
$T'$ & 0.0035 & 0.0733 & 0.357 & 0.704 & 3.50 \\
$\bar q$ & 5.25 & 6.42 & 13.5 & 14.2 & 14.4 \\
$\bar q'$ & 15.3 & 16.3 & 36.9 & 40.6 & 42.0 \\
$\bar g$ & 4.17 & 5.35 & 12.4 & 13.1 & 13.3 \\
$\bar g'$ & 17.4 & 18.5 & 47.8 & 53.3 & 55.5 \\ \botrule
\end{tabular} \label{tab_DoF}}
\end{table}

\begin{figure}[pt]
\centerline{\psfig{file={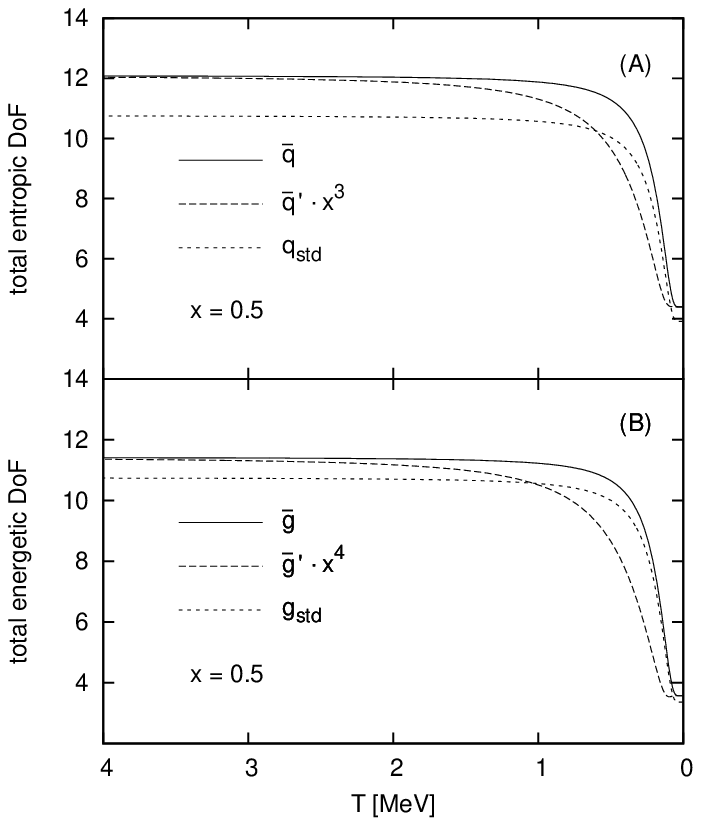},width=12.2cm}}
\caption{Total entropic (panel A) and energetic (panel B) degrees of freedom computed in ordinary and mirror sectors and for the standard. The mirror values have been multiplied by $x^3$ ($q$) and $x^4$ ($g$) to make them comparable with the ordinary ones, since $(T'/xT) \sim 1$. \label{fig_DoF_tot_x_05}}
\centerline{\psfig{file={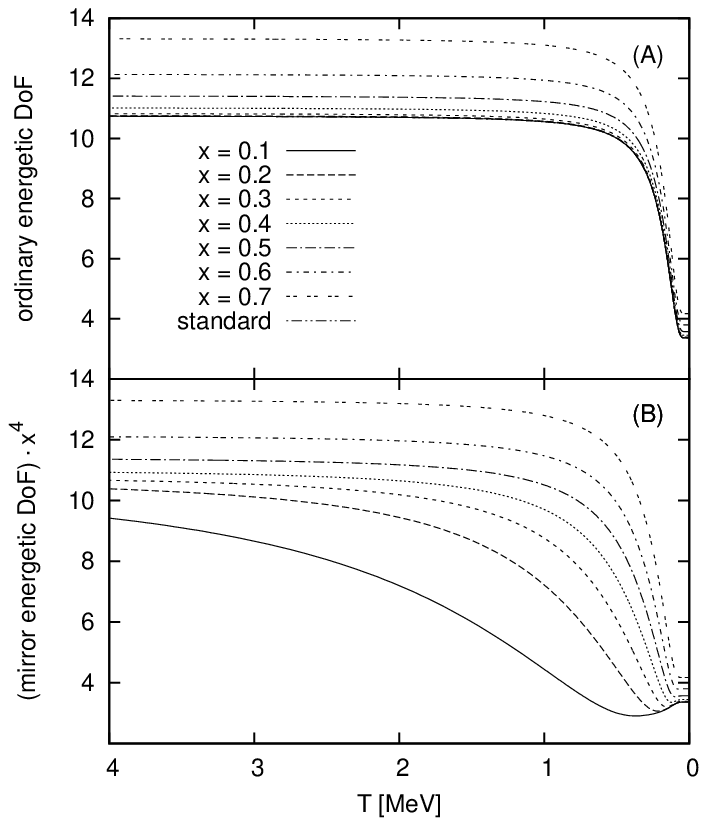},width=12.2cm}}
\caption{Total energetic degrees of freedom in the ordinary (panel A) and mirror (panel B) sectors computed for several values of $x$. In panel (A) the standard is shown for comparison. In panel (B) the mirror values have been multiplied by $x^4$ to make them comparable with the ordinary ones, since $(T'/xT) \sim 1$. \label{fig_ord_mir_DoF}}
\end{figure}

In Fig.~\ref{fig_DoF_tot_x_05}(A) $q_{\rm std}$, $\bar q$, $\bar q'$ are plotted for the intermediate value $x=0.5$; panel (B) of the same figure shows the corresponding values of $g$.
In the figure $\bar q'$ ($\bar g'$) has been scaled by a factor $x^3 $ ($x^4$); in this way the asymptotic values are the same than the ordinary sector ones because at the extremes $T'/T=x$ (see Fig.~\ref{fig_Tm_xT}).
We can see that $\bar q$, $\bar g$, $\bar q'$ and $\bar g'$ have similar trends, but, due to $T'<T$, the d.o.f. number in the mirror sector begins to decrease before (at higher $T$).

In Fig.~\ref{fig_ord_mir_DoF} the values of $\bar g$ in ordinary and mirror sectors are plotted in comparison with the standard for several values of $x$ (from $0.1$ to $0.7$ with step $0.1$).
As in Fig.~\ref{fig_DoF_tot_x_05}, the mirror values $\bar g'$ have been multiplied by $x^4$.
The predicted quartic dependence of $g$ on $x$ at the extremes is proved correct.
In panel (A) we note that ordinary d.o.f. with $x<0.3$ are practically identical to the standard case.
In addition, the plot shape does not change with $x$ in the ordinary sector, while it does in the mirror one.
This sector evolves with temperature $T' \sim xT < T$; therefore, for lower $x$ the number of d.o.f. begins to decrease below the asymptotic value at high $T$ before than for high $x$.
The change of the shape of the plots in panel (B) is related to the same physical processes responsible for the analogous effect present in Fig.~\ref{fig_Tm_xT}, that is due to the $e^+$-$e^-$ annihilation in the mirror sector.


\paragraph{Number of neutrino families.}
\label{sec_Nu_num}

\begin{figure}[pt]
\centerline{\psfig{file={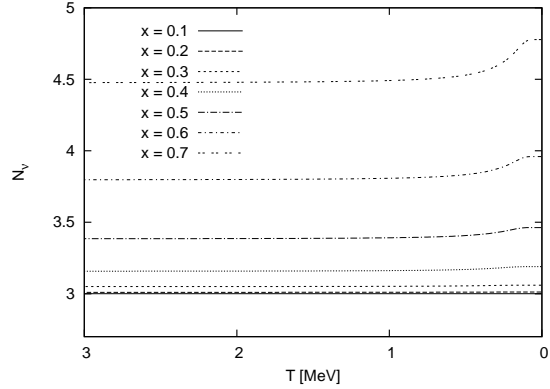},width=7.5cm}}
\caption{Effective number of neutrino families $N_{\nu}$ in the ordinary sector for several values of $x$. \label{fig_Nnu}}
\end{figure}

We know that the SM contains three neutrino species; the possible existence of a fourth neutrino has been investigated for a long time, also using BBN constraints.
This is why in literature one can often find bounds on the number of d.o.f. in terms of {\it effective} extra-neutrino number $\Delta N_{\nu} = N_{\nu} - 3$.
In general, the effective number of neutrinos $N_{\nu}$ is found assuming that all particles contributing to the Universe energy density, except electrons, positrons and photons, are neutrinos; in formula that means
\begin{eqnarray} \label{Nnu_conversion0}
  \bar g (T) = g_e(T) + g_{\gamma} +
    \frac{7}{8}\cdot 2 N_{\nu}\cdot\left(\frac{T_{\nu}}{T}\right)^4 \;,
\end{eqnarray}
which implies
\begin{eqnarray} \label{Nnu_conversion}
  N_{\nu} =
    \frac{\bar g (T) - g_e(T) - g_{\gamma}}{\dfrac{7}{8}\cdot 2}
    \cdot \left( \frac{T}{T_{\nu}} \right)^4 \;.
\end{eqnarray}
$N_{\nu}$ has been worked out using Eq.~(\ref{Nnu_conversion}) together with the results of previous numerical simulations; some data are reported in Table~\ref{tab_Nnu}, while plots for several $x$ values and temperatures from $0$ to $3$ MeV are shown in Fig.~\ref{fig_Nnu}.
We stress that the standard value $N_{\nu} = 3$ is the same at any temperatures, while {\it a distinctive feature of mirror scenario is that the number of neutrinos raises for decreasing temperatures}.
Anyway, this effect is not a problem; on the contrary it may be useful since recent cosmological data fits give indications for a number of neutrinos at recent times higher than at BBN epoch.
In Ref.~\refcite{Mangano:2006ur} the authors found $N_\nu = 5.2^{+2.7}_{-2.2}$ using recent CMB and LSS data.
We can use the aforementioned mirror feature together with the results of a previous work on CMB and LSS power spectra,\cite{Ciarcelluti:2003wm,Ciarcelluti:2004ip} where the author studied the dependence of the spectra on the parameters $x$ and $N_{\nu}$ for a flat Universe with mirror dark matter.
As can be easily evinced from Figs. 11, 12, 14, 15 of Ref.~\refcite{Ciarcelluti:2004ip}, an increase of the effective number of neutrinos is well mimicked by an increase of the parameter $x$, and the amounts of the respective increases are in accordance with what is required to justify the data of CMB and LSS.
Thus, considering the sum of these effects, i.e. the raise in $N_{\nu}$ before and after BBN, and the similarity of CMB and LSS spectra of mirror models and standard $N_{\nu}$ with the ones obtained without mirror sector but with larger $N_{\nu}$, the mentioned discrepancy naturally disappears.
Similarly, the effective number of neutrinos in the mirror sector can be worked out as
\begin{eqnarray}
\label{Nnu_conversion_mir}
  N'_{\nu} =
    \frac{\bar g' - g'_e(T') - g'_{\gamma}}{\dfrac{7}{8}\cdot 2} 
    \cdot \left( \frac{T'}{T'_{\nu}} \right)^4 \;.
\end{eqnarray}
Once again these values are higher than the ordinary ones, but now by a factor $x^{-4}$; they have been computed numerically and some special values are given in Table~\ref{tab_Nnu_mir}. For lower $x$ this number can become very high, inducing relevant consequences on the primordial nucleosynthesis in the mirror sector, that is highly dependent on this parameter.

\begin{table}[pt]
\tbl{Effective number of neutrinos in the ordinary sector for some special cases. Temperatures are in MeV.}
{\begin{tabular}{@{}lcccc@{}} \toprule
\hphantom{0}$T$ & $x=0.1$ & $x=0.3$ & $x=0.5$ & $x=0.7$ \\ \colrule
\multicolumn{5}{c}{ordinary sector} \\
\hphantom{0}0.005 & 3.00074 & 3.05997 & 3.46270 & 4.77751 \\
\hphantom{0}0.1\hphantom{00} & 3.00074 & 3.05997 & 3.46244 & 4.76829 \\
\hphantom{0}0.5\hphantom{00} & 3.00074 & 3.05997 & 3.40706 & 4.52942 \\
\hphantom{0}1\hphantom{0000} & 3.00071 & 3.05202 & 3.39166 & 4.49133 \\
\hphantom{0}5\hphantom{0000} & 3.00063 & 3.04989 & 3.38430 & 4.47563 \\ \botrule
\end{tabular} \label{tab_Nnu}}
\end{table}

\begin{table}[pb]
\tbl{Effective number of neutrinos in the mirror sector for some special cases. Temperatures are in MeV.}
{\begin{tabular}{@{}lcccc@{}} \toprule
\hphantom{0}$T'$ & $x=0.1$ & $x=0.3$ & $x=0.5$ & $x=0.7$ \\ \colrule
\multicolumn{5}{c}{mirror sector} \\
\hphantom{0}0.005 & 74011 & 917.0 & 121.4 & 33.83 \\
\hphantom{0}0.1\hphantom{00} & 62007 & 805.0 & 111.6 & 32.21 \\
\hphantom{0}0.5\hphantom{00}  & 61447 & 763.1 & 101.9 & 28.86 \\
\hphantom{0}1\hphantom{0000}  & 61435 & 761.8 & 101.4 & 28.66 \\
\hphantom{0}5\hphantom{0000}  & 61432 & 761.4 & 101.3 & 28.59 \\ \botrule
\end{tabular} \label{tab_Nnu_mir}}
\end{table}


\paragraph{Predictions for special temperatures.}

It is possible to obtain the asymptotic values of $N_{\nu}$ at $T \gg T_{D\nu}$ and $T \ll T_{{\rm ann} \, e^{\pm}}$ in a simple way starting from the standard values
\begin{eqnarray}
\label{g_extremes_std}
  g_{\rm std}(T \gg T_{D\nu}) = 10.75 \;,
\cr\cr
  g_{\rm std}(T \ll T_{{\rm ann} \, e^{\pm}}) \simeq 3.36 \;.
\end{eqnarray}
Without the mirror sector, we have, as expected
\begin{eqnarray} \label{Nnu_Delta_Nnu_standard}
  N_{\nu} (T \gg T_{D\nu})  &=& \frac{10.75 - 2 - \dfrac{7}{8} \cdot 4}
    {{\dfrac{7}{8}\cdot 2}} = 3 \;, \nonumber\\[1mm]
  N_{\nu} (T \ll T_{{\rm ann} \, e^{\pm}}) &=& \frac{3.36 - 2}
    {{\dfrac{7}{8}\cdot 2}} 
    \cdot \left( \frac{11}{4} \right)^\frac{4}{3} = 3 \;.
\end{eqnarray}
Instead, when the mirror sector is present, we can use the previously mentioned quadratic approximation $\bar g = g_{\rm std} (1+x^4)$ at the special temperatures we are considering, and hence
\begin{eqnarray} \label{Nnu_asynt}
  N_{\nu} (T \gg T_{D\nu})
    &=&\frac{10.75(1+x^4)-2-\dfrac{7}{8}\cdot 4}{{\dfrac{7}{8}\cdot 2}} = 3 + 6.14~x^4 \;, \nonumber\\[1mm]
  N_{\nu} (T \ll T_{{\rm ann} \, e^{\pm}}) 
    &=&\frac{3.36 (1+x^4)- 2}{{\dfrac{7}{8}\cdot 2}} \cdot
       \left( \frac{11}{4} \right)^\frac{4}{3} \simeq 3 + 7.40~x^4\;.
\end{eqnarray}
From the above equations we can see that the increase $\Delta N_{\nu}$ is
\begin{eqnarray} \label{N_nu_mir_approx}
  \Delta N_{\nu} 
    &=& N_{\nu} (T \ll T_{{\rm ann} \, e^{\pm}}) - N_{\nu} (T \gg T_{D\nu}) = \nonumber \\[1mm]
    &=& x^4 \cdot \frac{1}{\dfrac{7}{8}\cdot 2} \left[ 3.36 
      \left( \frac{11}{4} \right)^\frac{4}{3} - 10.75 \right] 
    \simeq 1.255 \cdot x^4 \;.
\end{eqnarray}
This leads to $N_{\nu}$ higher than the standard in the presence of the mirror world and to a further growth of this parameter at low temperatures.
If we assume a conservative limit on the effective number of neutrino families $N_{\nu} (T \gg T_{D\nu}) \le 4$, this implies $x \le 0.64$ and $N_{\nu} (T \ll T_{{\rm ann} \, e^{\pm}}) \le 4.26$.
For some interesting values of the parameter $x$ we obtain
\begin{eqnarray} \label{}
  x = 0.7 \hspace{0.4cm} &\Rightarrow& \hspace{0.4cm}
    N_{\nu}(T \gg T_{D\nu}) \simeq 4.5
    \hspace{0.4cm} \mathrm{and} \hspace{0.4cm}
    N_{\nu} (T \ll T_{{\rm ann} \, e^{\pm}}) \simeq 4.8 \;, \nonumber\\[1mm]
  x = 0.6 \hspace{0.4cm} &\Rightarrow& \hspace{0.4cm}
    N_{\nu} (T \gg T_{D\nu}) \simeq 3.8
    \hspace{0.4cm} \mathrm{and} \hspace{0.4cm}
    N_{\nu} (T \ll T_{{\rm ann} \, e^{\pm}}) \simeq 3.96 \;, \nonumber\\[1mm]
  x = 0.3 \hspace{0.4cm} &\Rightarrow& \hspace{0.4cm}
    N_{\nu} (T \gg T_{D\nu}) \simeq 3.05
    \hspace{0.23cm} \mathrm{and} \hspace{0.4cm}
    N_{\nu} (T \ll T_{{\rm ann} \, e^{\pm}}) \simeq 3.06 \;.
\end{eqnarray}
These rough estimates are in agreement with the asymptotic numerical values computed before and after BBN; nevertheless, the previous detailed analysis of the evolution of this quantity is crucial for studies of the nucleosynthesis in both sectors.


\section{Big Bang nucleosynthesis}
\label{sec-bbn}

The BBN provides one of the most stringent tests for physics beyond the Standard Model.
Presently it has only one free parameter, the baryon to photon ratio $\eta = n_b / n_\gamma$, also called baryonic asymmetry.
Since microphysics is the same, there should be a similar primordial nucleosynthesis also for mirror particles, but, as anticipated in the previous section, the different initial conditions for the mirror sector have a big influence on this process.
For mirror BBN there are two parameters: the mirror baryon to photon ratio $\eta' = n'_b / n'_\gamma$ and the ratio of temperatures $x$, that enters via the degrees of freedom, which determines the approximate Hubble expansion rate at a given temperature.
Using Eqs.~(\ref{Hubble}) and (\ref{gm-ast}) we obtain 
\begin{eqnarray} \label{Hubble-m}
H(T') &=& 1.66 \sqrt{g'(T')(1+x^{-4})} \frac{T'^2}{M_{\rm Pl}} \;.
\end{eqnarray}
Since $n'_\gamma = x^3 n_\gamma$, using the definition $\beta = \Omega'_{b} / \Omega_{b} = n'_{b} / n_{b}$ (the second equality is given to the fact that the mass of mirror baryons is the same as the ordinary ones), we obtain $\eta' = \beta x^{-3} \eta$.
With our assumptions (\ref{beta-bounds}) and (\ref{x-bound}) on $x$ and $\beta$, the inequality $\eta' > \eta$ is always verified.


\subsection{A simple analytical model for He$'$ abundance}
\label{bbn-he1}

It is useful first to recall the standard BBN. The relevant time scales are: the freeze out temperature of weak interactions, $T_W \approx 0.8$ MeV, and the ``deuterium bottleneck'' temperature, $T_N \simeq 0.07$ MeV ($t_N \sim 200$ s).
For $T$ above $T_W$ the weak interactions are rapid enough to keep the neutron-proton ratio in chemical equilibrium.
The neutron abundance,\footnote{
The mass fraction or abundance of a certain nuclear species is defined as
	\begin{eqnarray}
	\label{}
	X_A \equiv \frac{n_A A}{n_b} 
	\hspace{0.2cm} , \hspace{0.5cm} \sum_i X_i = 1 
	\end{eqnarray}
where $A$ is the atomic number, $n_A$ the number density, and $n_b$ the baryonic number density of the Universe.
Usually the abundance of $^4$He is alternatively refered as $Y=Y_4=X_{\rm He}$, so that for $^4$He$'$ we may use $Y'=Y'_4=X_{\rm He'}$.}
$X_n = n_n/n_b$, is then given by $X_n(T) = [1 + \exp(\Delta m/T)]^{-1}$, where $\Delta m \simeq 1.29$ MeV is the neutron-proton mass difference.
For $T < T_W$ the weak reaction rate $\Gamma_W \simeq G_F^2 T^5$ drops below the Hubble expansion rate $H(T) \simeq 5.5 \: T^2/M_{\rm Pl}$, the neutron abundance freezes out at the equilibrium value $X_{n}(T_{W})$ and it then evolves only due to the neutron decay: $X_{n}(t)=X_{n}(T_{W})\exp(-t/\tau)$, where $\tau=886.7$ s is the neutron lifetime. 
At temperatures $T > T_{N}$, the process $p+n \leftrightarrow d+\gamma$ is faster than the expansion of the Universe, and free nucleons and deuterium are in chemical equilibrium.  
The light element nucleosynthesis essentially begins when the system cools down to the temperature 
\begin{equation} \label{T_N}
T_{N} \simeq \frac{B_d}{-\ln(\eta)+1.5\ln\left(\dfrac{m_N}{T_N}\right)} \simeq 0.07 ~ {\rm MeV} \;, 
\end{equation}
where $B_{d}=2.22$ MeV is the deuterium binding energy, and $m_{N}$ is the nucleon mass. 
Thus, assuming that all neutrons end up in $^4$He, the primordial $^{4}$He mass fraction is
\begin{equation} \label{helium}
Y \simeq \frac{4 \left(\dfrac{n_n}{2}\right)}{n_n + n_p} 
  = 2X_{n}(t_N,T_W) = 
  \frac{2\exp\left(-\dfrac{t_N}{\tau}\right)}{1+\exp\left(\dfrac{\Delta m}{T_W}\right)} 
  \simeq 0.25 \;.
\end{equation}

As discussed, the presence of the mirror sector with a temperature $T' < T$ has almost no impact on the standard BBN in the limit $x < 0.64$, which in fact has been set by uncertainties of the present observational situation. 
In the mirror sector nucleosynthesis proceeds along the same lines. 
However, the impact of the ordinary world for the mirror BBN is dramatic!

Therefore, comparing the Hubble expansion rate $H(T')$ in Eq.~(\ref{Hubble-m}) with the weak reaction rate $\Gamma_W (T') \simeq G_F^2 T'^5$, we find a freeze-out temperature 
\begin{equation} \label{}
T'_W=(1+x^{-4})^{1/6} T_W \;, 
\end{equation}
which is larger than $T_{W}$, whereas the time scales as 
\begin{equation} \label{}
t'_W = \frac{t_W}{(1+x^{-4})^{5/6}} < t_W
\end{equation}
(obtained using Eq.~(\ref{Hubble-m}) and the relation $t \propto H^{-1}$). 
In addition, $\eta'$ is different from $\eta\simeq 5 \times 10^{-10}$. 
However, since $T_N$ depends on baryon density only logarithmically (see Eq.~(\ref{T_N})), the temperature $T'_N$ remains essentially the same as $T_N$, while the time $t'_N$ scales as 
\begin{equation} \label{}
t'_N = \frac{t_N}{(1+x^{-4})^{1/2}} \;. 
\end{equation}
Thus, for the mirror $^4$He mass fraction we obtain
\begin{equation} \label{m_helium}
Y' \simeq 2X'_n(t'_N) = \frac
  { 2\exp\left[- \dfrac{ t_N }{ \tau(1+x^{-4})^{1/2} }\right] }
  { 1+\exp\left[\dfrac{ \Delta m }{ T_W(1+x^{-4})^{1/6} } \right] } \;. 
\end{equation}
We see that $Y'$ is an increasing function of $x^{-1}$ and is always bigger than $Y$.  
In particular, for $x\rightarrow 0$ one has $Y' \rightarrow 1$.\footnote{
In reality, Eq.~(\ref{m_helium}) is not valid for small $x$, since in this case deuterium production through the reaction $n+p\leftrightarrow d+\gamma$ can become ineffective. 
It is possible to calculate that for $x < 0.3 \cdot (\eta'\cdot 10^{10})^{-1/2}$, the rate at which neutrons are captured to form the deuterium nuclei becomes smaller than the Hubble rate for temperatures $T' > T'_N$. 
In this case mirror nucleosynthesis is inhibited, because the neutron capture processes become ineffective before deuterium abundance grows enough to initiate the synthesis of the heavier elements.\cite{Berezhiani:2000gw}
We have to remark, however, that with our assumption $\beta 
> 1$, the condition $x < 0.3 \cdot (\eta'\cdot 10^{10})^{-1/2}$ is never fulfilled and the behaviour of $Y'_4$ is well approximated by Eq.~(\ref{m_helium}). }
Thus, we have reached the important conclusion that, if the dark matter of the Universe is represented by the baryons of the mirror sector, it should contain considerably bigger fraction of primordial $^4$He than the ordinary world. 
In particular, the helium fraction of mirror matter is between 20\% and 80-90\%, depending on the values of $x$ and $\eta '$.


\subsection{An accurate numerical model}

If we want a complete and accurate study of the primordial production of both ordinary and mirror elements, we need to numerically solve the equations governing the evolution of nuclides and nuclei.
This study is required, since the primordial chemical composition is an initial condition of the following evolution of the Universe.

As we have seen, the presence of the mirror sector can be parametrized in terms of extra d.o.f. number or extra neutrino families; therefore, since the physical processes involved in BBN are the same in both sectors, it is possible to use and modify a standard numerical code to work out the light elements production.
The choice is the Kawano code for BBN\cite{Kawano:1992ua,Wagoner:1972jh} since it is a well-tested and fast program and its accuracy is large enough for our purposes.
For the neutron lifetime we consider the value $\tau = 885.7$ s.

The number of d.o.f. enters the program in terms of neutrino species number; this quantity is a free parameter, but instead of using the same number during the whole BBN process, we use as input the variable $N_{\nu} (T, x)$ numerically computed following the procedure described in the previous section.
The only parameter of the mirror sector which affects ordinary BBN is $x$; the baryonic ratio $\beta$ does not induce any changes on the production of ordinary nuclides, but it plays a crucial role for the mirror nuclides production.

In Table~\ref{tab-bbn-ord} I report the final abundances (mass fractions) of the elements produced in the ordinary sector at the end of BBN process (at $T \sim 8\cdot 10^{-4}$ MeV) for a final baryon to photon ratio $\eta = 6.14 \cdot 10^{-10}$ and for several $x$ values, and compare them with a standard scenario.
We can easily infer that for $x < 0.3$ the light element abundances do not change more than a few percent, and the difference between the standard and $x=0.1$ is of order $10^{-4}$ or less.
For $x > 0.3$ the differences become more important.

\begin{table}[h]
\tbl{Elements produced in the ordinary sector. The last row includes all elements with atomic number larger than 7.}
{\begin{tabular}{@{}lccccc@{}} \toprule
  & {standard}
  & $ x=0.1 $
  & $ x=0.3 $
  & $ x=0.5 $
  & $ x=0.7 $   \\ \colrule
$n{\rm /H} \: (10^{-16})$ & 1.161\hphantom{0} & 1.161\hphantom{0} & 1.159\hphantom{0} & 1.505\hphantom{0} & 2.044\hphantom{0} \\
$p$\hphantom{000000} & 0.7518 & 0.7518 & 0.7511 & 0.7463 & 0.7326 \\
${\rm D/H} \: (10^{-5})$ & 2.554\hphantom{0} & 2.555\hphantom{0} & 2.575\hphantom{0} & 2.709\hphantom{0} & 3.144\hphantom{0} \\
${\rm T/H} \: (10^{-8})$ & 8.064\hphantom{0} & 8.065\hphantom{0} & 8.132\hphantom{0} & 8.588\hphantom{0} & 10.07\hphantom{000} \\
${\rm ^3He/H} \: (10^{-5})$ & 1.038\hphantom{0} & 1.038\hphantom{0} & 1.041\hphantom{0} & 1.058\hphantom{0} & 1.113\hphantom{0} \\
${\rm ^4He}$\hphantom{0000000} & 0.2483 & 0.2483 & 0.2491 & 0.2538 & 0.2675 \\
${\rm ^6Li/H} \: (10^{-14})$ & 1.111\hphantom{0} & 1.111\hphantom{0} & 1.124\hphantom{0} & 1.210\hphantom{0} & 1.499\hphantom{0} \\
${\rm ^7Li/H} \: (10^{-10})$ & 4.549\hphantom{0} & 4.548\hphantom{0} & 4.523\hphantom{0} & 4.356\hphantom{0} & 3.871\hphantom{0} \\
${\rm ^7 Be/H} \: (10^{-10})$ & 4.266\hphantom{0} & 4.266\hphantom{0} & 4.238\hphantom{0} & 4.051\hphantom{0} & 3.502\hphantom{0} \\
${\rm ^8Li + /H} \: (10^{-15})$ & 1.242\hphantom{0} & 1.242\hphantom{0} & 1.251\hphantom{0} & 1.306\hphantom{0} & 1.464\hphantom{0} \\ \botrule
\end{tabular} \label{tab-bbn-ord}}
\end{table}

Between the different possible non-standard BBN scenarios, it would be of relevance a more careful inspection at the effects of the presence of mirror particles on the predicted abundances of light elements.
A more extended investigation in this sense is required in order to evaluate if it could help solving the still present problems on standard BBN due to the lithium anomalies.

Even mirror baryons undergo nucleosynthesis via the same physical processes than the ordinary ones, thus we can use the same numerical code also for the mirror nucleosynthesis.
Mirror BBN is affected also by the second mirror parameter, that is the mirror baryon density (introduced in terms of the ratio $\beta = \Omega'_b / \Omega_b \sim 1 \div 5$), which raises the baryon to photon ratio $\eta' = \beta x^{-3} \eta$.

The results are reported in Table~\ref{tab-bbn-mir}, which is the analogous of Table~\ref{tab-bbn-ord} but for a mirror sector with $\beta$ = 5 and $\beta$ = 1.
We can see that BBN in the mirror sector is much more different from the standard than BBN in the ordinary sector. 
This is a consequence of the high ordinary contribution to the number of total mirror d.o.f., which scales as $\sim x^{-4}$ (while in the ordinary sector the mirror contribution is almost insignificant, since it scales as $\sim x^4$).
As we expect, at higher $x$ the mirror abundances are closer to the ordinary ones, because the mirror sector becomes hotter and thus more similar to the ordinary one (since $T' \sim xT$). 
Moreover, we note that in general lower $\beta$ implies final mass fractions closer to the standard, again as we may expect, since for $\beta=1$ the two sectors have the same baryonic densities.
Hence, numerical simulations confirm the previous simple analytical model: the mirror helium abundance should be much larger than that of the ordinary helium, and for $x <0.5$ the mirror helium gives a dominant mass fraction of the dark matter of the Universe.
In addition, mirror BBN produces much larger abundances of so-called metals (elements heavier than lithium), that have a large influence on the opacity of mirror matter, which has an important role in several astrophysical processes, as for example mirror star formation.
The interesting feature that mirror sector can be a helium dominated world has important consequences on mirror star formation and evolution,\cite{Berezhiani:2005vv} and other related astrophysical aspects.
In particular, I recall the reader's attention to the fact that the predicted dark matter composition, dominated by mirror helium, in combination with results on mirror stellar evolution,\cite{Berezhiani:2005vv} support the proposed interpretation of the DAMA/LIBRA annual modulation signal in terms of mirror dark matter.\cite{Foot:2008nw,Ciarcelluti:2008qk,Ciarcelluti:2010dm}

\begin{table}[h]
\tbl{Elements produced in the mirror sector.}
{\begin{tabular}{@{}lllll@{}} \toprule
  & $ x=0.1\: (\beta = 5) $
  & $ x=0.3\: (\beta = 5) $ 
  & $ x=0.5\: (\beta = 5) $
  & $ x=0.7\: (\beta = 5) $
\\ \colrule
$n{\rm /H} $     &\ 5.762 $\cdot 10^{-25}$ &\ 2.590 $\cdot 10^{-22}$ &\ 1.840 $\cdot 10^{-20}$ &\ 1.726 $\cdot 10^{-19}$ \\
$p $       &\ 0.1735                 &\ 0.3646                 &\ 0.4966                 &\ 0.5924                 \\
${\rm D/H} $     &\ 1.003 $\cdot 10^{-12}$ &\ 4.838 $\cdot 10^{-9}$  &\ 6.587 $\cdot 10^{-8}$  &\ 3.279 $\cdot 10^{-7}$  \\
${\rm T/H}$      &\ 9.679 $\cdot 10^{-21}$ &\ 1.238 $\cdot 10^{-13}$ &\ 2.108 $\cdot 10^{-11}$ &\ 3.722 $\cdot 10^{-10}$ \\
${\rm ^3He/H}$   &\ 3.282 $\cdot 10^{-6}$  &\ 3.740 $\cdot 10^{-6}$  &\ 4.172 $\cdot 10^{-6}$  &\ 4.691 $\cdot 10^{-6}$  \\
${\rm ^4He} $    &\ 0.8051                 &\ 0.6351                 &\ 0.5035                 &\ 0.4077                 \\
${\rm ^6Li/H}$   &\ 7.478 $\cdot 10^{-21}$ &\ 1.309 $\cdot 10^{-17}$ &\ 1.016 $\cdot 10^{-16}$ &\ 3.361 $\cdot 10^{-16}$ \\
${\rm ^7Li/H} $  &\ 1.996 $\cdot 10^{-7}$  &\ 3.720 $\cdot 10^{-8}$  &\ 1.535 $\cdot 10^{-8}$  &\ 7.962 $\cdot 10^{-9}$  \\
${\rm ^7 Be/H}$  &\ 1.996 $\cdot 10^{-7}$  &\ 3.720 $\cdot 10^{-8}$  &\ 1.535 $\cdot 10^{-8}$  &\ 7.962 $\cdot 10^{-9}$  \\
${\rm ^8Li +/H}$ &\ 4.354 $\cdot 10^{-9}$  &\ 5.926 $\cdot 10^{-11}$ &\ 3.827 $\cdot 10^{-12}$ &\ 3.949 $\cdot 10^{-13}$ \\ \colrule
  & $ x=0.1\: (\beta = 1) $
  & $ x=0.3\: (\beta = 1) $ 
  & $ x=0.5\: (\beta = 1) $
  & $ x=0.7\: (\beta = 1) $
\\ \colrule
$n/H $     &\ 8.888 $\cdot 10^{-17}$ &\ 1.915 $\cdot 10^{-16}$ &\ 2.058 $\cdot 10^{-16}$ &\ 2.076 $\cdot 10^{-16}$ \\
$p $       &\ 0.1772                 &\ 0.3675                 &\ 0.5028                 &\ 0.6017                 \\
$D/H $     &\ 1.331 $\cdot 10^{-6}$  &\ 7.094 $\cdot 10^{-6}$  &\ 1.352 $\cdot 10^{-5}$  &\ 2.235 $\cdot 10^{-5}$  \\
$T/H$      &\ 3.068 $\cdot 10^{-9}$  &\ 2.190 $\cdot 10^{-8}$  &\ 4.358 $\cdot 10^{-8}$  &\ 7.328 $\cdot 10^{-8}$  \\
$^3He/H$   &\ 5.228 $\cdot 10^{-6}$  &\ 6.880 $\cdot 10^{-6}$  &\ 8.232 $\cdot 10^{-6}$  &\ 9.719 $\cdot 10^{-6}$  \\
$^4He $    &\ 0.8226                 &\ 0.6326                 &\ 0.4974                 &\ 0.3984                 \\
$^6Li/H$   &\ 8.638 $\cdot 10^{-15}$ &\ 1.660 $\cdot 10^{-14}$ &\ 1.790 $\cdot 10^{-14}$ &\ 1.951 $\cdot 10^{-14}$ \\
$^7Li/H $  &\ 5.712 $\cdot 10^{-8}$  &\ 8.930 $\cdot 10^{-9}$  &\ 2.948 $\cdot 10^{-9}$  &\ 1.120 $\cdot 10^{-9}$  \\
$^7 Be/H$  &\ 5.711 $\cdot 10^{-8}$  &\ 8.878 $\cdot 10^{-9}$  &\ 2.891 $\cdot 10^{-9}$  &\ 1.064 $\cdot 10^{-9}$  \\
$^8Li +/H$ &\ 2.036 $\cdot 10^{-10}$ &\ 2.514 $\cdot 10^{-12}$ &\ 1.657 $\cdot 10^{-13}$ &\ 1.814 $\cdot 10^{-14}$ \\ \botrule
\end{tabular} \label{tab-bbn-mir}}
\end{table}


\subsection{BBN with photon--mirror photon kinetic mixing}

So far we have considered just gravitational effects of the existence of mirror matter. 
Now we study the implications of photon--mirror photon mixing for the early Universe.
In particular, we will check that this kinetic mixing is consistent with constraints from ordinary Big Bang nucleosynthesis as well as more stringent constraints from cosmic microwave background and large scale structure considerations.
We will then estimate, under some simple and plausible assumptions, the primordial He$'$ mass fraction.
Besides studies of mirror star formation and evolution, and the formation of dark matter structures at non-linear scales, this quantity is also important since the mirror dark matter interpretation of the direct detection experiments depends on it.
In fact, the velocity dispersion of the particles in the mirror matter halo depends on the particular particle species and on the abundance of mirror helium $Y'$, and satisfies\cite{Foot:2010th,Ciarcelluti:2010dm}
\begin{eqnarray}
v_0^2 (i) 
          = v_{\rm rot}^2 \frac{m_p}{m_i}\frac{1}{2 - \dfrac{5}{4}Y'} \:,
\end{eqnarray}
where the index $i$ labels the particle type ($i=e', {\rm H}', {\rm He}', {\rm O}', {\rm Fe}'...$), $v_{\rm rot} \approx 254$ km/s is the local rotational velocity for our galaxy, and $m_p$ is the proton mass.
While the DAMA experiments turn out to be relatively insensitive to $Y'$, other experiments, such as those using electron scattering, exhibit a greater sensitivity to $Y'$, and might ultimately be able to measure this parameter.\cite{Foot:2010hu,Foot:2009gk}


\subsubsection{Thermal evolution I: an approximate model}

In the mirror dark matter scenario, it is assumed there is a temperature asymmetry ($T' < T$) between the ordinary and mirror radiation sectors in the early Universe due to some physics at early times (for specific models, see Refs.~\refcite{Kolb:1985bf,Berezhiani:2000gw,Berezinsky:1999az,Berezhiani:1995am,Hodges:1993yb}).
This is required in order to explain ordinary BBN, which suggests that $T'/T \lesssim 0.7$.
In addition, analyses\cite{Ignatiev:2003js}\cdash\cite{Ciarcelluti:2004ip} based on numerical simulations of CMB and LSS suggest $T'/T \lesssim 0.3$ (as we will see in next sections).
However, if photon--mirror photon kinetic mixing exists, it can potentially thermally populate the mirror sector.
For example, Carlson and Glashow\cite{Carlson:1987si} derived the approximate bound $\epsilon \lesssim 3 \times 10^{-8}$ from requiring that the mirror sector does not come into thermal equilibrium with the ordinary sector before BBN.
We expect the kinetic mixing to populate the mirror sector, but with $T'<T$.
Assuming an effective initial condition $T' \ll T$, we can estimate the evolution of $T'/T$ in the early Universe as a function of $\epsilon$, and thereby check the compatibility of the theory with the BBN and CMB/LSS constraints on $T'/T$. 

Photon--mirror photon kinetic mixing can populate the mirror sector in the early Universe, via the process $e^+ e^- \to {e^+}' {e^-}'$.
The cross section for $e^+ e^- \to {e^+}' {e^-}'$ is
\begin{eqnarray}
\sigma = {\frac{4\pi \alpha^2 \epsilon^2}{3s}} \:,
\label{sigma}
\end{eqnarray}
where $s\simeq \langle 2E \rangle^2$ is the Mandelstam $s$ quantity.
This implies a production rate for $e'$ and a generation of energy density in the mirror sector
\begin{eqnarray}
\left(\frac{d\rho'}{dt}\right)_{\rm generation} = {\frac{dn_{e'}}{dt}} \langle E \rangle 
\simeq 2\,n_{e^+}\sigma n_{e^-} \cdot 3.15~T \:,
\label{drhodt}
\end{eqnarray}
where $n_{e'}$ includes both ${e^+}'$ and ${e^-}'$, and $\langle E \rangle \simeq 3.15~T$ for a Fermi-Dirac distribution.

The ${e^+}',{e^-}'$ will interact with each other via mirror weak and mirror electromagnetic interactions, populating the $\gamma', \nu_e', \nu_\mu', \nu_\tau'$, and thermalizing to a common mirror sector temperature $T'$.
The energy density in the mirror sector is then expressed by the second of Eqs.~(\ref{rho}), 
with $g' = 43/4=10.75$ for $1 \lesssim T'(\rm MeV) \lesssim 100$.
Thus, when $g'$ is constant, we can write 
\begin{eqnarray}
{\frac{d \rho'}{d t}} \simeq {\frac{\pi^2}{30}}~g'~{\frac{d T'^4}{d t}} \;.
\label{drho}
\end{eqnarray}
Using the expression (\ref{drhodt}) for the rate of mirror energy density generation, and taking the ratio $T'/T$ in order to cancel the time dependence due to the expansion of the Universe (redshift), we have
\begin{eqnarray}
{\frac{d \left(\dfrac{T'}{T}\right)^4}{d t}} \simeq
{\frac{20}{3.15~\pi}}~{\frac{1}{g'}}~\alpha^2 \epsilon^2 n_{e^+} n_{e^-} 
T^{-5} \simeq A~\epsilon^2~T \;,
\label{dT4dt}
\end{eqnarray}
with $A \simeq (1.91 / \pi^5) \alpha^2 \simeq 3.32 \cdot 10^{-7}$, and where in the last equality we used the expression for densities of relativistic fermions in thermal equilibrium\cite{Kolb:1990vq}
\begin{eqnarray}
n_{e^+} = n_{e^-} = {\frac{3}{2}}{\frac{\zeta(3)}{\pi^2}}~T^3 \;.
\end{eqnarray}

\noindent In radiation dominated epoch we can express the time as\cite{Kolb:1990vq}
\begin{eqnarray} \label{timeTemp}
t = 0.301 \: g^{-1/2} {\frac{M_{\rm Pl}}{T^2}} \;,
\end{eqnarray}
where $M_{\rm Pl} \simeq 1.22 \times 10^{22}$ MeV is the Planck mass.
For $g = 10.75$, Eq.~(\ref{timeTemp}) implies
\begin{eqnarray}
{\frac{dT}{dt}}
= -{\frac{1}{0.6}}~g^{1/2}\,M_{\rm Pl}^{-1}\,T^3
\simeq -5.46~{\frac{T^3}{M_{\rm Pl}}}~.
\label{dTdt}
\end{eqnarray}
Using together Eqs.~(\ref{dT4dt}) and (\ref{dTdt}), we obtain
\begin{eqnarray}
\frac{d \left(\dfrac{T'}{T}\right)^4}{dT}
\simeq -{\frac{A}{5.46}} M_{\rm Pl} ~ \epsilon^2 {\frac{1}{T^2}}
\simeq -B \, \epsilon^2 {\frac{1}{T^2}} \;,
\label{dT4dT}
\end{eqnarray}
where $B=(A / 5.46) M_{\rm Pl} \simeq 0.74 \cdot 10^{15}$ MeV.
Finally, integrating Eq.~(\ref{dT4dT}), with the initial condition $T' = 0$ at $T = T_i$, we derive the expression for the $T'/T$ ratio as a decreasing function of $T$
\begin{eqnarray}
{\frac{T'}{T}} 
&\simeq& 0.52 \cdot 10^4 \,\epsilon^{1/2} \left({\frac{1}{T(\rm MeV)}}
- {\frac{1}{T_i(\rm MeV)}} \right)^{1/4} \simeq \nonumber \\[1mm]
&\simeq& 0.164 ~\epsilon_{-9}^{1/2} \left({\frac{1}{T(\rm MeV)}} \right)^{1/4} \;.
\label{T'overT}
\end{eqnarray}
The last equivalence is obtained considering $T_i \gg 1$ MeV and defining $\epsilon_{-9} = \epsilon /10^{-9}$.
Clearly in all these computations we have neglected the change of the $T'/T$ ratio due to the $e^+$-$e^-$ annihilation processes in both sectors (as described in detail in Sec.~\ref{thermuniv}), and the transfer from the mirror sector to the ordinary one (since $T' < T$).

Note that the $T'/T$ ratio freezes out when $T \lesssim 2\,m_e$ since the number density of $e^{\pm}$ becomes Boltzmann suppressed and the process $e^+ e^- \to {e^+}' {e^-}'$ can no longer effectively heat the mirror sector.
In addition, after ${e^+}'$-${e^-}'$ annihilation the effective number of mirror degrees of freedom decreases, and thus we estimate
\begin{eqnarray}
\frac{T'}{T} \simeq 0.2~\epsilon^{1/2}_{-9}\
~~~~{\rm for} ~~\ T \lesssim 1 \,{\rm MeV} \;.
\label{new4}
\end{eqnarray}
We plot this expression in Fig.~\ref{mirbbn1} together with the related He$'$ mass fraction (see Sec.~\ref{prim-mir-He}).
Constraints from ordinary BBN suggest that $\delta N_{\nu} \lesssim 0.5 \Rightarrow T'/T < 0.6$.
A more stringent constraint arises from CMB and LSS, which suggest\cite{Ignatiev:2003js}\cdash\cite{Ciarcelluti:2004ip} $T'/T \lesssim 0.3$ and implies, from Eq.~(\ref{new4}), that $\epsilon \lesssim 3\times 10^{-9}$.
In the limit $T'/T \rightarrow 1$ we find the mentioned old upper bound obtained in Ref.~\refcite{Carlson:1987si}.
However, this is an extrapolation beyond the validity of approximations, since, when $T'$ approaches $T$, the transfer of energy from the ordinary to the mirror sector clearly becomes more and more inefficient.
Nevertheless, far from this limit, this simple analysis shows that a photon--mirror photon kinetic mixing of strength $\epsilon \sim 10^{-9}$ (that is able to explain the DAMA/LIBRA annual modulation signal consistently with the results of the other direct detection experiments) is consistent with constraints from ordinary BBN and CMB/LSS data.


\subsubsection{Thermal evolution II: a more accurate model}
\label{thermevol2}

A more precise result for the thermal evolution of the mirror sector can be obtained using a more accurate model for the transfer of energy from the ordinary to the mirror sector, as in Ref.~\refcite{Ciarcelluti:2008qk}.
In this case it is not possible to solve analytically, but we need to compute a numerical solution.

The generation of energy density in the mirror sector due to the process $e^+ e^- \to {e^+}' {e^-}'$ is now
\begin{eqnarray}
{\frac{\partial \rho'}{\partial t}} =
n_{e^+} n_{e^-} \langle \sigma v_{\rm M\o l} {\cal E} \rangle \;,
\end{eqnarray}
where ${\cal E}$ is the energy transferred in the process, $v_{\rm M\o l}$ is the M\o ller velocity (see e.g. Ref.~\refcite{Gondolo:1990dk}), and again $n_{e^-} \simeq n_{e^+} \simeq (3\zeta(3) / 2\pi^2) T^3$.

It is useful to consider the quantity $\rho'/\rho$, in order to cancel the time dependence due to the expansion of the Universe.
Using expressions~(\ref{rho}) and the time-temperature relation~(\ref{timeTemp}) with $g = 10.75$, we find that
\begin{eqnarray}
\frac{d\left(\dfrac{\rho'}{\rho}\right)}{dT} = \frac{-n_{e^-}n_{e^+} \langle \sigma
v_{\rm M\o l} {\cal E} \rangle}
{\dfrac{\pi^2 g T^4}{30}} \ \frac{0.6 M_{\rm Pl}}{\sqrt{g}T^3} \ .
\label{drhomrhodt1}
\end{eqnarray}

Let us focus on $\langle \sigma v_{\rm M\o l} {\cal E} \rangle$.
This quantity is
\begin{equation}
\langle \sigma v_{\rm M\o l} {\cal E} \rangle =
\frac{
  \displaystyle\int \sigma v_{\rm M\o l} (E_1 + E_2)
  {\frac{1}{1 + e^{E_1/T}}}
  {\frac{1}{1 + e^{E_2/T}}} \mathrm{d}^3 p_1 \, \mathrm{d}^3 p_2 }{
  \displaystyle\int {\dfrac{1}{1 + e^{E_1/T}}}
  {\dfrac{1}{1 + e^{E_2/T}}} \mathrm{d}^3 p_1 \, \mathrm{d}^3 p_2 } \;,
\end{equation}
where we have neglected Pauli blocking effects.
If one makes the simplifying assumption of using Maxwellian statistics instead of Fermi-Dirac statistics then one can show (see the Appendix of Ref.~\refcite{Ciarcelluti:2008qk}) that in the massless electron limit
\begin{equation}
\langle \sigma v_{\rm M\o l} {\cal E} \rangle = {\frac{2\pi \alpha^2 \epsilon^2}{3T}} \;,
\label{ms}
\end{equation}
and Eq.~(\ref{drhomrhodt1}) reduces to
\begin{eqnarray}
{\frac{d\left(\dfrac{\rho'}{\rho}\right)}{dT}} = {\frac{-A}{T^2}} \;,
\label{rho2}
\end{eqnarray}
where
\begin{eqnarray}
A = \frac{27\zeta(3)^2 \alpha^2 \epsilon^2 M_{\rm Pl}}{\pi^5 g\sqrt{g}} \;.
\end{eqnarray}
This is essentially the same as Eq.~(\ref{dT4dT}) obtained in the previous approximation.
Note that the ${e^{\pm}}'$ will thermalize with $\gamma'$.
However, because most of the ${e^{\pm}}'$ are produced in the low $T' \lesssim$ 5 MeV region, mirror weak interactions are too weak to significantly populate the $\nu'_{e,\mu,\tau}$ (i.e. one can easily verify {\it a posteriori} that the evolution of $T'/T$ for the parameter space of interest is such that $G_F^2T'^5 \ll (\sqrt{g}T^2 / 0.3 M_{\rm Pl})$.
Thus to a good approximation the radiation content of the mirror sector consists of ${e^{\pm}}', \gamma'$ leading to $g' = 11/2$ and hence $\rho'/\rho = (g'/g)(T'^4/T^4)$, with $g'/g \approx 22/43$.
 
Assuming the initial condition $T' = 0$ at $T = T_i$, Eq.~(\ref{rho2}) has the analytic solution
\begin{eqnarray}
\frac{T'}{T} = \left(\frac{g}{g'}A\right)^{1/4} \left[ \frac{1}{T} - \frac{1}{T_i}\right]^{1/4} \;.
\label{TmT2}
\end{eqnarray}

\begin{figure}[pb]
\centerline{\psfig{file={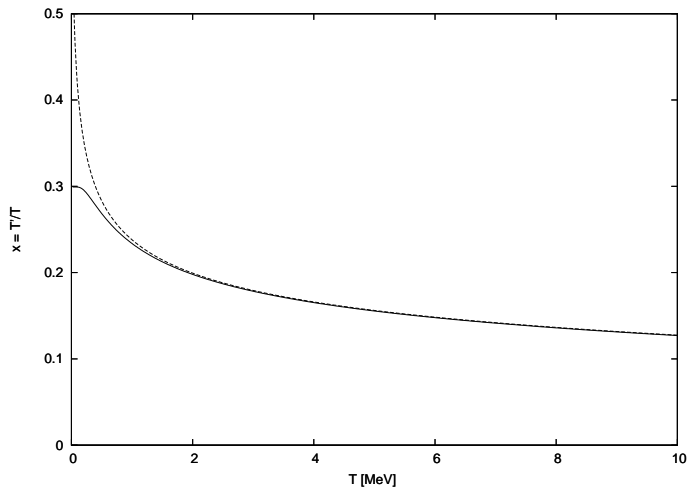},width=6.5cm,angle=270}}
\vspace*{8pt}
\caption{Evolution of $x = T'/T$ for $\epsilon = 8.5 \times 10^{-10}$.
The solid line is the numerical solution including the effects of the electron mass, while the dashed line is the analytic result obtained using Eq.~(\ref{TmT2}), which holds in the massless electron limit.
As expected the two solutions agree in the $T \gtrsim 1 $ MeV region, where the effects of the electron mass should be negligible. \label{TmTepsi2}}
\end{figure}

Let us now include the effects of the electron mass.
With non-zero electron mass, the evolution of $T'/T$ cannot be solved analytically, but Eq.~(\ref{drhomrhodt1}) can be solved numerically.
Note that the number density is
\begin{eqnarray}
n_{e^-} = \frac{1}{\pi^2} \int^{\infty}_{m_e} \frac{ \sqrt{E^2 - m_e^2} E}
{1 + \exp(E/T)}  \ dE
\label{bla}
\end{eqnarray}
and, as discussed in the Appendix of Ref.~\refcite{Ciarcelluti:2008qk},
\begin{eqnarray}
\langle \sigma v_{\rm M\o l} {\cal E} \rangle = &&
\frac{1}{8m^4_e T^2 K_2^2 (m_e/T)} \cdot \nonumber \\
&& \cdot \int_{4m_e^2}^{\infty} ds \sigma (s -
4m_e^2) \sqrt{s} \int_{\sqrt{s}}^{\infty} dE_+ e^{-E_+/T} E_+ \sqrt{\frac{E_+^2}{s} - 1} \;,
\label{bla2}
\end{eqnarray}
where the cross section is
\begin{eqnarray}
\sigma = \frac{4\pi}{3} \alpha^2 \epsilon^2 \frac{1}{s^3} (s + 2m_e^2)^2
\ .
\end{eqnarray}
Numerically solving Eq.~(\ref{drhomrhodt1}) with the above inputs (i.e. numerically solving the integrals Eq.~(\ref{bla}) and Eq.~(\ref{bla2}) at each temperature step), we find that\footnote{
For simplicity we have neglected the effect of heating of the photons via $e^{+}$-$e^-$ annihilations. 
Note that the same effect occurs for the mirror photons which are heated by the annihilations of ${e^+}' {e^-}'$, so that $x_f$ is approximately unchanged by this effect.}
\begin{eqnarray}
\epsilon  \simeq 8.5\times 10^{-10}  \left( \frac{x_f}{0.3} \right)^2 \;,
\label{df3}
\end{eqnarray}
where $x_f$ is the final value ($T \to 0$) of $x = T'/T$.
In Fig.~\ref{TmTepsi2}, we plot the evolution of $T'/T$, for $\epsilon = 8.5 \times 10^{-10}$, considering both the numerical solution including the effects of the electron mass and the analytic result obtained using Eq.~(\ref{TmT2}), which holds in the massless electron limit.
As expected, the two solutions agree in the $T \lesssim 1 $ MeV region, where the effects of the electron mass should be negligible.

In deriving this result we have made several simplifying approximations. The most significant of these are the following:
a) Using Maxwellian statistics instead of Fermi-Dirac statistics to simplify the estimate of $\langle \sigma v_{\rm M\o l} {\cal E} \rangle $. 
Using Fermi-Dirac statistics should decrease the interaction rate by around 8\%.
b) We have neglected Pauli blocking effects. 
Including Pauli blocking effects will slightly reduce the interaction rate since some of the ${e^{\pm}}'$ states are filled thereby reducing the available phase space. 
We estimate that the effect of the reduction of the interaction rate due to Pauli blocking will be around $\lesssim$ 10\%.
c) We have assumed that negligible $\nu'_{e,\mu,\tau}$ are produced via mirror weak interactions from the ${e^{\pm}}'$.
Production of $\nu'_{e,\mu,\tau}$ will slightly decrease the $T'/T$ ratio.
The effect of this is equivalent to reducing the interaction rate by around $\lesssim$ 10\%.
Taking these effects into account, we revise Eq.~(\ref{df3}) to
\begin{eqnarray}
\epsilon  = (1.0\pm 0.10)\times 10^{-9}  \left( \frac{x_f}{0.3} \right)^2 \ .
\label{result}
\end{eqnarray}
Since successful large scale structure studies\cite{Ignatiev:2003js}\cdash\cite{Ciarcelluti:2004ip} suggest a rough bound $x_f \lesssim 0.3$, then our result, Eq.~(\ref{result}), suggests a corresponding rough bound $\epsilon \lesssim 10^{-9}$.
This upper bound is similar to what previously find with the simpler approximation of Eq.~(\ref{new4}), and again confirm that the current proposed interpretation of direct detection experiments in terms of mirror dark matter is compatible with constraints coming from early Universe analysis.


\subsubsection{Primordial mirror Helium}
\label{prim-mir-He}

Let us now consider the implications of kinetic mixing for mirror BBN.
It has already been discussed in previous sections that, compared with the ordinary matter sector, we expect a larger mirror helium mass fraction if $T' < T$, as currently required.
Essentially, this is because the expansion rate of the Universe is faster at earlier times, which implies that the freeze out temperature of mirror weak interactions will be higher than that in the ordinary sector.
Given our above calculations of the $T'/T$ evolution, we can estimate the mirror helium mass fraction as a function of $\epsilon$ within our model, which we now discuss.

We follow the same procedure used in Sec.~\ref{bbn-he1}, recalling that in the mirror sector we have the same relation (\ref{helium}) than in the ordinary one, except that we change $t_N \rightarrow t'_N$ and $T_W \rightarrow T_W'$.
This is possible since $T_N$ depends only logarithmically on the baryon to photon ratio, 
so that we still have $T_N' \sim T_N$, even though the mirror baryon to photon ratio is much larger than the ordinary one.
We obtain the freeze out temperature from the equality $\Gamma_W(T_W') = H(T_W')$, where $\Gamma$ and $H$ are expressed as follows
\begin{eqnarray}
\Gamma_W(T') &\simeq& G_F^2 \,T'^5 \label{gamm'} \;, \\
H(T') &\simeq&
1.66~g^{1/2}\left[1+\left( \frac{T'}{T} \right)^{-4}
\right]^{1/2}{\frac{T'^2}{M_{\rm Pl}}} \simeq \nonumber \\[1mm]
&\simeq& 1.66 ~g^{1/2}\left(1+B^{-4/3}\,\epsilon^{-8/3}\,
T'^{4/3}\right)^{1/2}{\frac{T'^2}{M_{\rm Pl}}} \simeq \nonumber \\[1mm]
&\simeq& {\frac{1.66 ~g^{1/2}\,B^{-2/3}}{M_{\rm Pl}}}\,\epsilon^{-4/3}\,
{T'}^{8/3} \;.
\label{Ht'}
\end{eqnarray}
For Eq.~(\ref{Ht'}) we used Eq.~(\ref{T'overT}) with  $B=0.74 \cdot 10^{15}$ MeV and considered that $B^{-4/3}\,\epsilon^{-8/3}\,T'^{4/3}$ is much larger than 1 for $\epsilon$ values of interest.
Imposing the equality of expressions (\ref{gamm'}) and (\ref {Ht'}) we obtain
\begin{eqnarray}
{\frac{\Gamma_W(T_W')}{H(T_W')}}
&\simeq& {\frac{G_F^2 ~M_{\rm Pl}}{1.66 ~g^{1/2}}} \,B^{2/3}
\,\epsilon^{4/3} \,{T_W'}^{7/3} = 1 \;, \\[1mm]
T_W' &\simeq& B^{-2/7} \,\epsilon^{-4/7} \,T_W^{9/7} \;,
\label{TW'}
\end{eqnarray}
where for the second expression we used $T_W \simeq \left(1.66 ~g^{1/2} / G_F^2 ~M_{\rm Pl}\right)^{1/3}$, as computed from the equation $\Gamma_W(T_W) = H(T_W)$.

The quantity $t_N'$ can be estimated straightforwardly
\begin{eqnarray}
t_N'
\simeq 0.3 ~g^{-1/2}{\left[1+\left(\frac{T'}{T}\right)^{-4}\right]^{-1/2}}{\frac{M_{\rm Pl}}{T_N^2}}
\simeq B^{1/2} \epsilon \,T_N^{-1/2}\,t_N \;,
\label{tN'}
\end{eqnarray}
where we used $t_N \simeq 0.3 ~g^{-1/2} M_{\rm Pl} / T_N^2$ and $B^{-1}\epsilon^{-2}T_N \gg 1$.

Finally, using Eqs.~(\ref{TW'}) and (\ref{tN'}), we can estimate the mirror helium mass fraction
\begin{eqnarray}
Y' \simeq 2\,X_n' (T_W', t_N')
\simeq {\frac{2\,\exp(-t_N'/\tau)}{1+\exp(\Delta m/T_W')}}
\simeq {\frac{2\,\exp\left[{-\dfrac{B^{1/2} ~T_N^{-1/2}}{\tau}} t_N ~\epsilon \right]}
{1+\exp\left[{\dfrac{\Delta m~B^{2/7}}{T_W^{9/7}}}~\epsilon^{4/7}\right]}} \;.
\label{YHe'}
\end{eqnarray}
Using the typical values $T_W \simeq 0.8$ MeV, $T_N \simeq 0.07$ MeV, $\Delta m \simeq 1.29$ MeV, $B \simeq 0.74 \cdot 10^{15}$ MeV, $\tau \simeq 886.7$ s, $t_N \simeq 200$ s, we obtain
\begin{eqnarray}
Y' \simeq \frac{2\,\exp\left[-2.3 \cdot 10^7 ~\epsilon \right]}
{1+\exp\left[0.3 \cdot 10^5 ~\epsilon^{4/7} \right]}
\simeq \frac{2\,\exp\left[-2.3 \cdot 10^{-2} ~\epsilon_{-9} \right]}
{1+\exp\left[0.22 ~\epsilon_{-9}^{4/7} \right]} \;.
\end{eqnarray}
Thus, for $\epsilon \simeq 10^{-9}$, we find $Y' \simeq 0.87$.

\begin{figure}[pt]
\centerline{\psfig{file={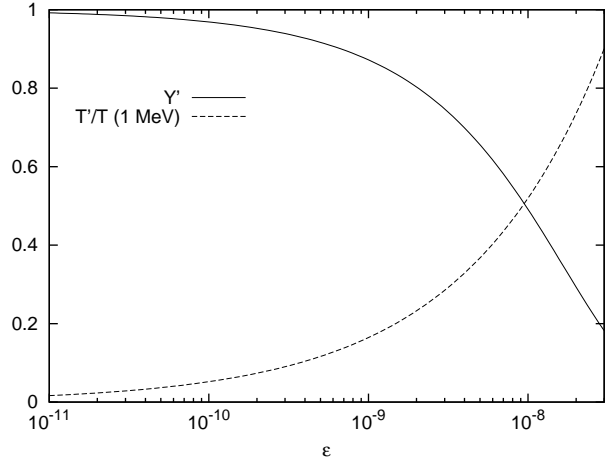},width=9.0cm}}
\caption{The mirror helium mass fraction versus the photon--mirror photon kinetic mixing strength $\epsilon$.
The ratio of mirror and ordinary temperatures at $T=1$ MeV is also plotted. \label{mirbbn1}}
\end{figure}

In Fig.~\ref{mirbbn1} we plot the predicted value of $Y'$, together with the ratio $T'/T$ at 1 MeV, as a function of $\epsilon$, in the interesting region around $\epsilon \sim 10^{-9}$.
We see that a decrease of the strength of the photon--mirror photon kinetic mixing induces a lower mirror temperature, and thus a larger primordial mirror helium abundance (since mirror BBN happens at a larger cosmic expansion rate).

A more precise computation of the primordial mirror helium abundance can be done using the calculation of the $T'/T$ evolution in Sec.~\ref{thermevol2}, and numerically solving the equations for the rates of reactions governing the equilibrium between protons and neutrons.

We assume the same initial condition, $T'\ll T$, as in Sec.~\ref{thermevol2}, with a thermal evolution driven by the photon--mirror photon kinetic mixing, which populates the mirror radiation just with ${e^{\pm}}'$ and $\gamma'$.
We may describe the evolution of the ratio of the temperatures in the two sectors with the approximate analytical expression (\ref{TmT2}), which is valid for $T' \gtrsim 1$ MeV and for $T \lesssim 100$ MeV, where we multiply $A$ by the factor $\omega \approx 0.8$ in order to take into account the reduction in the rate from our approximations (Maxwellian
statistics instead of Fermi-Dirac, neglecting Pauli-blocking factors, massless electrons).
Assuming $T_i \gg 100$ MeV, then Eq.~(\ref{TmT2}) reduces to
\begin{eqnarray}
\frac{T'}{T} \simeq \frac{0.25}{(T/{\rm MeV})^{1/4}} \sqrt{\frac{\epsilon}{10^{-9}}} \;.
\label{ana2}
\end{eqnarray}

In our model the mirror sector starts with a temperature much lower than the ordinary one, and later the photon--mirror photon kinetic mixing increases only the temperature of mirror electron-positrons and photons, since neutrinos are decoupled.
We may thus assume $T_{\nu'} \ll T'$, where $T' = T_{\gamma'} \simeq T_{e'}$,
which is a reasonable approximation for the $\epsilon$ values of interest.
In this scenario the only reactions we need to consider to compute the He$'$ abundance $Y$ are 
\begin{eqnarray}
n' + {e^+}' \to p' + \bar \nu' ~~~~~~ {\rm and} ~~~~~~ p' + {e^-}' \to n' + \nu' \;.
\label{reactions}
\end{eqnarray}
We may neglect the neutron decay $n' \to p' + {e^-}' + \bar \nu'$, since the process is much slower than the time for primordial mirror nucleosynthesis, that we estimate to happen in the first few seconds of the Universe.
The reaction rates of the processes (\ref{reactions}) can be adapted from the standard relations present in Weinberg's book,\cite{Weinberg:GravAndCosm} in which we neglect the Pauli blocking effect on neutrinos because $T_{\nu'} \ll T'$
\begin{eqnarray}
&&\lambda_{n'\rightarrow p'} = \lambda(n'+{e^+}' \to p' + \bar \nu') = B\int_0^\infty E_{\nu'}^2 p_{e'}^2 
dp_{e'} [e^{E_{e'}/T'} +
1]^{-1}  \;,
\nonumber \\[1.5mm]
&&\lambda_{p'\rightarrow n'} = \lambda(p'+{e^-}' \to n' + \nu') = B\int_{(Q^2-m_e^2)^{1/2}}^\infty E_{\nu'}^2 p_{e'}^2 dp_{e'} 
[e^{E_{e'}/T'} + 1]^{-1} \;, ~~~~
\label{1}
\end{eqnarray}
where 
\begin{eqnarray}
B = \frac{G_{\rm wk}^2 (1+3g_A^2) \cos^2\theta_C}{2 \pi^3 \hbar} \;,
\end{eqnarray}
$G_{\rm wk} = 1.16637 \times 10^{-5} \ {\rm GeV^{-2}}$ is the weak coupling constant, measured from the rate of the decay process $\mu^+ \rightarrow e^+ + \nu_e + \bar\nu_\mu$, $g_{\rm A} = 1.257$ is the axial vector coupling of beta decay, measured from the rate of neutron decay, and $\theta_{\rm C}$ is the Cabibbo angle, with $\cos \theta_{\rm C} = 0.9745$, measured from the rate of $^{14}$O beta decay and other $0^+ \rightarrow 0^+$ transitions, and $E_{e'}(p_{e'})=(p_{e'}^2+m_{e}^2)^{1/2}$, $E_{\nu'}(p_{\nu'})=p_{\nu'}$.
For $n'+{e^+}' \to p' + \bar \nu', \ E_{\nu'} - E_{e'} = Q$,
while for $p' + {e^-}' \to n' + \nu', \ E_{e'} - E_{\nu'} = Q$,
where $Q \equiv m_n - m_p = 1.29$ MeV.
The extremals of integrals in Eqs.~(\ref{1}) are fixed considering that integrations are taken over all allowed positive values of $p_{e'}$.

We may rewrite Eqs.~(\ref{1}) by substituting $q=-E_{\nu'}$ for $\lambda_{n'\rightarrow p'}$ and $q=E_{\nu'}$ for $\lambda_{p'\rightarrow n'}$, thus obtaining
\begin{eqnarray}
\lambda_{n'\rightarrow p'} = 
&&B \int_{-\infty}^{-m_e-Q} q^2 (q+Q)^2 \left[1 - \frac{m_e^2}{(q+Q)^2}\right]^{1/2} \;
[e^{-(q+Q)/T'} + 1]^{-1} \: dq \;,
\nonumber \\[1.5mm]
\lambda_{p'\rightarrow n'} =
&&B \int_0^\infty q^2 (q+Q)^2 \left[1 - \frac{m_e^2}{(q+Q)^2}\right]^{1/2} \;
[e^{(q+Q)/T'} + 1]^{-1} \: dq \;.
\label{12}
\end{eqnarray}

The differential equation for the ratio $X_{n'}$ of neutrons to nucleons is
\begin{eqnarray}
\frac{dX_{n'}}{dt} = \lambda_{p'\rightarrow n'} (1 - X_{n'}) -
\lambda_{n'\rightarrow p'} X_{n'} \;.
\label{eq:Xn}
\end{eqnarray}
Note that $Y' \simeq 2X_{n'}$ since, as mentioned before, we can neglect $n'$ decay, and thus all available mirror neutrons go into He$'$.

\begin{figure}[pt]
\centerline{\psfig{file={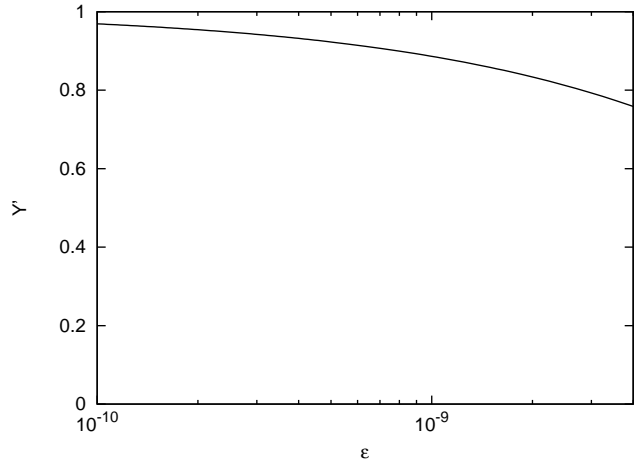},width=9.0cm}}
\caption{Mass abundance of primordial mirror helium He$'$ versus the strength of the photon--mirror-photon kinetic mixing ($\epsilon$). 
\label{fig:He-epsi}}
\end{figure}

We have solved the above equations numerically, using the usual time-temperature relation for radiation dominated epoch (\ref{timeTemp}),
where $g$ takes into account only the degrees of freedom of ordinary particles, since the contribution of mirror particles is negligible given the initial condition $T' \ll T$.
We used the initial condition $X_{n'}(0)=0.5$ in Eq.~(\ref{eq:Xn}), and followed the evolution until $X_{n'}$ reaches the asymptotic value.
The results are shown in Fig.~\ref{fig:He-epsi}, where we plot the obtained mass fraction of mirror helium versus the strength of the photon--mirror photon kinetic mixing ($\epsilon$), for the parameter range of interest for cosmology.
As expected, we obtain high values of the primordial He$'$ mass fraction, with $ Y' \gtrsim 0.8$ for $\epsilon \lesssim 3 \times 10^{-9}$.
For the preferred value emerging from the analysis of the DAMA signal, $\epsilon \simeq 10^{-9}$, we obtain $ Y'\simeq 0.9$, which means that the dark matter is largely mirror helium dominated.


\subsubsection{Heavier elements}

We can make a rough estimate of the primordial mass fraction of mirror elements of carbon mass and heavier. 
These are produced essentially via three-body interactions, the most important of which is the triple alpha process in the mirror sector
\begin{eqnarray}
^{4}{\rm He'} + \:^{4}{\rm He'} + \:^{4}{\rm He'} \to \:^{12}{\rm C'} + \gamma' \;. 
\end{eqnarray}
The rate for this process can be obtained from the rate for the corresponding process in the ordinary matter sector,\cite{Wagoner:1966pv} and is given by
\begin{eqnarray}
\frac{dX_{\rm C'}}{dt} = 
  && \frac{1}{32} [Y_{\rm He'}(t)]^3 \cdot 1.2\times 10^{-11} \rho_{b'}^2 \: T'^{-3} \cdot \nonumber \\[1mm]
  && \cdot [\exp(-0.37 \: T'^{-1}) + 30.3 \exp(-2.4 \: T'^{-1})] \;,
\end{eqnarray}
where $X_{\rm C'}$ is the mass fraction of C$'$, $T'$ is in MeV units, $\rho_{b'}$ is the mirror baryon density in g/cm$^3$ and the rate is in sec$^{-1}$.
He$'$ cannot be produced in significant proportion until $T' \lesssim 1$ MeV, which, from Eq.~(\ref{ana2}), implies that $T \lesssim 7$ MeV (assuming $T_i > 100$ MeV).
Using $\rho_{b'} \approx 0.2 (T/{\rm MeV})^3 \ {\rm g/cm}^3$, then we estimate that the total mass fraction of C$'$ that can be produced is expected to be small, $X_{\rm C'} < 10^{-8}$.

Anyway, this is just the primordial chemical abundance. 
Of course, light nuclei are expected to be processed into heavier nuclei by stellar nucleosynthesis on successive populations of mirror stars.
In this process of heavy element enrichment of the mirror dark matter interstellar medium a crucial role is played by the high fraction of He$'$ inside mirror stars.
In fact, accurate studies have shown\cite{Berezhiani:2005vv} that the stellar evolution is more rapid for a higher initial He$'$ content, and for $Y' \approx 0.9$ it can be orders of magnitude faster than the standard case of $Y = 0.25$ in ordinary stars of the same masses.
This means that the enrichment of heavy mirror elements in the halo of the galaxy can be plausibly efficient enough to explain the relatively large abundances of He$'$, O$'$ and other heavy elements suggested by the direct detection experiments.\cite{Foot:2010hu}


\section{Evolution of primordial perturbations}
\label{sec-evol-pert}


We have shown that mirror baryons could provide a significant contribution to the energy density of the Universe and thus they could constitute a relevant component of dark matter. 
Immediate questions arise: how does the mirror baryonic dark matter behave? what are its differences from the more familiar dark matter candidates as the cold dark matter (CDM), the hot dark matter (HDM), {\it etc.}?
In this section we discuss the problem of the cosmological structure formation in the presence of mirror baryons as a dark matter component. 
We focus on the structure formation theory in linear regime, analyzing the trends of relevant scales (sound speed, Jeans length and mass, Silk mass) in both sectors and comparing them with the CDM case, and finally showing the temporal evolution of perturbations in all the components of a Mirror Universe (ordinary and mirror photons and baryons, and possibly CDM).


\subsection{\bf  Mirror baryonic structure formation}
\label{baryo_struct_form}

We extend the linear structure formation theory (for the standard scenario, see Refs.~\refcite{Padma:book-sf} and \refcite{Tsagas:2002sd}) 
to the case of dark matter with a non-negligible mirror baryonic component.

In a Mirror Universe we assume that a mirror sector is present, so that the matter is made of ordinary baryons (the only certain component), non-baryonic (dark) matter, and mirror baryons. 
Thus, it is necessary to study the structure formation in all these three components. 
We proceed in this way: first of all, we recall the situation for ordinary baryons, and then we compute the same quantities for mirror baryons, comparing them with each other and with the CDM case.

In general, when dealing with the pre-recombination plasma, we distinguish between two types of perturbations, namely between ``isoentropic'' ({\it adiabatic}) and ``entropic'' ({\it isocurvature} or {\it isothermal}) modes,\cite{Zeldovich:spu1967} while after matter-radiation decoupling perturbations evolve in the same way regardless of their original nature. 

Here we study only adiabatic perturbations, which are today the preferred perturbation modes, and we leave out the isocurvature modes, which could also have a contribution, but certainly cannot be the dominant component.\cite{Enqvist:2000hp,Trotta:2006ww}
We recall that an adiabatic perturbation satisfies the {\it condition for adiabaticity}
\begin{equation}
\delta_m = {\frac{3}{4}} \delta_r \;, 
\label{adcond}
\end{equation}
which, defining as usual $\delta \equiv \left( {\delta\rho} / \rho \right)$, relates perturbations in matter ($ \delta_m $) and radiation ($ \delta_r $) components.

We will now consider cosmological models where baryons, ordinary or mirror, are the dominant form of matter.
Thus, it is crucial to study the interaction between baryonic matter and radiation during the plasma epoch in both sectors, and the simplest way of doing it is by looking at models containing only these two components.

According to the Jeans theory,\cite{Jeans:pt1902,Jeans:book} the relevant scale for the gravitational instabilities is characterized by the Jeans scale (length and mass), which now needs to be defined in both the ordinary and mirror 
sectors. 
Then, we define the ordinary and mirror Jeans lengths as
\begin{equation}
\lambda_{\rm J} \simeq v_{\rm s} \sqrt{\frac{\pi}{ G\rho_{\rm dom} } } ~~,~~~~~~~~
\lambda'_{\rm J} \simeq v'_{\rm s} \sqrt{\frac{\pi}{ G\rho_{\rm dom} } } \;, 
\label{Jl1}
\end{equation}
where $\rho_{\rm dom}$ is the density of the dominant species, and $ v_{\rm s} $, $ v'_{\rm s} $ are the sound speeds.
The Jeans masses are
\begin{equation}
M_{\rm J} = {\frac{4}{3}}\pi\rho_{b}\left(\frac{\lambda_{\rm J}}{2}\right)^3 
 = {\frac{\pi}{6}}\rho_{b}\left({\lambda_{\rm J}}\right)^3 ~~,~~~~~~~~
M'_{\rm J} = {\frac{4}{3}}\pi\rho'_{b}\left(\frac{\lambda'_{\rm J}}{2}\right)^3 \;,
\label{Jm1}
\end{equation}
where the density is now that of the perturbed component (ordinary or mirror baryons).


\subsubsection{Evolution of the adiabatic sound speed}
\label{evol_Vs}

Looking at the expressions of the Jeans length (\ref{Jl1}), it is clear that the key issues are the evolutions of the sound speeds in both sectors, since they determine the scales of gravitational instabilities. 
Using the definition of adiabatic sound speed, we obtain for the two sectors
\begin{equation} \label{assom}
  v_{\rm s} = {\left(\frac{\partial p}{\partial \rho} \right)}^{1/2} = w^{1/2} ~~,~~~~~~~~
  v_{\rm s}'= {\left(\frac{\partial p'}{\partial \rho'} \right)}^{1/2} = (w')^{1/2} \;,
\end{equation}
where $ w $ and $ w' $ are relative respectively to the ordinary and mirror equations of state $ p = w \rho $ and $ p' = w' \rho' $.

First of all, we consider the standard case of a Universe made of only one sector. 
In a mixture of radiation and baryonic matter the total density and pressure are respectively $ \rho = \rho_\gamma + \rho_b $ and 
$ p \simeq p_\gamma = \rho_\gamma /3 $ (remind that $ p_b \simeq 0 $). 
Hence, the adiabatic sound speed is given by
\begin{equation}
v_{\rm s} = \left({\frac{\partial p}{\partial \rho}}\right)^{1/2}
  \simeq {\frac{1}{\sqrt{3}}}\left(1+{\frac{3\rho_b} 
  {4\rho_\gamma}}\right)^{-{1/2}} \;, \label{vsord}
\end{equation}
where we have used the adiabatic condition (\ref{adcond}).
In particular, using the scaling laws $ \rho_m \propto \rho_{0m}(1+z)^3 $ and $ \rho_\gamma \propto \rho_{0\gamma}(1+z)^4 $, together with the definition of matter-radiation equality (\ref{z-eq}) (where we now consider only baryons and photons), we obtain
\begin{equation} \label{ass2} 
v_{\rm s}(z) \simeq {\frac{1}{\sqrt3}}
 \left[ 1 +{\frac{3}{4}} \left({\frac{1+z_{\rm eq}}{1+z}}\right)\right]^{-1/2} \;. 
\end{equation}
In fact, the relation above is valid only for an ordinary Universe, and it is an approximation, for small values of $x$ and the mirror baryon density (remember that $\beta = {\Omega_b' / \Omega _b}$), of the more general equation for a Universe made of two sectors of baryons and photons, obtained using Eqs.~(\ref{vsord}) and (\ref{z-eq_2}) and given by
\begin{equation} \label{ass3} 
v_{\rm s}(z) \simeq {\frac{1}{\sqrt3}}
\left[ 1 +{\frac{3}{4}}  \left(\frac{1+x^{4}}{1+\beta } \right) 
\left( {\frac{1+z_{\rm eq}}{1+z}} \right) \right]^{-1/2} \;.
\end{equation}
In the most general case, the matter is made not only of ordinary and mirror baryons, but also of some other form of dark matter, then the factor $1+\beta $ is replaced by $1+\beta +\beta _{DM}$, where $\beta _{DM} = {\left(\Omega _m - \Omega _b - \Omega _b' \right) / \Omega _b}$. 
The presence of the term $ [(1 + x^4) / (1 + \beta) ] $ in equation (\ref {ass3}) is linked to the shift of matter-radiation equality epoch (\ref {shifteq}), thus it only balance this effect, without changing the value of the sound speed computed using Eq.~(\ref {ass2}).

The mirror plasma contains more baryons and less photons than the ordinary one, $ \rho'_b = \beta \rho_b $ and $ \rho'_\gamma = x^4 \rho_\gamma $.  
Then, using Eqs.~(\ref {assom}) and (\ref {z-eq_2}), we have
\begin{equation} \label{mirsound} 
v'_{\rm s}(z) \simeq 
  {\frac{1}{\sqrt3}} \left(1+ {\frac{3\rho'_b}{4\rho'_\gamma}}\right)^{-1/2} 
  \approx {\frac{1}{\sqrt3}} \left[ 1 +{\frac{3}{4}} \left({\frac{1+x^{-4}}
  {1+\beta ^{-1}}}\right) \left({\frac{1+z_{\rm eq}}{1+z}}\right)\right]^{-1/2} \;. 
\end{equation}
Let us consider for simplicity the case when dark matter of the Universe is entirely due to mirror baryons, $ \Omega_m \simeq \Omega'_b $ (i.e., $ \beta \gtrsim 5 $). 
Hence, for the redshifts of cosmological relevance, $ z \sim z_{\rm eq} $, we have $ v'_{\rm s} \sim 2x^2 /3 $, which is always less than $ v_{\rm s} \sim 1/\sqrt{3} $ (some example: if $ x = 0.7 $, $ v_{\rm s}' \approx 0.5 \cdot v_{\rm s} $; if $ x = 0.3 $, $ v_{\rm s}' \approx 0.1 \cdot v_{\rm s} $). 
In expression (\ref{mirsound}) it is crucial the presence of the factor $[(1+x^{-4}) / (1+\beta ^{-1})]$, which is always larger than 1 (given the bounds (\ref{beta-bounds}) and (\ref{x-bound}) on the mirror parameters), so that $ v'_{\rm s} < v_{\rm s} $ during all the history of the Universe, and only in the limit of very low scale factors, $ a \ll a_{\rm eq} $, we obtain $ v'_{\rm s} \simeq v_{\rm s} \simeq 1/\sqrt{3} $. 
As we will see in the following, this has important consequences on structure formation scales.

Now we define $ a_{b\gamma } $ as the scale factor corresponding to the redshift 
\begin{equation}
(1+z_{b\gamma }) = (a_{b\gamma})^{-1} = 
  \frac{\Omega _b}{\Omega _\gamma} = 3.9\cdot 10^4 (\Omega _b h^2) \;. 
\end{equation}
Since $ 1+z_{\rm rec} \simeq 1100 $, ordinary baryon-photon equipartition occurs before recombination only if $ \Omega_b h^2>0.026 $ (which seems unlikely, given its current estimates). 
According to Eq.~(\ref{shiftzbg}), in the mirror sector the scale of baryon-photon equality $ a_{b\gamma}' $ is dependent on $ x $ and it transforms as
\begin{equation} \label{shiftabg} 
a_{b\gamma}' 
  = {\frac{\Omega_{\gamma}'}{\Omega_b'}} 
  \simeq {\frac{\Omega_{\gamma} \, x^4}{\Omega_b \, \beta}} 
  = a_{b\gamma} {\frac{x^4}{\beta} } < a_{b\gamma} \;.
\end{equation}
If we remember the definition of the quantity $ x_{\rm eq} \approx 0.046 (\Omega_m h^2)^{-1} $, we have that for $ x > x_{\rm eq} $ the decoupling occurs after equipartition (as in the ordinary sector for $ \Omega_b h^2>0.026 $), while for $ x < x_{\rm eq} $ it occurs before (as for $ \Omega_b h^2 < 0.026 $). 

Regardless of which sector we are considering, in the radiation era $ \rho_\gamma \gg \rho_b $, ensuring that $ v_{\rm s} \simeq 1/\sqrt{3} $. 
In the interval between equipartition and decoupling, when $ \rho_b \gg \rho_\gamma $, Eq.~(\ref{vsord}) gives $ v_{\rm s} \simeq \sqrt{4 \rho_\gamma / 3 \rho_b}\propto a^{-1/2} $. 
After decoupling there is no more pressure equilibrium between baryons and photons, and $ v_{\rm s} $ is just the velocity dispersion of a gas of hydrogen and helium, $ v_{\rm s} \propto a^{-1} $. 
If $ \Omega_b h^2 < 0.026 $ or $ x < x_{\rm eq} $ (according to what sector we consider), photon-baryon equipartition occurs after decoupling, and the intermediate situation does not arise. 

It follows that, by taking care to interchange $ a_{b\gamma} $ with $ a_{b\gamma}' $ and $ a_{\rm dec} $ with $ a_{\rm dec}' $, we have for the sound speed the same trends with the scale factor in both sectors, though with the aforementioned differences in the values. 
The situation whit $ x > x_{\rm eq} $ is resumed below
\begin{equation}
v'_{\rm s}(a) \propto\left\{\begin{array}{l}
 {\rm const.} \hspace{11mm} a<a'_{b\gamma} \;,\\[1mm]
 a^{-1/2} \hspace{12mm} a'_{b\gamma}<a<a'_{\rm dec} \;,\\[1mm]
 a^{-1}  \hspace{15mm} a>a'_{\rm dec} \;.\\\end{array}\right.  \label{vsz}
\end{equation}

If we recall that the matter-radiation equality for a single sector (ordinary) Universe, $ (a_{\rm eq})_{\rm ord} $, is always bigger than that for a two sectors (mirror) one, $ (a_{\rm eq})_{\rm mir} $, according to
\begin{equation} \label{shiftaeq} 
(a_{\rm eq})_{\rm mir} 
  = {\frac{\left( 1+x^4 \right)}{\left( 1+\beta \right)} } ~ (a_{\rm eq})_{\rm ord} 
  < (a_{\rm eq})_{\rm ord} \;,
\end{equation}
together with our hypothesis $ x < 1 $ (from the BBN bound) and $ \beta > 1 $ (cosmologically interesting situation, i.e., significant mirror baryonic contribution to the dark matter), we obtain the useful inequality always verified in a Universe made of ordinary and mirror sectors 
\begin{equation} \label{hiera1} 
a_{b\gamma}' < a_{\rm eq} < a_{b\gamma} \;.
\end{equation}

It's very important to remark (see Ref.~\refcite{Padma:book-sf}) that at decoupling $ v_{\rm s}^2 $ drops from $ (p_\gamma / \rho _b) $ to $ (p_b /\rho _b) $. 
Since $ p_\gamma \propto n_\gamma T $ while $ p_b \propto n_b T $ with $ (n_\gamma /n_b) \simeq 10^9 \gg 1 $ in the ordinary sector, this is a large drop in $ v_{\rm s} $ and consequently in $ \lambda _{\rm J} $. 
More precisely, $ v_{\rm s}^2 $ drops from the value $ (1/3)(\rho _\gamma / \rho _b) = (1/3)(\Omega _\gamma / \Omega _b)(1+z_{\rm dec}) $ to the value $ (5/3) (T_{\rm dec}/m_b) = (5/3) (T_0/m_b)(1+z_{\rm dec}) $, with a reduction factor 
\begin{equation}
F_1 (\Omega_b h^2 > 0.026) 
= {\frac{(v_{\rm s}^2)^{\rm (+)}_{\rm dec}}{(v_{\rm s}^2)^{\rm (-)}_{\rm dec}}} 
= 6.63 \cdot 10^{-8} (\Omega _b h^2) \;,
\end{equation}
where $ (v_{\rm s}^2)^{\rm (-)}_{\rm dec} $ and $ (v_{\rm s}^2)^{\rm (+)}_{\rm dec} $ indicate the sound speed respectively just before and after the decoupling.
If we consider now the mirror sector and the drop in $ (v_{\rm s}')^2 $ at decoupling, we find for the reduction factor
\begin{equation} \label{f1p}
F_1' (x > x_{\rm eq}) = \beta x^{-3} F_1 \;.
\end{equation}
In the case $ \Omega_b h^2 < 0.026 $, $ v_{\rm s}^2 $ drops directly from $ (1/3) $ to $ (5/3) (T_{\rm dec}/m_b) = (5/3) (T_0/m_b)(1+z_{\rm dec}) $ with a suppression
\begin{equation}
F_2 (\Omega_bh^2 < 0.026) = 
  {\frac{(v_{\rm s}^2)^{\rm (+)}_{\rm dec}}{(v_{\rm s}^2)^{\rm (-)}_{\rm dec}}} 
  = 1.9 \cdot 10^{-9} \;.
\end{equation}
It's easy to find that in the mirror sector the reduction factor in the case $ a_{b\gamma}' > a_{\rm dec}' $ is the same as in the ordinary one
\begin{equation} \label{f2p}
F_2' (x < x_{\rm eq}) = F_2 \;.
\end{equation}
Some example: for $ x = 0.7 $, $ F_1' \approx 2.9 \beta F_1 $; for $ x = 0.5 $, $ F_1' = 8 \beta F_1 $; for $ x = 0.3 $, $ F_1' \approx 37 \beta F_1 $. 
We remark that, if $ \beta \geq 1 $, $ F_1' $ is at least about an order of magnitude larger than $ F_1 $. 
In fact, after decoupling $ (v_{\rm s}')^2 = (5/3) (T_{\rm dec}' / m_b) = (5/3) (T_{\rm dec} / m_b) = (v_{\rm s})^2$ (since $T_{\rm dec}' = T_{\rm dec} $), and between equipartition and recombination $ (v_{\rm s}')^2 < (v_{\rm s})^2 $. 
The relation above means that the drop is smaller in the mirror sector than in the ordinary one.
Obviously, before equipartition $ (v_{\rm s}')^2 = (v_{\rm s})^2 = 1/3 $, and this is the reason why the parameter $ F_2 $ is the same in both sectors.

\begin{figure}[pb]
\centerline{\psfig{file={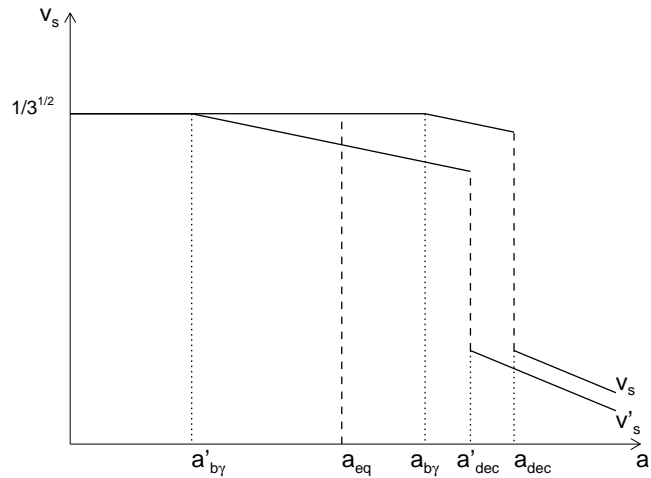},width=8.5cm}}
\vspace*{8pt}
\caption{The trends of the mirror sound speed ($v_{\rm s}'$) as a function of the scale factor, compared with the ordinary sound speed ($v_{\rm s}$). 
The ordinary model has $ \Omega_b h^2 = 0.08 $, while the mirror model has $ x = 0.6 $ and $ \beta = 2 $. 
All the key epochs are also reported: photon-baryon equipartition and decoupling in both sectors, and the matter-radiation equality. \label{scalord2}}
\end{figure}

In Fig.~\ref{scalord2} we plot the trends with scale factor of the mirror sound speed, in comparison with the ordinary one. 
The ordinary model is a typical one with $ \Omega_b h^2 > 0.026 $, while the mirror model has $ x = 0.6 $ and $ \beta = 2 $ (this means that mirror baryonic density is twice the ordinary one, chosen in these models about four times its current estimation in order to better show the general behaviour). 
In the same figure we show also the aforementioned relative positions of the key epochs (photon-baryon equipartition and decoupling) for both sectors, together with the matter-radiation equality.
If we reduce the value $ \Omega_b h^2 $, $ a_{b\gamma} $ goes toward higher values, while $ a_{\rm dec} $ remains fixed, so that for $ \Omega_b h^2 < 0.026 $ decoupling happens before equipartition and the intermediate regime, where $ v_{\rm s} \propto a^{-1/2} $, disappears.
Analogously, if we reduce $ x $, $ a'_{\rm dec} $ shifts to lower values, until for $ x < x_{\rm eq} $ it occurs before the mirror equipartition $ a'_{b\gamma} $, so that the intermediate regime for $ v'_{\rm s} $ 
disappears.

\begin{figure}[pt]
\centerline{\psfig{file={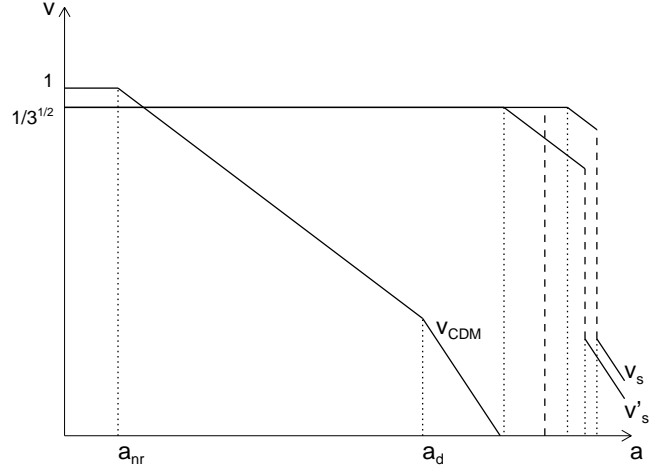},width=8.5cm}}
\vspace*{8pt}
\caption{The trends of the mirror ($ v_{\rm s}' $) and ordinary ($ v_{\rm s} $) sound speed compared with the velocity dispersion of a typical non baryonic cold dark matter candidate of mass $ \sim $ 1 GeV ($ v_{CDM} $); $ a_{\rm nr} $ and $ a_{\rm d} $ indicate the scale factors at which the dark matter particles become non relativistic or decouple. The ordinary and mirror models are the same as in Fig.~\ref{scalord2}, but the horizontal scale is expanded by some decade in order to show the CDM velocity. \label{scalord3}}
\end{figure}

In Fig.~\ref{scalord3} the same ordinary and mirror sound speeds are plotted together with the velocity dispersion of a typical non baryonic cold dark matter candidate of mass $ \sim $ 1 GeV. 
Note that the horizontal scale is expanded by some decade compared to Fig.~\ref{scalord2}, because the key epochs for the CDM velocity evolution (the epochs when the particles become non relativistic, $ a_{\rm nr} $, and when they decouple, $ a_{\rm d} $) occur at much lower scale factors.

\subsubsection{Evolution of the Jeans length and the Jeans mass}

Recalling the definitions (\ref{Jl1}) and (\ref{Jm1}), and using the results of Sec.~\ref{evol_Vs} relative to the sound speed, we can compute the evolution of the mirror Jeans length and mass and compare them with the analogous quantities for ordinary baryons and CDM.

We find for the evolution of the adiabatic Jeans length and mass of ordinary baryons in the case $ \Omega_b h^2 > 0.026 $
\begin{equation}
\lambda_{\rm J}\propto\left\{\begin{array}{l}
 a^2\,\\[1mm]
 a^{3/2}\,\\[1mm]
 a\,\\[1mm]
 a^{1/2} \,
 \\\end{array}\right.
\hspace{12mm}M_{\rm J}\propto{\frac{\lambda _{\rm J}^3}{a^3}}
\propto\left\{\begin{array}{l}
 a^3\hspace{20mm}a<a_{\rm eq}\;,\\[1mm]
 a^{3/2}\hspace{17mm}a_{\rm eq}<a<a_{b\gamma }\;,\\[1mm]
 {\rm const.}\hspace{14mm}a_{b\gamma }<a<a_{\rm dec}\;,\\[1mm]
 a^{-3/2}\hspace{15mm}a_{\rm dec}<a \;.
\end{array}\right.  \label{aJl}
\end{equation}

\noindent Otherwise, if $ \Omega_b h^2 \leq 0.026 $, $a_{b\gamma } > a_{\rm dec}$, and there is no intermediate phase $a_{b\gamma } < a < a_{\rm dec}$. 

In the mirror sector it's no more sufficient to interchange $ a_{b\gamma } $ with $ a_{b\gamma }' $ and $ a_{\rm dec} $ with $ a_{\rm dec}' $, as made for the sound speed, because from relation (\ref{hiera1}) we know that in the mirror sector the photon-baryon equipartition happens before the matter-radiation equality (due to the fact 
that we are considering a mirror sector with more baryons and less photons than the ordinary one). 
It follows that, due to the shifts of the key epochs, the intervals of scale factor for the various trends are different. 
As usual, there are two different possibilities, $ x > x_{\rm eq} $ and $ x < x_{\rm eq} $ (which correspond roughly to $ \Omega_b h^2 > 0.026 $ and $ \Omega_b h^2 < 0.026 $ in an ordinary Universe), where, as discussed in Sec.~\ref{evol_Vs}, for the second one the intermediate situation is absent.

Using the results of Sec.~\ref{evol_Vs} for the sound speed, we find the evolution of the adiabatic Jeans length and mass in the case $ x > x_{\rm eq} $
\begin{equation}
\lambda_{\rm J}' \propto \left\{\begin{array}{l}
 a^2\,\\[1mm]
 a^{3/2}\,\\[1mm]
 a\,\\[1mm]
 a^{1/2} \,
 \\\end{array}\right.
\hspace{12mm}M_{\rm J}' 
\propto {\frac{(\lambda _{\rm J}')^3}{a^3}}\propto\left\{\begin{array}{l}
 a^3 \hspace{20mm} a < a_{b\gamma }' \;,\\[1mm]
 a^{3/2} \hspace{17mm} a_{b\gamma }' < a < a_{\rm eq} \;,\\[1mm]
 {\rm const.} \hspace{14mm} a_{\rm eq} < a < a_{\rm dec}' \;,\\[1mm]
 a^{-3/2} \hspace{15mm} a_{\rm dec}' < a \;.
\end{array}\right.  \label{aJlm}
\end{equation}

\begin{figure}[pb]
\centerline{\psfig{file={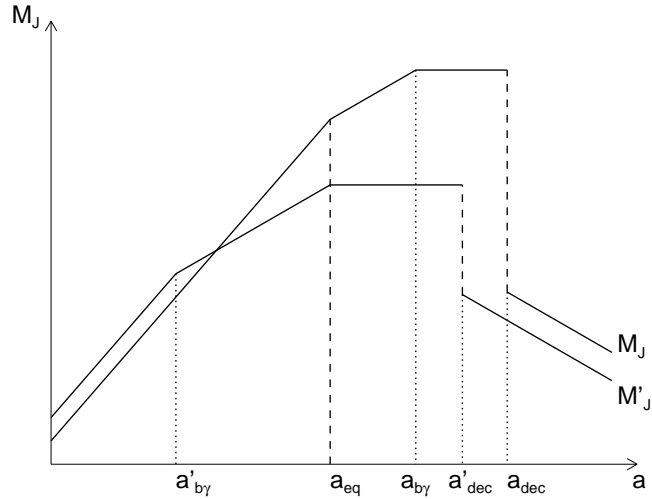},width=8.5cm}}
\vspace*{8pt}
\caption{The trends of the mirror Jeans mass ($ M'_{\rm J} $) as a function of the scale factor, compared with the ordinary Jeans mass ($ M_{\rm J} $).  The ordinary model has $ \Omega_b h^2 = 0.08 $, while the mirror model has $ x = 0.6 $ and $ \beta = 2 $. The horizontal scale is the same as in Fig.~\ref{scalord2}. We remark that the same behaviours of the ordinary sector are present in the mirror sector for different intervals of scale factor.
\label{scalord8}}
\end{figure}

We plot in Fig.~\ref{scalord8} with the same horizontal scale the trends of the mirror Jeans mass compared with those for the ordinary sector; the parameters of both mirror and ordinary models are the ones previously used, i.e. $ \Omega_b h^2 = 0.08 $, $ x = 0.6 > x_{\rm eq} $ and $ \beta = 2 $. 

In the ordinary sector the largest value of the Jeans mass is just before decoupling (see Ref.~\refcite{Padma:book-sf}), in the interval $ a_{b\gamma } < a < a_{\rm dec} $ , where
\begin{equation}
M_{\rm J}(a \lesssim a_{\rm dec}) 
  = 1.47 \cdot 10^{14} M_\odot \left( 1 + \beta \right)^{-3/2} 
  \left(\Omega_{b} h^2 \right)^{-2} \;,
\end{equation}
that for a hypothetical $ \Omega_b \simeq 0.1 h^{-2} $ and $ \beta = 0 $ is $ \sim 10^{16} M_\odot $. 
Just after decoupling we have
\begin{equation}
M_{\rm J}(a \gtrsim a_{\rm dec}) 
  = 2.5 \cdot 10^{3} M_\odot \left( 1 + \beta \right)^{-3/2} 
  \left(\Omega_{b} h^2 \right)^{-1/2} \;,
\label{mj1after}
\end{equation}
that for $ \Omega_b \simeq 0.1 h^{-2} $ and $ \beta = 0 $ is $ \sim 10^{4} M_\odot $. 
This drop is very sudden and large, changing the Jeans mass by $ F_1^{3/2} \simeq 1.7 \cdot 10^{-11} (\Omega _b h^2)^{3/2} $. 
Otherwise, if $\Omega_b h^2 \leq 0.026$, 
$ a_{b\gamma } > a_{\rm dec} $, there is no intermediate phase 
$ a_{b\gamma } < a < a_{\rm dec} $, and 
$ M_{\rm J}(a \lesssim a_{\rm dec}) $ is larger
\begin{equation}
M_{\rm J}(a \lesssim a_{\rm dec}) 
  \simeq 3.1 \cdot 10^{16} M_\odot \left( 1 + \beta \right)^{-3/2} 
  \left(\Omega_{b} h^2 \right)^{-1/2} \;,
\end{equation}
while after decoupling it takes the value in Eq.~(\ref{mj1after}), so that the drop is larger, $ F_2^{3/2} \simeq 8.3 \cdot 10^{-14} $.
We observe that, with the assumptions $ \Omega_b = 1 $ (totally excluded by observations) and $ \beta = 0 $, $ M_{\rm J,max} $ (which is the first scale to become gravitationally unstable and collapse soon after decoupling) has the size of a supercluster of galaxies.

If we now consider the expression (\ref {shiftabg}), we have
\begin{equation}
{\frac{a'_{b\gamma }}{a_{\rm eq}} }
  = \left(\frac{ 1 + \beta }{ \beta } \right) 
  \left(\frac{ x^4 }{ 1 + x^4 } \right) \;,
\end{equation}
which can be used to express the value of the mirror Jeans mass in the interval $ a_{\rm eq} < a < a_{\rm dec}' $ (where $ M_{\rm J}' $ takes the maximum value) in terms of the ordinary Jeans mass in the corresponding ordinary interval $ a_{b\gamma} < a < a_{\rm dec} $. 
We obtain
\begin{equation}
M_{\rm J}'(a \lesssim a_{\rm dec}') \approx 
  \beta^{-1/2} \left( {\frac{x^4}{1 + x^4} } \right)^{\rm 3/2} 
  \cdot M_{\rm J}(a \lesssim a_{\rm dec}) \;,
\end{equation}
which, for $ \beta \geq 1 $ and $x < 1$, means that the Jeans mass for the mirror baryons is lower than for the ordinary ones over the entire permitted ($\beta$, $x$) parameter space, with implications for the structure formation process. 
If, e.g., $ x = 0.6 $ and $ \beta = 2 $, then $ M_{\rm J}' \sim 0.03 \; M_{\rm J} $. 
We can also express the same quantity in terms of $ \Omega_b $, $ x $ and $ \beta $, in the case that all the dark matter is in the form of mirror baryons, as
\begin{equation} \label{mj_mir_1}
M_{\rm J}'(a \lesssim a_{\rm dec}') \approx 
  3.2 \cdot  10^{14} M_\odot \;
  \beta^{-1/2} ( 1 + \beta )^{-3/2} \left( \frac{x^4}{1+x^4} \right)^{3/2} 
  ( \Omega_{b} h^2 )^{-2} \;.
\end{equation}

If we remember Eq.~(\ref {f1p}), we obtain that for the mirror model the drop in the Jeans mass at decoupling is $ (F_1')^{3/2} = \beta^{3/2} x^{-9/2} (F_1)^{3/2} $, which, given our bounds on $ x $ and $ \beta $, is larger than $ (F_1)^{3/2} $. 
We give here some numerical example: for $ x = 0.7 $, $ (F_1')^{3/2} \approx 5 \beta^{3/2} (F_1)^{3/2} $; for $ x = 0.6 $ and $ \beta = 2 $ (the case of Fig.~\ref{scalord8}), $ (F_1')^{3/2} \approx 28 (F_1)^{3/2} $.

It's important to stress that these quantities are strongly dependent on the values of the free parameters $ x $ and $ \beta $, which shift the key epochs and change their relative positions. 
We can describe some case useful to understand the general behaviour, but if we want an accurate solution of a particular model, we must unambiguously identify the different regimes and solve in detail the appropriate equations.

\begin{figure}[pb]
\centerline{\psfig{file={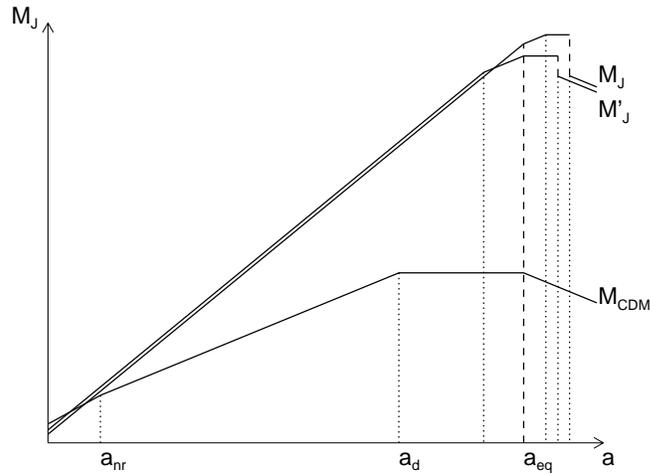},width=8.5cm}}
\vspace*{8pt}
\caption{The trends of the mirror ($ M'_{\rm J} $) and ordinary ($ M_{\rm J} $) Jeans mass compared with those of a typical non baryonic cold dark matter candidate of mass $ \sim 1 \, {\rm GeV} $ ($ M_{CDM} $); $ a_{\rm nr} $ and $ a_{\rm d} $ indicate the scale factors at which the dark matter particles become non relativistic or decouple, respectively. The models are the same as in Fig.~\ref{scalord8}, but the horizontal scale is expanded by some decade to show the CDM Jeans mass, as in Fig.~\ref{scalord3}.
\label{scalord9}}
\end{figure}

In Fig.~\ref{scalord9} we plot the trends of the mirror and ordinary Jeans mass compared with those of a typical non baryonic cold dark matter candidate of mass $ \sim $ 1 GeV. 
Apart from the usual expansion of the horizontal scale, due to the much lower values of the CDM key epochs as compared to the baryonic ones, a comparison of the mirror scenario with the cold dark matter one shows that the maximal value of the CDM Jeans mass is several orders of magnitude lower than that for mirror baryons.
This implies that a very large range of mass scales, which in a mirror baryonic scenario oscillate before decoupling, in a cold dark matter scenario would grow unperturbed during all the time (for more details see Sec.~\ref{mirevolpert}).

For the case $ x < x_{\rm eq} $, both $ a_{b\gamma}' $ and $ a_{\rm dec}' $ are smaller than the previous case $ x > x_{\rm eq} $, while the matter-radiation equality remains practically the same; as explained in Sec.~\ref{mir-dm}, the mirror decoupling (with the related drop in the associated quantities) happens before the matter-radiation equality, and the trends of the mirror Jeans length and mass are the following
\begin{equation}
\lambda_{\rm J}' \propto \left\{\begin{array}{l}
 a^2\,\\[1mm]
 a^{3/2}\,\\[1mm]
 a\,\\[1mm]
 a^{1/2} \,
 \\\end{array}\right.
\hspace{12mm}M_{\rm J}' \propto {\frac{(\lambda _{\rm J}')^3}{a^3}}
\propto\left\{\begin{array}{l}
 a^3 \hspace{20mm} a < a_{b\gamma }' \;,\\[1mm]
 a^{3/2} \hspace{17mm} a_{b\gamma }' < a < a_{\rm dec}' \;,\\[1mm]
 {\rm const.} \hspace{14mm} a_{\rm dec}' < a < a_{\rm eq} \;,\\[1mm]
 a^{-3/2} \hspace{15mm} a_{\rm eq} < a \;.
\end{array}\right.
\end{equation}

\begin{figure}[pb]
\centerline{\psfig{file={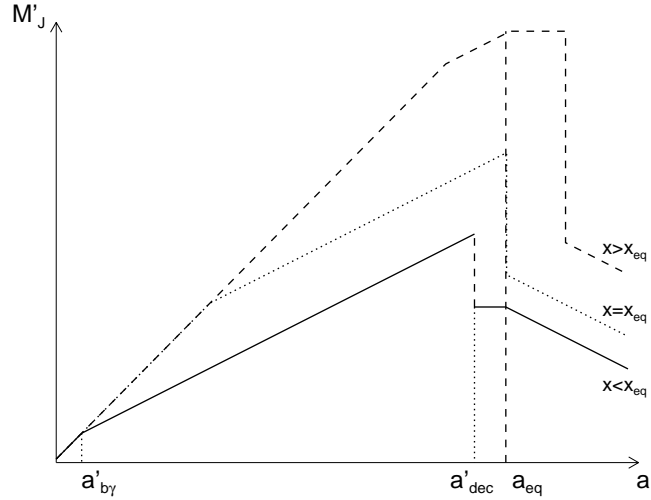},width=8.5cm}}
\vspace*{8pt}

\caption{The trends of the mirror Jeans mass for the cases $ x < x_{\rm eq} $ (solid line), $ x = x_{\rm eq} $ (dotted) and $ x > x_{\rm eq} $ (dashed). The model with $ x > x_{\rm eq} $ is the same as in Fig.~\ref{scalord8}, the others are obtained changing only the value of $ x $ and keeping constant all other parameters. As clearly shown in the figure, the only key epoch which remains almost constant in the three models is the matter-radiation equality; the mirror baryon-photon equipartition and decoupling indicated are relative to the model with $ x < x_{\rm eq} $. The change in the trends when $ x $ becomes lower than $ x_{\rm eq} $ is also evident, due to the fact that $ a_{\rm dec}' $ becomes lower than $ a_{\rm eq} $. \label{scalord10}}
\end{figure}

In this case we obtain for the highest value of the Jeans mass just before decoupling the expression
\begin{equation} \label{mj_mir_2}
M_{\rm J}'(a \lesssim a_{\rm dec}') \approx 
  3.2 \cdot  10^{14} M_\odot \; 
  \beta^{-1/2} ( 1 + \beta )^{-3/2} 
  \left( \frac{x}{x_{\rm eq}} \right)^{3/2} \left( \frac{x^4}{1+x^4} \right)^{3/2} 
  ( \Omega_{b} h^2 )^{-2} \;.
\end{equation}
In case $ x = x_{\rm eq} $, the expressions (\ref{mj_mir_1}) and (\ref{mj_mir_2}), respectively valid for $ x \ge x_{\rm eq} $ and $ x \le x_{\rm eq} $, are coincident, as we expect.
If we consider the differences between the highest mirror Jeans mass for the particular values $ x = x_{\rm eq}/2 $, $ x = x_{\rm eq} $ and $ x = 2 x_{\rm eq} $, we obtain the following relations
\begin{equation}
M_{\rm J,max}'(x_{\rm eq}/2) \approx \left(\frac{1}{2} \right)^{15/2} 
  \left[ \frac{1 + x_{\rm eq}^4} {1 + \left(\dfrac{x_{\rm eq}}{2} \right)^4} \right]^{3/2} 
  M_{\rm J,max}'(x_{\rm eq}) 
  \approx 0.005 \: M_{\rm J,max}'(x_{\rm eq}) \;,
\end{equation}
\begin{equation}
M_{\rm J,max}'(2x_{\rm eq}) \approx 2^6 
  \left[ \frac{1 + x_{\rm eq}^4} {1 + (2 x_{\rm eq})^4} \right]^{3/2} 
  M_{\rm J,max}'(x_{\rm eq}) 
  \approx 64 \: M_{\rm J,max}'(x_{\rm eq}) \;.
\end{equation}
In Fig.~\ref{scalord10} we plot the mirror Jeans mass for the three different possibilities: $ x < x_{\rm eq} $, $ x > x_{\rm eq} $ and $ x = x_{\rm eq} $ (the transition between the two regimes), keeping constant all other parameters.
In these three models the matter-radiation equality is the only key epoch which remains almost constant. 
The change in the trends when $ x $ becomes lower than $ x_{\rm eq} $, due to the fact that $ a_{\rm dec}' $ becomes lower than $ a_{\rm eq} $, generates an evident decrease of the Jeans mass.

\subsubsection{Evolution of the Hubble mass}
\label{subsec_evol_MH2}

The trends of the Hubble length and mass are expressed, as usually, by
\begin{equation}
\lambda_{\rm H} \propto \left\{\begin{array}{l}
 a^2\,\\[1mm]
 a^{3/2}\,
 \\\end{array}\right.
\hspace{12mm}M_{\rm H} \propto {\frac{(\lambda_{\rm H})^3}{a^3}}
\propto\left\{\begin{array}{l}
 a^3 \hspace{20mm} a < a_{\rm eq} \;,\\[1mm]
 a^{3/2} \hspace{17mm} a > a_{\rm eq} \;.
\end{array}\right.
\end{equation}

It should be emphasized that, as for the ordinary baryons, during the period of domination of photons ($ a < a'_{b\gamma } $) the mirror baryonic Jeans mass is an order of magnitude larger than the Hubble mass. 
In fact, following definitions we find
\begin{equation}
{\frac{M'_{\rm J}}{M_{\rm H}}} 
= {\frac{(\lambda'_{\rm J})^3}{\lambda_{\rm H}^3}} 
\simeq 26 \;.
\end{equation}

We plot the trends of the Hubble mass in Figs.~\ref {mirmasssca1} and \ref{mirmasssca2} together with other fundamental mass scales.


\subsubsection{Dissipative effects: collisional damping}
\label{disseff-colldamp}

A peculiar feature of the mirror baryonic scenario is that mirror baryons undergo the collisional damping as ordinary ones.
This dissipative process modify the purely gravitational evolution of perturbations. 
The physical phenomenon is the interaction between baryons and photons before the recombination, and the consequent dissipation due to viscosity and heat conduction.
Around the time of recombination the perfect fluid approximation breaks down, and the perturbations in the photon-baryon plasma suffer from collisional damping. 
As decoupling is approached, the photon mean free path increases and photons can diffuse from the overdense into the underdense regions, thereby smoothing out any inhomogeneities in the photon-baryon plasma. 
This effect is known as Silk damping.\cite{Silk:n1967}

In order to obtain an estimate of the effect, we follow Ref.~\refcite{Kolb:1990vq} for ordinary baryons, and then we extend to mirror baryons.
We consider the photon mean free path
\begin{equation}
\lambda _{\gamma} = {\frac{1}{X_{e}{\rm n}_{e}\sigma_{\rm T}}} 
\simeq 10^{29}a^3X_{e}^{-1}\left(\Omega_{b}h^2\right)^{-1}\,{\rm cm} \;,  \label{pmp}
\end{equation}
where $ X_{e} $ is the electron ionization factor, $ {\rm n}_{e} \propto a^{-3} $ is the number density of the free electrons and $ \sigma_{\rm T} $ is the cross section for Thomson scattering. 
Clearly, photon free streaming should completely damp all baryonic perturbations with wavelengths smaller than $ \lambda_\gamma $. 
Damping, however, occurs on scales much larger than $ \lambda_\gamma $ since the photons slowly diffuse from the overdense into the underdense regions, dragging along the still tightly coupled baryons. 
Integrating up to decoupling time we obtain the total distance traveled by a typical photon
\begin{equation}
\lambda _{\rm S}= 
  \sqrt{ {\frac{3}{5}} {\frac{ (\lambda _\gamma)_{\rm dec} t_{\rm dec} }
  {a^2_{\rm dec}} } } \,\,
  \simeq\,\,3.5\left(\Omega_b h^2\right)^{-3/4}\,{\rm Mpc} \;,
\label{lS}
\end{equation}
and the associated mass scale, the Silk mass, given by
\begin{equation}
M_{\rm S} 
  = {\frac{4}{3}}\pi\rho_b\left({\frac{\lambda _{\rm S}} 
  {2}}\right)^3\simeq6.2\times10^{12}
  \left(\Omega_b h^2\right)^{-5/4}\,{\rm M}_{\odot} \;,\label{Sm}
\end{equation}
which, assuming $ \Omega_b h^2 \simeq 0.02 $, gives $ M_{\rm S} \simeq 8 \times10^{14}~{\rm M}_{\odot} $.
This dissipative process causes that fluctuations on scales below the Silk mass are completely washed out at the time of recombination and no structure can form on these scales. 
This has consequences on large scale structure power spectrum, where small scales have very little power.

\begin{figure}[pb]
\centerline{\psfig{file={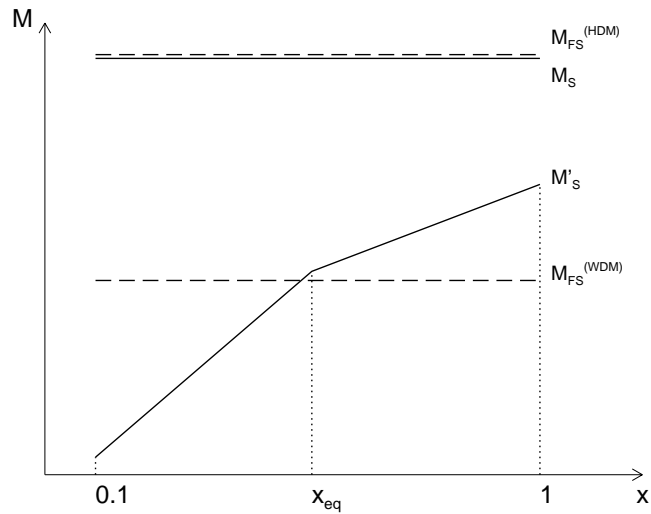},width=8.5cm}}
\vspace*{8pt}
\caption{The trend of the mirror Silk mass ($ M'_{\rm S} $) over a cosmologically interesting range of $ x $, which contains $ x_{eq} $ (we considered $ \Omega_m h^2 \simeq 0.15 $, so $ x_{\rm eq} \simeq 0.3 $). The axis are both logarithmic. We show for comparison also the values of the ordinary Silk mass ($ M_{\rm S} $) and of the free streaming mass ($ M_{\rm FS} $) for typical HDM and WDM candidates. \label{mirsilksca1}}
\end{figure}

In the mirror sector too, obviously, the photon diffusion from the overdense to underdense regions induces a dragging of charged particles and washes out the perturbations at scales smaller than the mirror Silk scale 
\begin{equation}
\lambda'_{\rm S} \simeq 3 f(x)(\beta \, \Omega_b h^2)^{-3/4} \;{\rm Mpc} \;, 
\label{mirsilklamb1}
\end{equation}
where $f(x)=x^{5/4}$ for $x > x_{\rm eq}$ and $f(x) = (x/x_{\rm eq})^{3/2} x_{\rm eq}^{5/4}$ for $x < x_{\rm eq}$. 
Thus, the density perturbation scales running the linear growth after the matter-radiation equality epoch are limited by the length $ \lambda'_{\rm S} $. 
The smallest perturbations that survive the Silk damping will have the mass 
\begin{equation} \label{ms_m}
M'_{\rm S} \sim \left(\frac{f(x)}{2}\right)^3 (\beta \, \Omega_b h^2)^{-5/4} 10^{12}~ M_\odot \;,
\end{equation}
which should be less than $ 2 \times 10^{12} ~ M_\odot $ in view of the BBN bound $ x < 0.64 $.  
Interestingly, for $ x \sim x_{\rm eq} $ we obtain, for the current estimate of $ \Omega_m h^2 $ and if all the dark matter is made of mirror baryons, $ M'_S \sim 10^{10} ~M_\odot $, a typical galaxy mass.

At this point it is very interesting a comparison between different damping scales, collisional (ordinary and mirror baryons) and collisionless (non-baryonic dark matter). 
We know that for hot dark matter (as a neutrino with mass $ \sim $10 eV) $ M_{\rm FS}^\nu \sim 10^{15} ~ M_\odot $, for a typical warm dark matter candidate with mass $ \sim $1 keV, $ M_{\rm FS}^{WDM} \sim 10^{9} - 10^{10} ~ M_\odot $, while for a cold dark matter candidate with mass $ \sim $1 GeV the free streaming scale is negligibly small, and it has practically no dissipation. 
From Eq.~(\ref{ms_m}) it is evident that the dissipative scale for mirror Silk damping is analogous to that for WDM free streaming. 
Consequently, the cutoff effects on the corresponding large scale structure power spectra are similar, though with important differences due to the presence of oscillatory features, which makes them distinguishable one from the other (see next sections). 
In Fig.~\ref{mirsilksca1} we show this comparison together with the trend of the mirror Silk mass over a cosmologically interesting range of $ x $.


\subsubsection{Scenarios}
\label{mirror_bar_struc}

\begin{figure}[pb]
\centerline{\psfig{file={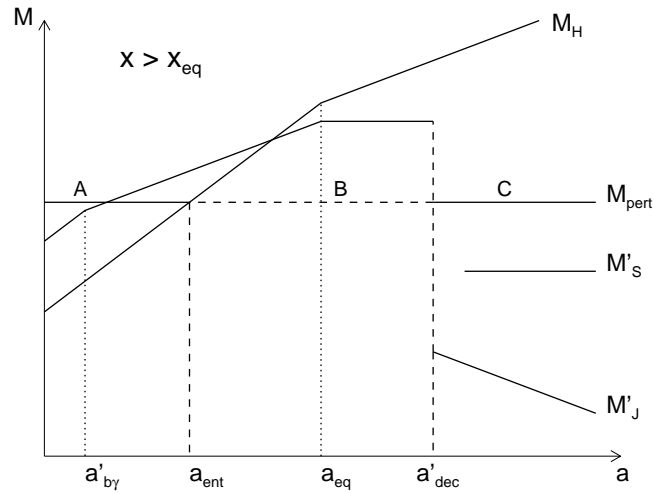},width=8.5cm}}
\vspace*{8pt}
\caption{Typical evolution of a perturbed scale $ M_{\rm pert} $ in adiabatic mirror baryonic dark matter scenario with $ x > x_{\rm eq} $. 
The figure shows the Jeans mass $ M_{\rm J}' $, the Silk mass $ M_{\rm S}' $ and the Hubble mass $ M_{\rm H} $. 
The time of horizon crossing of the perturbation is indicated by $ a_{\rm ent} $. 
The three evolutionary stages are also indicated: during stage A ($ a < a_{\rm ent} < a_{\rm eq} $) the mode grows as $ \delta_b \propto a^2 $; throughout stage B ($ a_{\rm ent} < a < a_{\rm dec}' $) the perturbation oscillates; finally, in stage C ($ a > a_{\rm dec}' $) the mode becomes unstable again and grows as $ \delta_b \propto a $. 
Fluctuations with size smaller than $ M_{\rm S}' $ are wiped out by photon diffusion. \label{mirmasssca1}}
\end{figure}

\begin{figure}[pb]
\centerline{\psfig{file={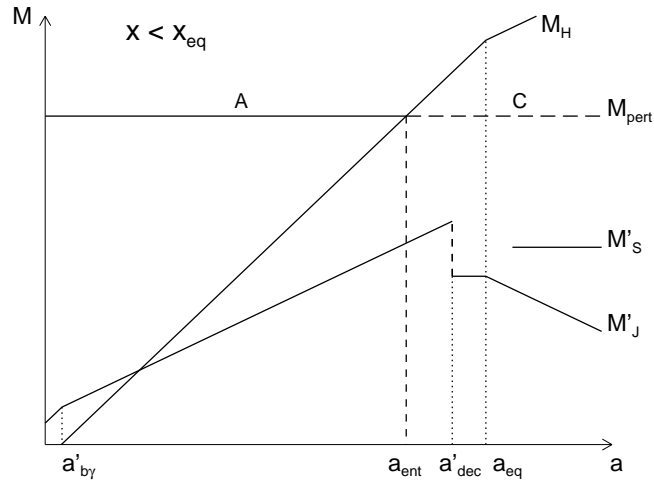},width=8.5cm}}
\vspace*{8pt}
\caption{Typical evolution of a perturbed scale $ M_{\rm pert} $ in adiabatic mirror baryonic dark matter scenario with $ x < x_{\rm eq} $. 
The value of $ M_{\rm pert} $ is the same as in Fig.~\ref{mirmasssca1}. 
The time of horizon crossing of the perturbation is indicated by $ a_{\rm ent} $. 
The figure shows the Jeans mass $ M_{\rm J}' $, the Silk mass $ M_{\rm S}' $ and the Hubble mass $ M_{\rm H} $. 
Unlike the case $ x > x_{\rm eq} $ (shown in the previous figure), now there are only the two evolutionary stages A ($ a < a_{\rm ent} $) and C ($ a > a_{\rm ent} $). 
Fluctuations with size smaller than $ M_{\rm S}' $ are wiped out by photon diffusion, but in this case the Silk mass is near to the maximum Jeans mass. \label{mirmasssca2}}
\end{figure}

After the description of the fundamental scales for structure formation, let us now collect all the informations and discuss the mirror scenarios. 
They are essentially two, according to the value of $ x $, which can be higher or lower than $ x_{\rm eq} $, and are shown respectively in Figs.~\ref{mirmasssca1} and \ref{mirmasssca2}.

Typically, adiabatic perturbations for mirror baryons with sizes larger than the maximum value of the Jeans mass, which is $ M_{\rm J}'(a_{\rm eq}) $ for $ x > x_{\rm eq} $ and $ M_{\rm J}'(a_{\rm dec}') $ for $ x < x_{\rm eq} $, 
experience uninterrupted growth. 
In particular, they grow as $ \delta_b \propto a^2 $ before matter-radiation equality and as $ \delta_b \propto a $ after equality. 
Fluctuations on scales in the mass interval $ M_{\rm S}' < M < M_{\rm J,max} $ grow as $ \delta_b \propto a^2 $ 
while they are still outside the Hubble radius. 
After entering the horizon and until recombination these modes oscillate like acoustic waves. 
The amplitude of the oscillation is constant before equilibrium but decreases as $ a^{-1/4} $ between equipartition and recombination. 
After decoupling the modes become unstable again and grow as $ \delta_b \propto a $. 
Finally all perturbations on scales smaller than the value of the Silk mass are dissipated by photon diffusion. 

Given this general behaviour, the schematic evolution of an adiabatic mode with a reference mass scale $ M_{\rm pert} $, with $ M_{\rm S}' < M_{\rm pert} < M_{\rm J}'(a_{\rm eq}) $, is shown in Fig.~\ref{mirmasssca1} for $ x > x_{\rm eq} $.
We distinguish between three evolutionary stages, called A, B and C, depending on the size of the perturbation and 
on the cosmological parameters $ \Omega_b h^2 $, $ x $ and $ \beta $, which determine the behaviour of the mass scales, and in particular the key moments (time of horizon crossing and decoupling) and the dissipative Silk scale. 
During stage A, i.e. before the horizon crossing ($ a < a_{\rm ent} < a_{\rm eq} $), the mode grows as $ \delta_b \propto a^2 $; throughout stage B ($ a_{\rm ent} < a < a_{\rm dec}' $) the perturbation enters the horizon, baryons and photons feel each other, and it oscillates; finally, in stage C ($ a > a_{\rm dec}' $), the photons and baryons decouple, and the mode becomes unstable again growing as $ \delta_b \propto a $. 
We remark that fluctuations with sizes larger than $M_{\rm J}'(a_{\rm eq})$ grow uninterruptedly (because after horizon crossing the photon pressure cannot balance the gravity), changing the trend from $ a^2 $ before MRE to $ a $ after it, while those with sizes smaller than $ M_{\rm S}' $ are completely washed out by photon diffusion.

After decoupling, all surviving perturbations (those with $ M_{\rm pert} > M'_{\rm S} $) grow steadily until their amplitude becomes so large that the linear theory breaks down and one needs to employ a different type of analysis. 

If we look, instead, at the schematic evolution of an adiabatic mode with the same reference mass scale $ M_{\rm pert} $ but for $ x < x_{\rm eq} $, as reported in Fig.~\ref{mirmasssca2}, we immediately notice the lower values of the maximum Jeans mass and the Silk mass, which are similar. 
Therefore, for the plotted perturbative scale there are now only the two stages A and C. 
In general, depending on its size, the perturbation mass can be higher or lower than the Silk mass (and approximately also higher or lower than the maximum Jeans mass), so modes with $ M_{\rm pert} > M'_{\rm S} $ grow continuously before and after their horizon entry, while modes with $ M_{\rm pert} < M'_{\rm S} $ are completely washed out.

We find that $ M'_{\rm J} $ becomes smaller than the Hubble horizon mass $ M_{\rm H} $ starting from a redshift\cite{Berezhiani:2000gw}
\begin{equation}
z_{\rm c}' \sim 3750 \: x^{-4} \: (\Omega_m h^2) \;, 
\end{equation}
which is about $ z_{\rm eq} $ for $ x=0.64 $, but it sharply increases for smaller values of $ x $, as shown in Fig.~\ref{figzz}. 
We can recognize this behaviour also watching at the intersections of the lines for $ M'_{\rm J} $ and $ M_{\rm H} $ in Figs.~\ref{mirmasssca1} and \ref{mirmasssca2}. 
Thus, density perturbation scales which enter horizon at $ z \sim z_{\rm eq} $ have masses larger than $ M'_J $ and thus undergo uninterrupted linear growth immediately after $ t_{\rm eq} $. 
Smaller scales for which $ M'_{\rm J} > M_{\rm H} $ would instead first oscillate. 
Therefore, the large scale structure formation is not delayed even if the mirror decoupling did not occur yet, i.e. even if $ x > x_{\rm eq} $. 

When compared with non baryonic dark matter scenarios, the main feature of the mirror baryonic scenario is that the mirror baryon density fluctuations should undergo the strong collisional damping around the time of mirror recombination, which washes out the perturbations at scales smaller than the mirror Silk scale. 
It follows that density perturbation scales which undergo the linear growth after the MRE epoch are limited by the length $ \lambda'_{\rm S}$. 
This could help in avoiding the excess of small scales (of few Mpc) in the CDM power spectrum without tilting the spectral index.
To some extent, the cutoff effect is analogous to the free streaming damping in the case of warm dark matter (WDM), but there are important differences. 
The point is that, alike usual baryons, the mirror baryonic dark matter shows acoustic oscillations with an impact on the large scale structure (LSS) power spectrum.\cite{Berezhiani:2003wj}\cdash\cite{Ciarcelluti:2004ij,Ciarcelluti:2004ip} 
In addition, the oscillations of mirror baryons, transmitted via gravity to the ordinary baryons, could cause observable anomalies in the CMB angular power spectrum.
This effect can be observed only if the mirror baryon Jeans scale $ \lambda'_{\rm J} $ is larger than the Silk scale of ordinary baryons, which sets a principal cutoff for CMB oscillations.
This would require enough large values of $ x $, and, together with the possible effects on the large scale power spectrum, it can provide a direct test for the mirror baryonic dark matter in CMB and LSS observations.
(For a complete discussion of the CMB and LSS power spectra for a Mirror Universe see next sections.)

{\em Clearly, for small $ x $ the mirror matter recombines before the MRE moment, and thus it behaves as the CDM as far as the large scale structure is concerned.} 
However, there still can be crucial differences at smaller scales which already went non-linear, like galaxies. 
In our scenario, dark matter in galaxies and clusters can contain both CDM and mirror components, or can be even constituted entirely by the mirror baryons. 

One can question whether the mirror matter distribution in halos can be different from that of the CDM. 
Simulations show that the CDM forms triaxial halos with a density profile too clumped toward the center, and overproduces the small substructures within the halo.
Since mirror baryons constitute a kind of collisional dark matter, it may potentially avoid these problems, at least the one related with the excess of small substructures.

Throughout the above discussion, we have assumed that the matter density of the Universe is close to unity. 
If, instead, the matter density is small and a vacuum density contribution is present, we have to add that the Universe may become curvature dominated starting from some redshift $ z_{\rm curv} $. 
Given the current estimate $ \Omega_\Lambda \simeq 0.7 $, this transition has yet occurred and the growth of perturbations has stopped around $ z_{\rm curv} $, when the expansion became too rapid for it.


\def \mirevolpert{Evolution of perturbations}
\subsection{\mirevolpert}
\label{mirevolpert}

Here we finally consider the temporal evolution of perturbations as function of the scale factor $ a $. 
All the plots are the results of numerical computations obtained using a Fortran code originally written for the ordinary Universe and modified to account for the mirror sector.

We used the synchronous gauge and the evolutionary equations presented in Ref.~\refcite{Ma:1995ey}. 
The difference in the use of other gauges is limited to the gauge-dependent behaviour of the density fluctuations on scales larger than the horizon. 
The fluctuations can appear as growing modes in one coordinate system and as constant mode in another, that is exactly what occurs in the synchronous and the conformal Newtonian gauges.

In the figures we plot the evolution of the components of a Mirror Universe, namely the cold dark matter,\footnote{
As non baryonic dark matter we consider only the cold dark matter, which is at present the standard choice in cosmology.
} the ordinary baryons and photons, and the mirror baryons and photons, changing some parameter to evaluate their influence. 

\begin{figure}[pb]
\centerline{\psfig{file={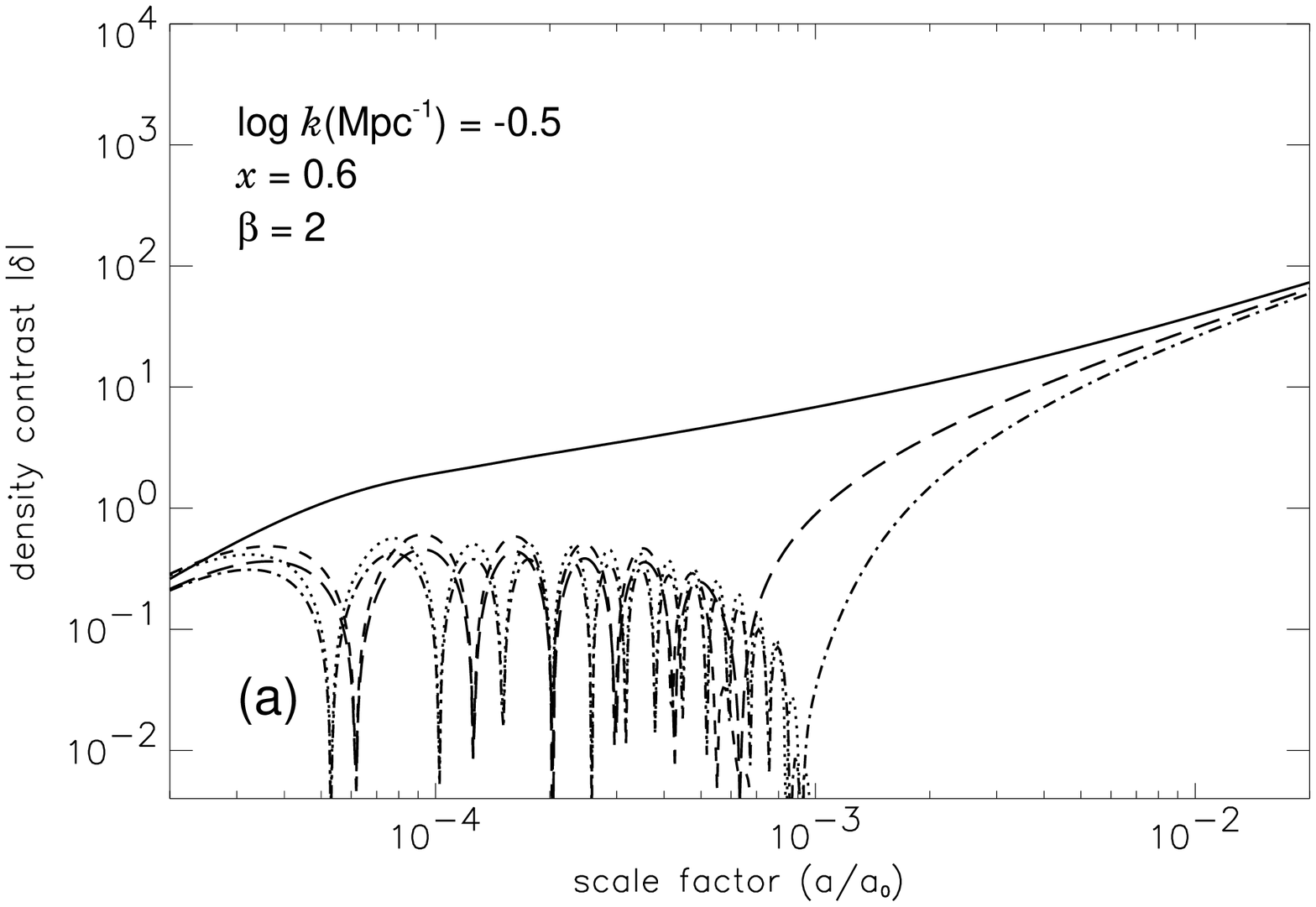},width=9.0cm}}
\centerline{\psfig{file={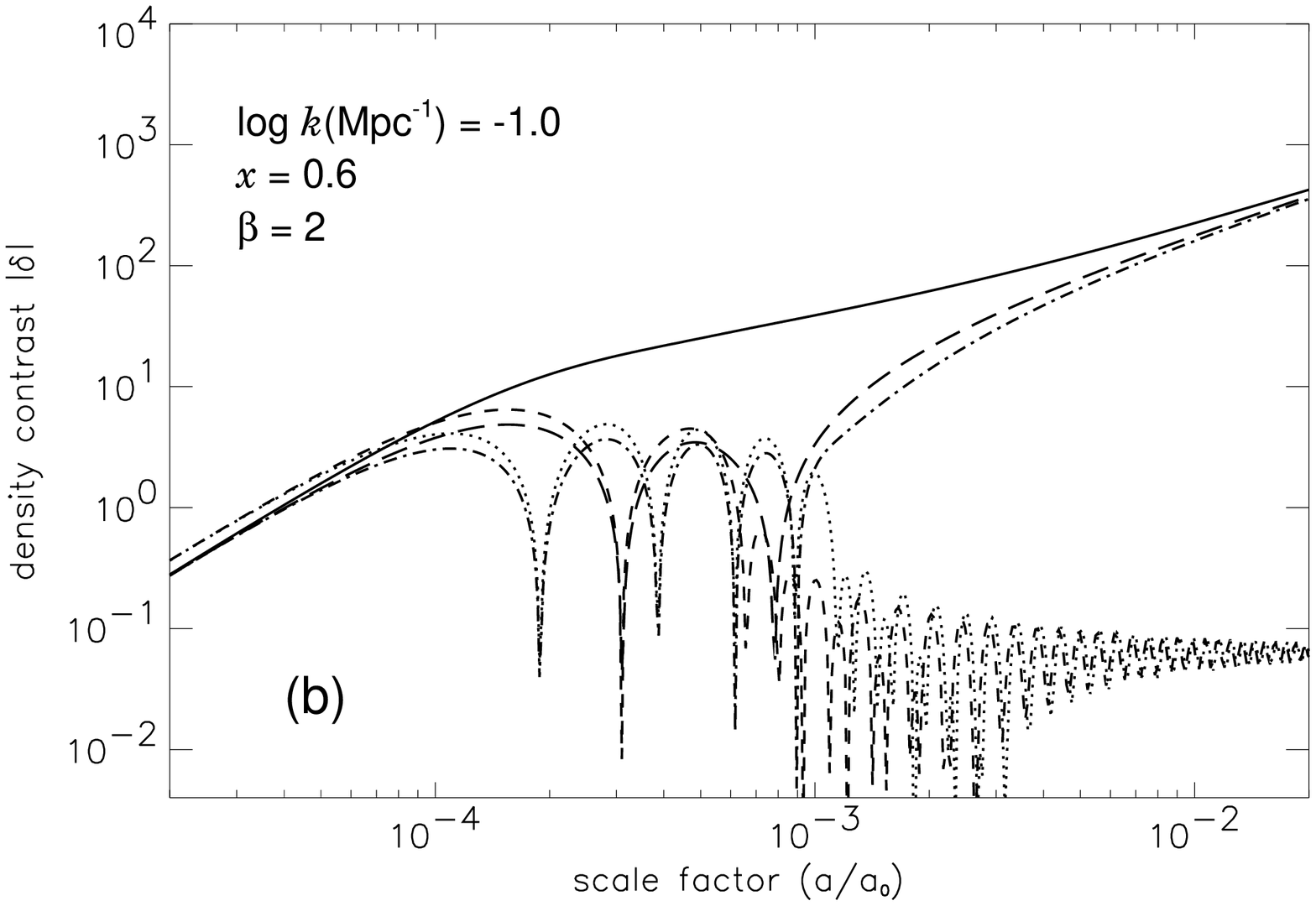},width=9.0cm}}
\vspace*{4pt}
\caption{Evolution of perturbations for the components of a Mirror Universe: cold dark matter (solid line), ordinary baryons and photons (dot-dashed and dotted) and mirror baryons and photons (long dashed and dashed). 
The model is a flat Universe with $ \Omega_m = 0.3 $, $ \Omega_b h^2 = 0.02 $, $ \Omega'_b h^2 = 0.04 $ ($ \beta = 2 $), $ h = 0.7 $, $ x = 0.6 $, and plotted scales are $ \log k ({\rm Mpc}^{-1}) = -0.5 $  ($ a $) and $ -1.0 $ ($ b $). \label{evol-x06-b2-k0510}}
\end{figure}

\begin{figure}[pb]
\centerline{\psfig{file={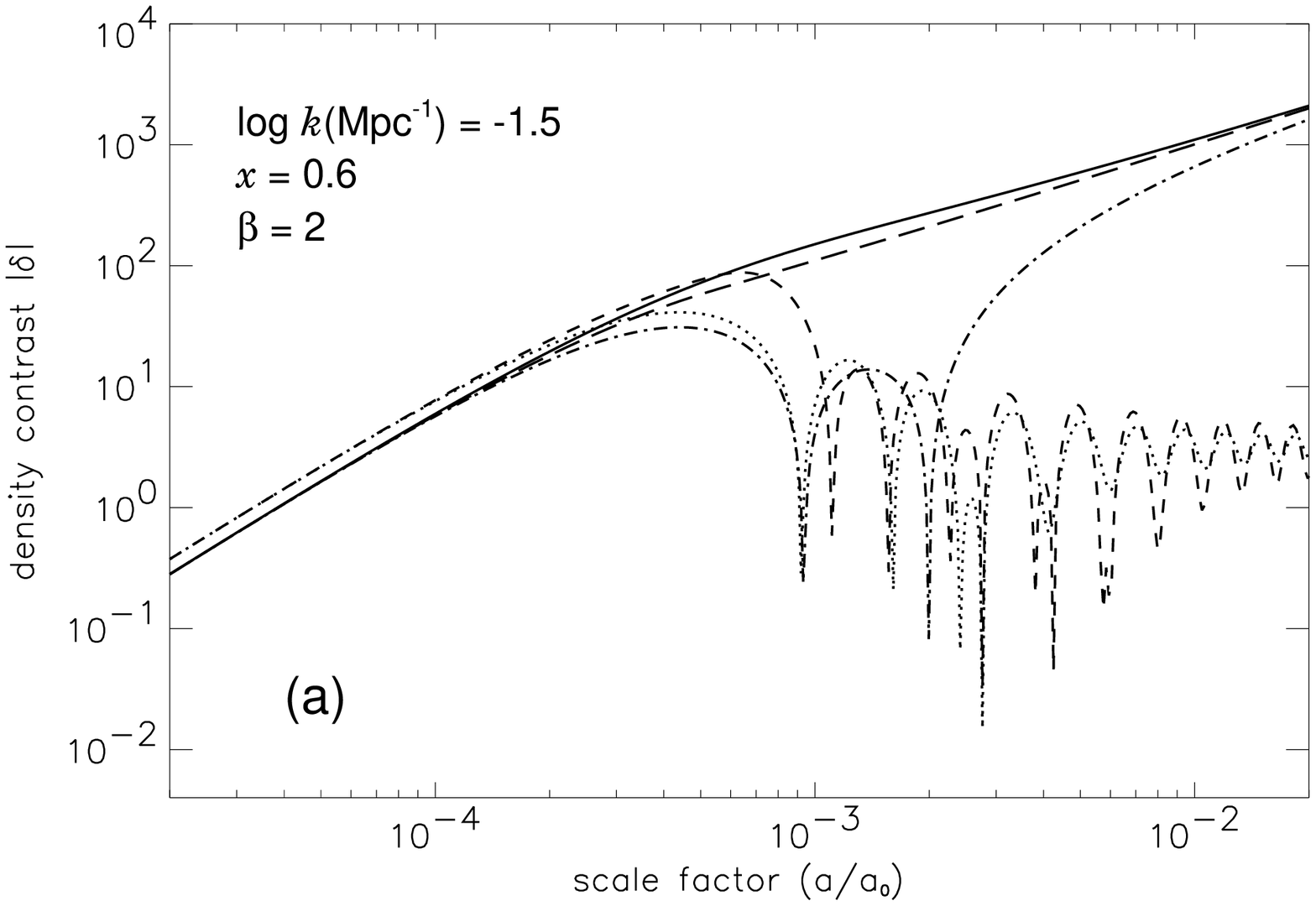},width=9.0cm}}
\centerline{\psfig{file={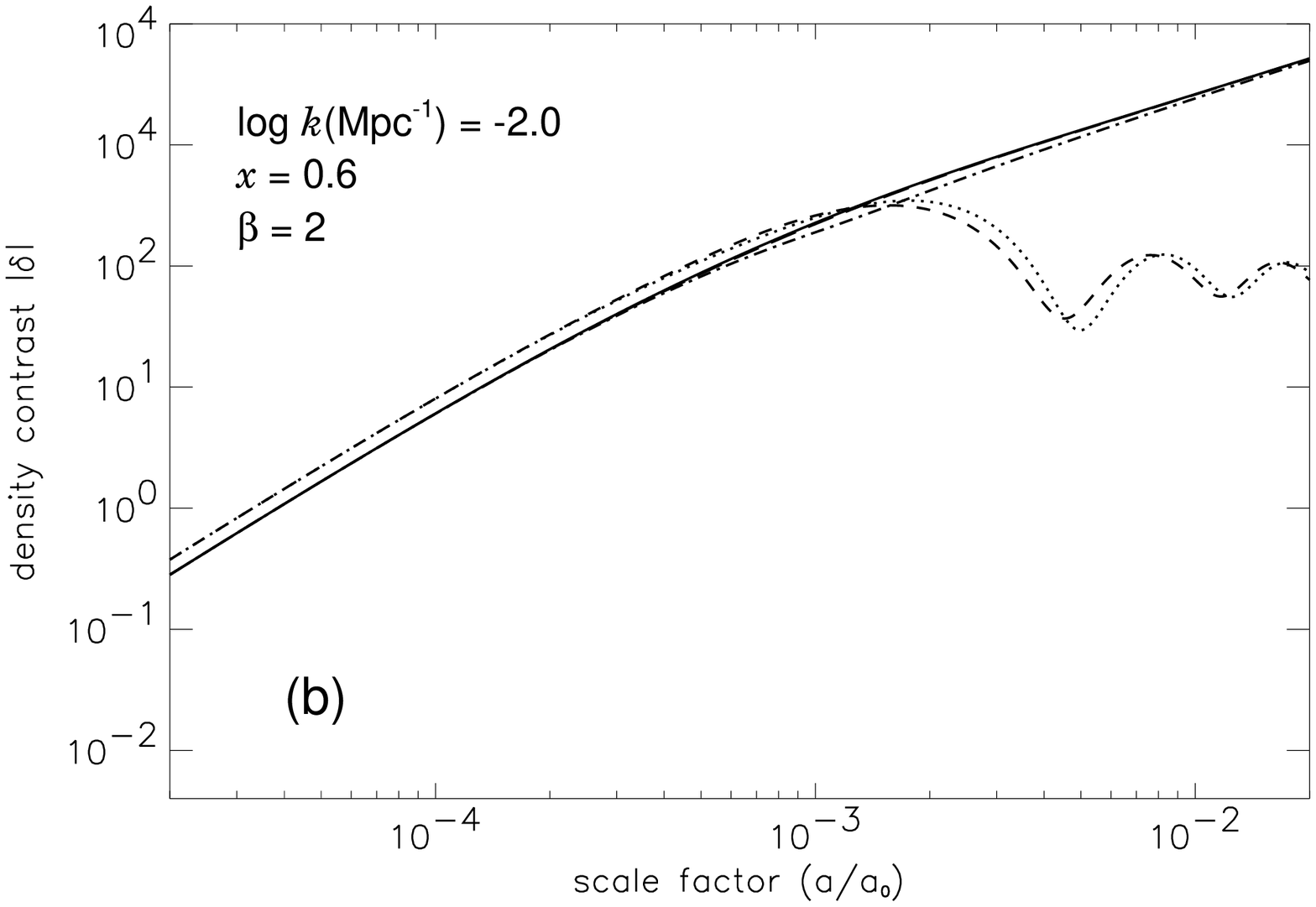},width=9.0cm}}
\vspace*{4pt}
\caption{The same as in Fig.~\ref{evol-x06-b2-k0510}, but for scales $ \log k ({\rm Mpc}^{-1}) = -1.5 $  ($ a $) and $ -2.0 $ ($ b $). \label{evol-x06-b2-k1520}}
\end{figure}

First of all, we comment Fig.~\ref{evol-x06-b2-k0510}(b), which is the most useful to recognize the general features of the evolution of perturbations. 
Starting from the smallest scale factor, we see that all three matter components and the two radiative components grow with the same trend (as $ a^2 $), but the radiative ones have a slightly higher density contrast (with a constant rate until they are tightly coupled); this is simply the consequence of considering adiabatic perturbations, which are linked in their matter and radiation components by the adiabatic condition (\ref{adcond}). 
This is the situation when the perturbation is out of horizon, but, when it crosses the horizon, around $ a \sim 10^{-4} $, things drastically change. 
Baryons and photons, in each sector separately, become causally connected, feel each other, and begin to oscillate for the competitive effects of gravity and pressure. 
Meanwhile, the CDM density perturbation continues to grow uninterruptedly, at first reducing his rate from $ a^2 $ to $ \ln a $ (due to the rapid expansion during the radiation era), and later, as soon as MRE occurs (at $ a \sim 3 \times 10^{-3} $ for the considered model), increasing proportionally to $ a $. 
The oscillations of baryons and photons continue until their decoupling, which in the mirror sector occurs before than in the ordinary one (scaled by the factor $ x $). 
This moment is marked in the plot as the point where the lines for the two components move away one from the other. 
From this point, the photons in both sectors continue the oscillations until they are completely damped, while the mirror and ordinary baryons rapidly fall into the potential wells created by the cold dark matter and start growing as $ a $. 
We remark that it's important the way in which the oscillation reaches the decoupling; if it is expanding, first it slows down (as if it continues to oscillate, but disconnected from the photons), and then it compresses driven by the gravity; if, otherwise, it is compressing, it directly continues the compression and we see in the plot that it immediately stops to oscillate. 
In this figure we have the first behaviour in the mirror sector, the second one in the ordinary sector.

\begin{figure}[pb]
\centerline{\psfig{file={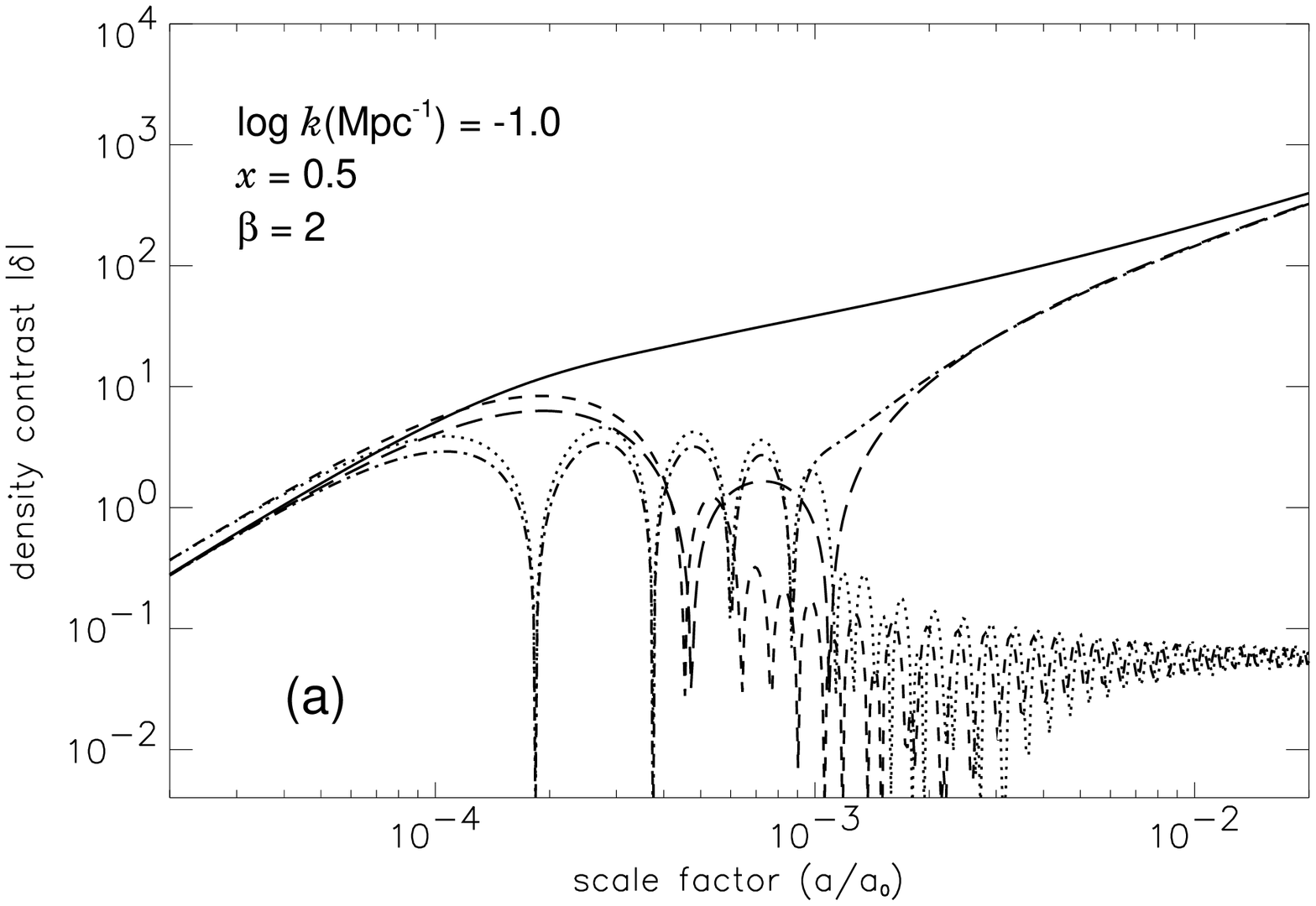},width=9.0cm}}
\centerline{\psfig{file={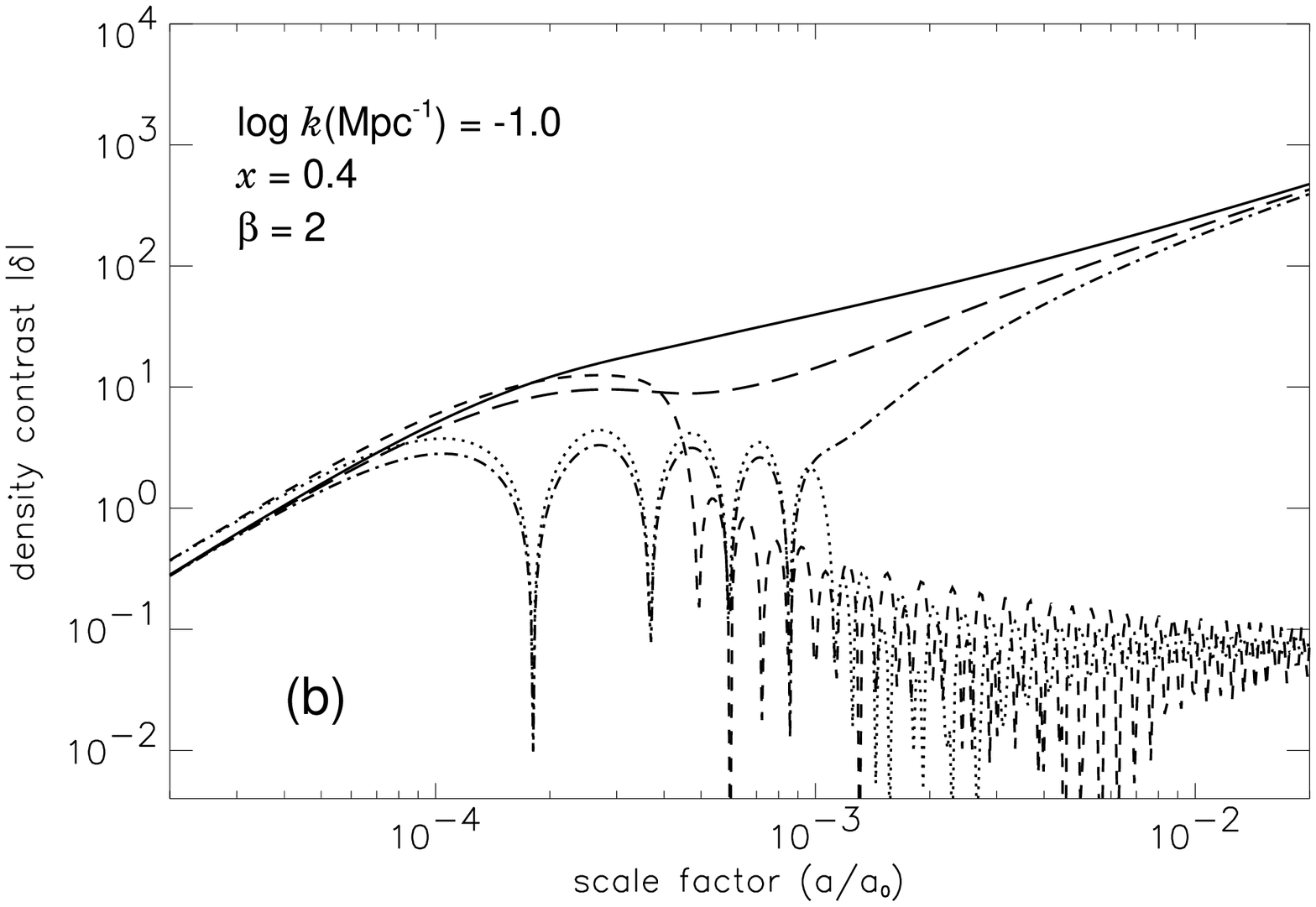},width=9.0cm}}
\vspace*{4pt}
\caption{Evolution of perturbations for the components of a Mirror Universe: cold dark matter (solid line), ordinary baryons and photons (dot-dashed and dotted) and mirror baryons and photons (long dashed and dashed). The model is a flat Universe with $ \Omega_m = 0.3 $, $ \Omega_b h^2 = 0.02 $, $ \Omega'_b h^2 = 0.04 $ ($ \beta = 2 $), $ h = 0.7 $, $ x = 0.5 $ ($ a $) or $ 0.4 $ ($ b $), and plotted scale is $ \log k ({\rm Mpc}^{-1}) = -1.0 $. \label{evol-x0504-b2-k10}}
\end{figure}

In Figs.~\ref{evol-x06-b2-k0510} and \ref{evol-x06-b2-k1520} we compare the behaviours of different scales for the same model. 
The scales are given by $ \log k (\rm Mpc^{-1}) = -0.5, -1.0, -1.5, -2.0 $, where $ k = 2 \pi / \lambda $ is the wave number.
The effect of early entering the horizon of small scales (those with higher $ k $) is evident. 
Going toward bigger scales the superhorizon growth continue for a longer time, delaying more and more the beginning of acoustic oscillations, until it occurs out of the coordinate box for the bigger plotted scale ($ \log k = -3.0 $).
Starting from the scale $ \log k = -1.5 $, the mirror decoupling occurs before the horizon entry of the perturbations, and the evolution of the mirror baryons density is similar to that of the CDM. 
The same happens to the ordinary baryons too, but for $ \log k \lesssim -2.0 $ (since they decouple later), while the evolution of mirror baryons is yet indistinguishable from that of the CDM. 
For bigger scales ($ \log k \lesssim -2.5 $), that are not shown, the evolutions of all three matter components are identical.

\begin{figure}[pb]
\centerline{\psfig{file={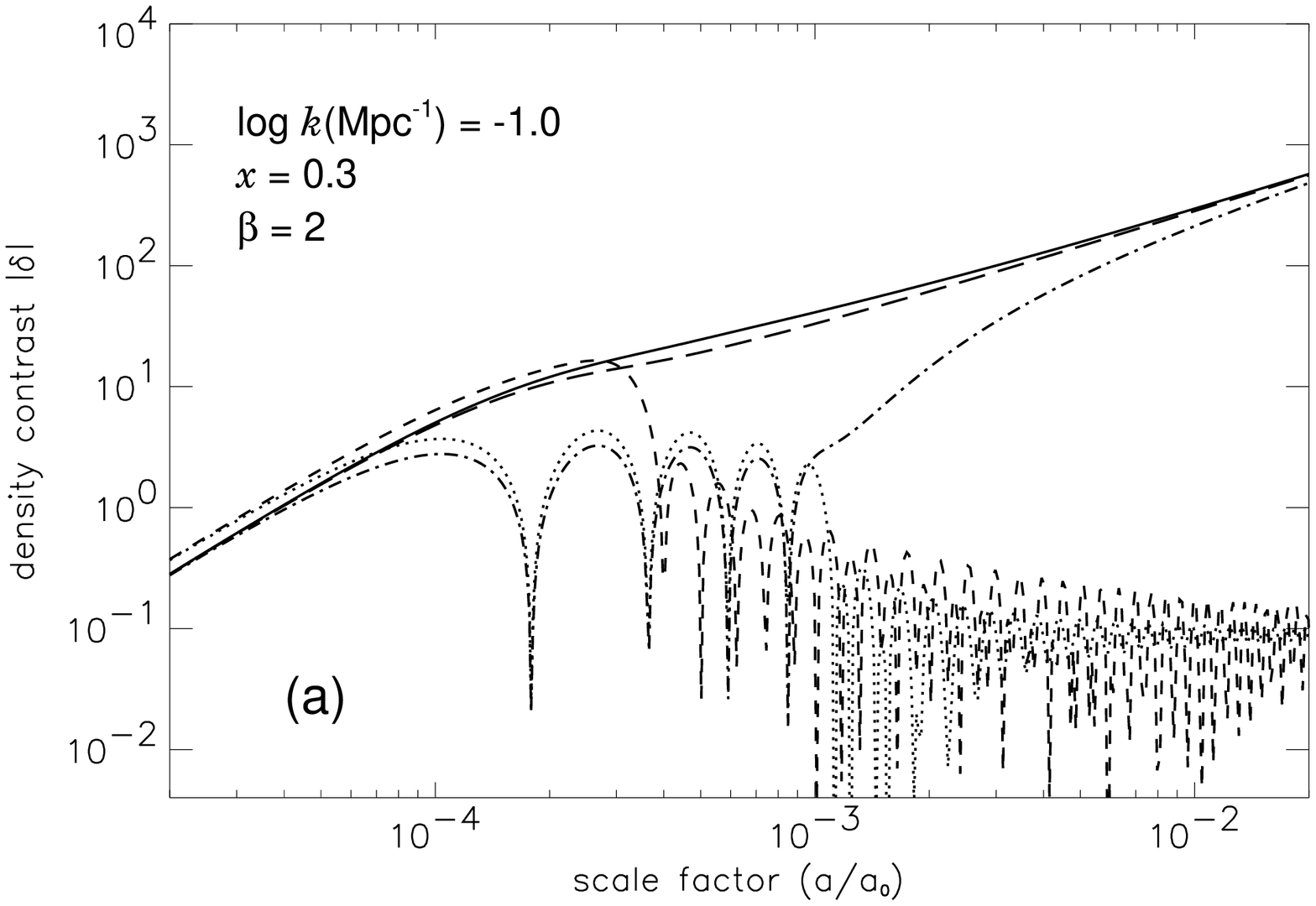},width=9.0cm}}
\centerline{\psfig{file={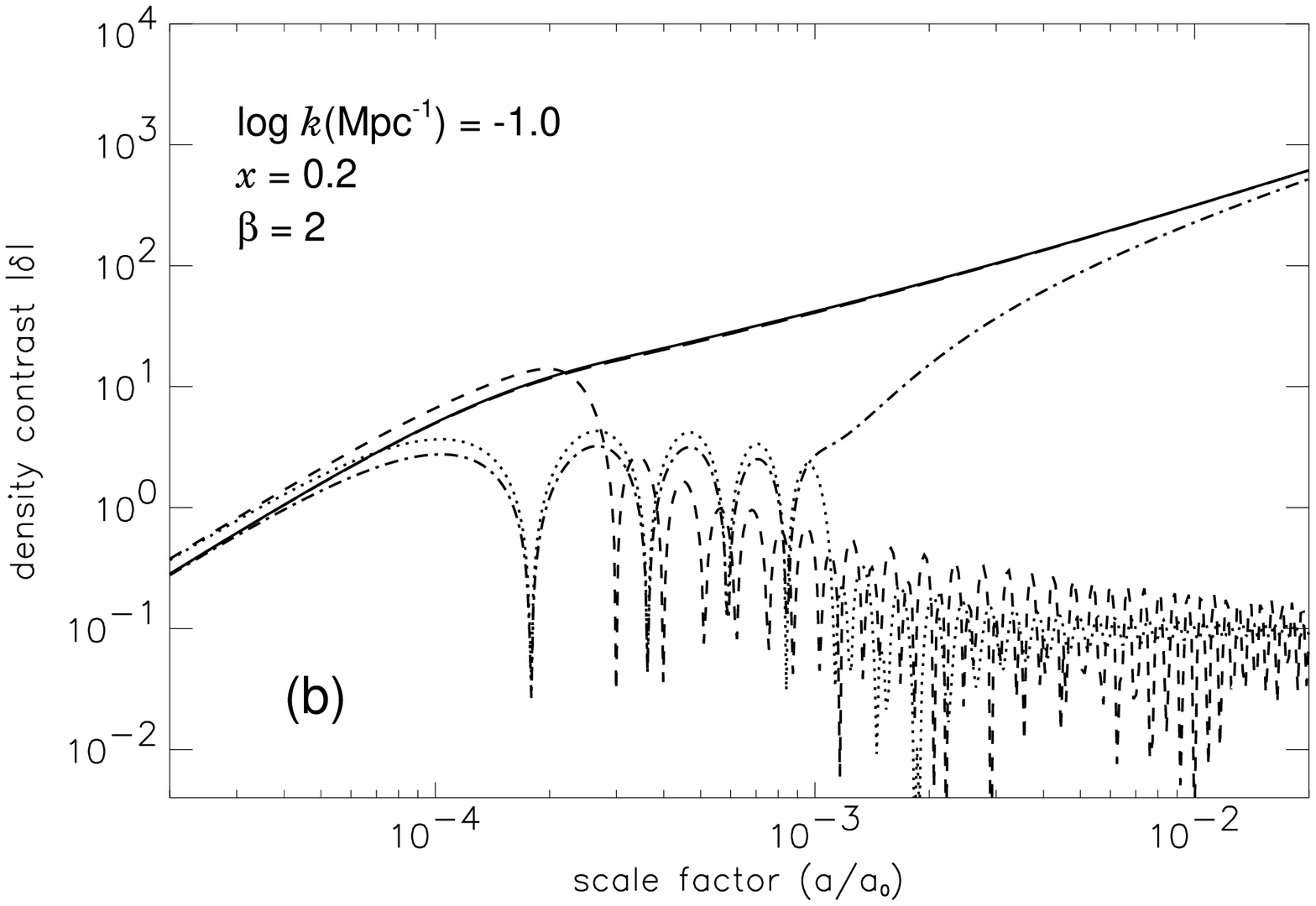},width=9.0cm}}
\vspace*{4pt}
\caption{The same as in Fig.~\ref{evol-x0504-b2-k10}, but for $ x = 0.3 $ ($ a $) and $ 0.2 $ ($ b $). \label{evol-x0302-b2-k10}}
\end{figure}

\begin{figure}[pb]
\centerline{\psfig{file={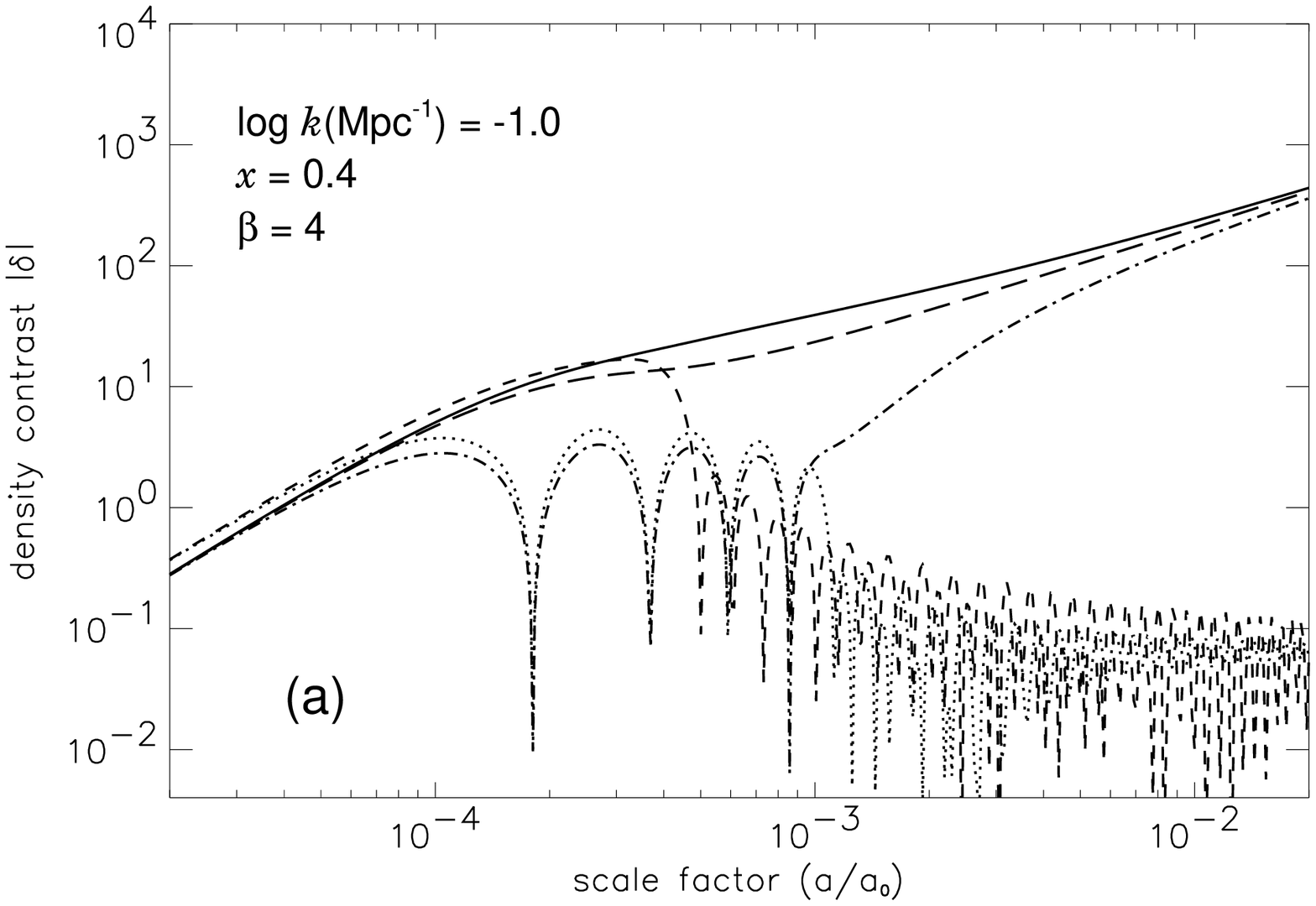},width=9.0cm}}
\vspace{-4.5cm}
\centerline{\psfig{file={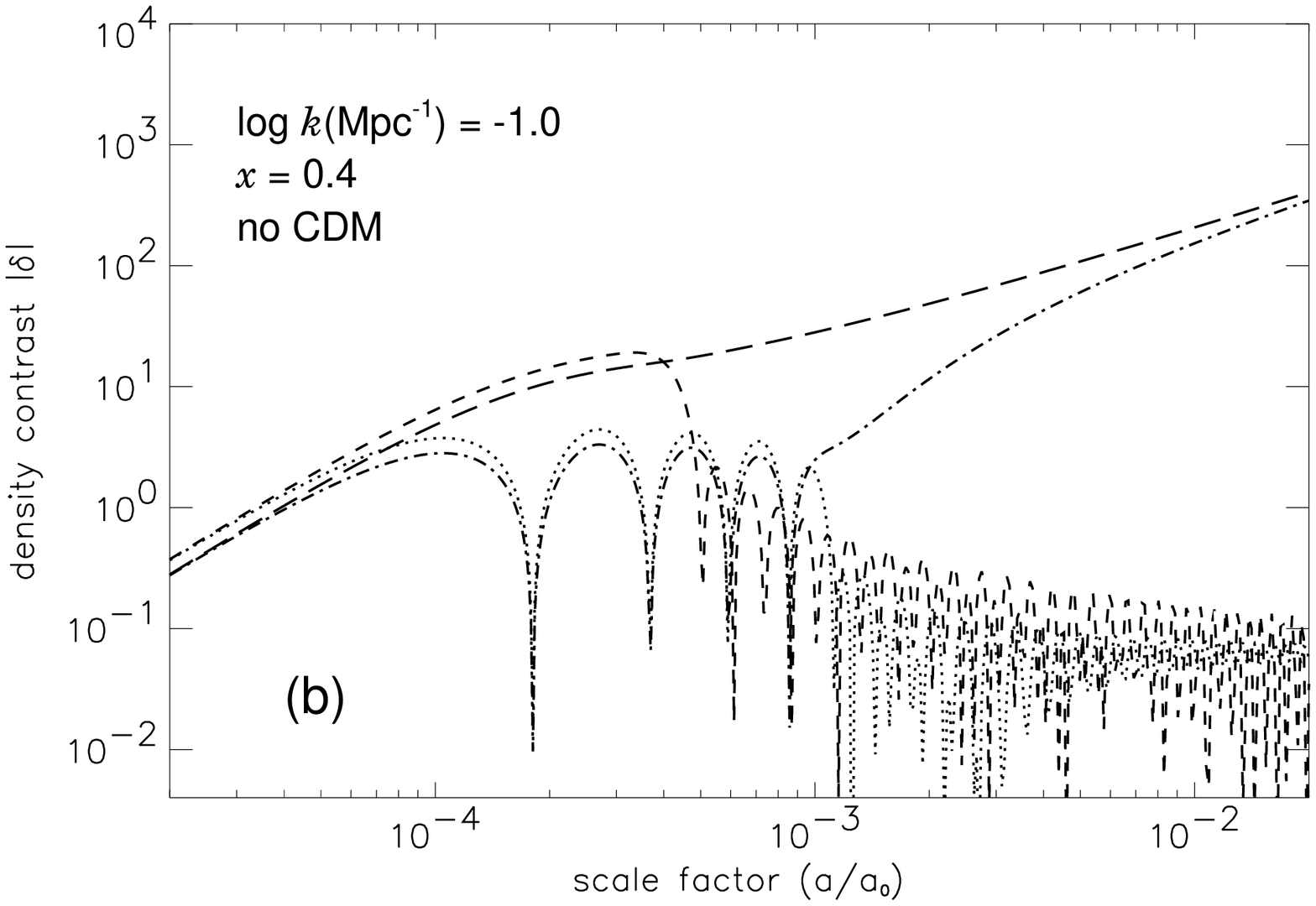},width=9.0cm}}
\vspace*{4pt}
\caption{Evolution of perturbations for the components of a Mirror Universe: cold dark matter (solid line), ordinary baryons and photons (dot-dashed and dotted) and mirror baryons and photons (long dashed and dashed). The models are flat with $ \Omega_m = 0.3 $, $ \Omega_b h^2 = 0.02 $, $ \Omega'_b = 4 \Omega_b $ ($ a $) or $ (\Omega_m - \Omega_b) $ (no CDM) ($ b $), $ h = 0.7 $, $ x = 0.4 $, and $ \log k ({\rm Mpc}^{-1}) = -1.0 $. \label{evol-x04-b4nocdm-k10}}
\end{figure}

As previously seen, the decoupling is a crucial point for structure formation, and it assumes a fundamental role specially in the mirror sector, where it occurs before than in the ordinary one: mirror baryons can start before 
growing perturbations in their density distribution. 
For this reason it's important to analyze the effect of changing the mirror decoupling time, obtained changing the value of $ x $ and leaving unchanged all other parameters, as it is possible to do using Figs.~\ref{evol-x06-b2-k0510}(b), \ref{evol-x0504-b2-k10} and \ref{evol-x0302-b2-k10} for $ x = 0.6, 0.5, 0.4, 0.3, 0.2 $ and the same scale $ \log k = -1.0 $.
It is evident the shift of the mirror decoupling toward lower values of $ a $ when reducing $ x $, according to the law (\ref{z'_dec}), which states a direct proportionality between the two. 
In particular, for $ x < x_{\rm eq} \approx 0.3 $ mirror decoupling occurs before the horizon crossing of the perturbation, and mirror baryons mimic more and more the CDM, so that for $ x \simeq 0.2 $ the perturbations in the two components are indistinguishable. 
For the ordinary sector apparently there are no changes, but at a more careful inspection we observe some difference due to the different amount of relativistic mirror species (proportional to $ x^4 $), which slightly shifts the matter-radiation equality. 

Obviously, these are cases where the CDM continues to be the dominant form of dark matter, and drives the growth of perturbations, given its continuous increase. 
In any case, if the dominant form of dark matter is made of mirror baryons the situation is practically the same, as visible comparing Figs.~\ref{evol-x0504-b2-k10}(b) and \ref{evol-x04-b4nocdm-k10}(a) (where we see only slight differences on the CDM and mirror baryons behaviours in the central region of the plots), since mirror baryons decouple before than ordinary ones and fall into the potential wells of the CDM, reinforcing them.

Finally, in the interesting case where mirror baryons constitute {\em all} the dark matter, they drive the evolution of perturbations. 
In fact, in Fig.~\ref{evol-x04-b4nocdm-k10}(b) we clearly see that the density fluctuations start growing in the mirror matter and the visible baryons are involved later, after being recombined, when they rewrite the spectrum of already developed mirror structures. 
This is another effect of a mirror decoupling occurring earlier than the ordinary one: the mirror matter can drive the growth of perturbations in ordinary matter, and provide the rapid growth soon after recombination, that is necessary to take into account of the evolved structures that we see today.

After all these considerations, it is evident that the case of mirror baryons is very interesting for structure formation, because they are collisional between themselves but collisionless for the ordinary sector, or, in other words, they are self-collisional. 
In this situation baryons and photons in the mirror sector are tightly coupled until decoupling, and structures cannot grow before this time, but the mirror decoupling happens before the ordinary one, thus structures have enough 
time to grow according to the limits imposed by CMB and LSS (something not possible in a purely ordinary baryonic scenario). 
Another important feature of the mirror dark matter scenario is that, if we consider small values of $ x $, the evolution of primordial perturbations is very similar to the CDM case, but with a fundamental difference: there exists a cutoff scale due to the mirror Silk damping, which kills the small scales, overcoming the problems of the CDM scenario with the excessive number of small satellites.


\section{Cosmic microwave background and large scale structure}

In the last decade the study of the the Cosmic Microwave Background (CMB) and the Large Scale Structure (LSS) is providing a great amount of observational data, and their continuous improvement provides powerful cosmological instruments that could help us to understand the nature of the dark matter of the Universe. 
In this context, the analysis of the CMB and LSS power spectra for a mirror baryonic dark matter scenario plays a key role as a cosmological test of the mirror theory. 

In this section we show the CMB and LSS spectra obtained in presence of mirror matter, together with the study of their consistence with observational data, that provides useful bounds on the mirror parameter space, stronger than those based on the analysis of BBN alone.\cite{Berezhiani:2003wj,Ciarcelluti:2003wm,Ciarcelluti:2004ip}


The models are numerically computed by means of the same program used for the computation of the temporal evolution of primordial perturbations (presented in the  previous section), assuming adiabatic scalar primordial perturbations, a flat space-time geometry, and different mixtures of ordinary and mirror matter and radiation, and cold dark matter.
Starting from an ordinary reference model, we study the influence of the mirror sector varying the two parameters that describe it for a given ordinary sector, replacing a fraction of CDM (or the entire dark matter) with mirror matter. 
Thus, we add to the usual ones two new mirror parameters: the ratio of the temperatures in the two sectors $ x = {T' / T} $ and the mirror baryons density $ \Omega'_b $ (also expressed via the ratio $ \beta $ of the baryonic densities in the two sectors).
The total and vacuum densities are fixed by our choice of a flat geometry: $ \Omega_0 = 1 $ and $ \Omega_\Lambda = 1 - \Omega_m $.


The shapes, heights and locations of peaks and oscillations in the photons and matter power spectra are predicted by all cosmological models based on the inflationary scenario, and their features represent specific signatures of the chosen cosmological parameters and of the composition of the Universe, which in turn can be accurately determined by precise measurements of these patterns. 
In particular, the exact form of the CMB and LSS power spectra is greatly dependent on assumptions about the matter content of the Universe. 

A detailed analysis of the dependence and the sensitivity of the mirror CMB and LSS power spectra on other cosmological parameters (matter density, ordinary baryon density, Hubble parameter, spectral index of scalar fluctuations, number of extra-neutrino species) was performed in Refs.~\refcite{Ciarcelluti:2003wm} and \refcite{Ciarcelluti:2004ip}, where the author compared the dependences on the number of massless neutrino species $ N_\nu $ and on the ratio of temperatures $ x $.
As expected, since we remember that the relativistic mirror particles can be parametrized in terms of effective number of extra-neutrino species, their effects on the CMB are similar, but different for the LSS.


\subsection{The cosmic microwave background for a Mirror Universe}
\label{cmb_2}

\begin{figure}[pb]
\centerline{\psfig{file={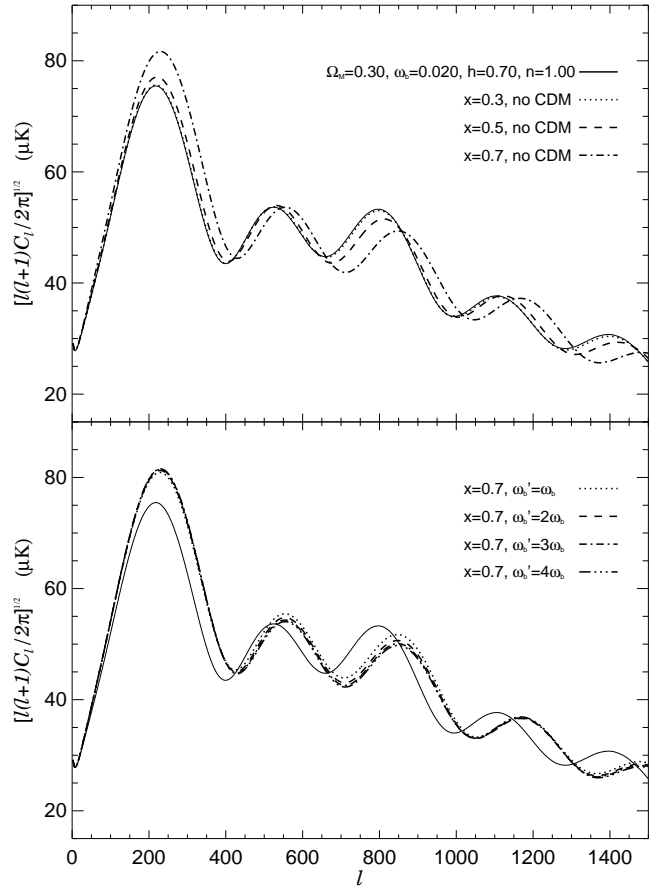},width=8.5cm}}
\vspace*{4pt}
\caption{CMB angular power spectrum for different values of $ x $ and $ \omega_b' = \Omega_b' h^2 $, compared with a standard model (solid line). {\sl Top panel.} Mirror models with the same parameters as the ordinary one, and with $ x = 0.3, 0.5, 0.7 $ and $ \omega_b' = \Omega_m h^2 - \omega_b $ (no CDM) for all models. {\sl Bottom panel.} Mirror models with the same parameters as the ordinary one, and with $ x = 0.7 $ and $ \omega_b' = \omega_b, 2 \omega_b, 3 \omega_b, 4 \omega_b $. \label{cmblssfig1}}
\end{figure}

As anticipated when we studied the structure formation in the previous section, we expect that the existence of a mirror sector influences the cosmic microwave background radiation observable today; in this section we want to evaluate this effect.

We choose a starting standard model and add a mirror sector simply removing cold dark matter and adding mirror baryons. 
The values of the parameters for this reference model are: $ \Omega_0 = 1 $, $ \Omega_m = 0.30 $, $ \Omega_\Lambda = 0.70 $, $ \omega_b = \Omega_b h^2 = 0.02 $, $ n = 1.00 $, $ h = 0.70 $, with only cold dark matter (no massive neutrinos) and scalar adiabatic perturbations (with spectral index $ n $).
This reference model is not the result of a fit, but is arbitrarily chosen consistent with the current knowledge of the cosmological parameters; however, this is not a shortcoming, because here we want only to put in evidence the differences from a representative reference model. 

From this starting point, first of all we substitute all the cold dark matter with mirror baryonic dark matter and evaluate the CMB angular power spectrum varying $ x $ from 0.3 to 0.7 (around the upper limit set by the BBN bounds). 
This is shown in top panel of Fig.~\ref{cmblssfig1}, where mirror models are plotted together with the reference model. 
The first evidence is that the deviation from the standard model is not linear in $ x $: it grows more for bigger $ x $ and for $ x \lesssim 0.3 $ the power spectra are practically coincident. 
This is important, because it means that a Universe where all the dark matter is made of mirror baryons could be indistinguishable from a CDM model if we analyze the CMB only. 
We see the largest separation from the reference model for $ x = 0.7 $, but it would increase for hypothetical larger values of $ x $. 
The height of the first acoustic peak grows for $ x \gtrsim 0.3 $, while the position remains nearly constant. 
For the second peak the opposite occurs, i.e. the height remains practically constant, while the position shifts toward higher multipoles $ l $; for the third peak, instead, we have a shift both in height and position (the absolute shifts are similar to the ones for the first two peaks, but the height now decreases instead of increasing).
Observing also other peaks, we recognize a general pattern: except for the first one, odd peaks change both height and location, even ones change location only.

In bottom panel of Fig.~\ref{cmblssfig1} we show the intermediate case of a mixture of CDM and mirror matter. 
We consider  $ x = 0.7$, a high value which permits us to see well the differences obtained changing $ \omega_b' $ from  $ \omega_b $ to $ 4 \omega_b $. 
The dependence on the amount of mirror baryons is lower than on the ratio of temperatures $ x $. 
In fact, the position of the first peak is nearly stable for all mirror models (except for a very low increase of height for growing $ \omega_b' $), while differences appear for other peaks. 
In the second peak the position is shifted as in the case without CDM independently of $ \omega_b' $, while the height is inversely proportional to $ \omega_b' $ with a separation appreciable for $ \omega_b' \lesssim 3 \omega_b $. 
For the third peak the behaviour is the same as for the case without CDM, with a slightly stronger dependence on $ \omega_b' $, while for the other peaks there is a weaker dependence on $ \omega_b' $. 
A common feature is that the heights of the peaks are not linearly dependent on the mirror baryonic density, while their positions are practically insensitive to $ \omega_b' $ but depend only on $ x $.

The dependence of the peaks on $ x $ and $ \omega_b' $, together with other parameters, is analyzed in more details in Refs.~\refcite{Ciarcelluti:2003wm} and \refcite{Ciarcelluti:2004ip}.


\subsubsection{The mirror cosmic microwave background radiation}
\label{cmb_3}

In the same way as ordinary photons at decoupling from baryons formed the CMB we observe today, also mirror photons at their decoupling formed a mirror cosmic microwave background radiation, which unfortunately we cannot observe because they don't couple with the ordinary baryons of which we are made.
Indeed, there is in principle the possibility that mirror CMB photons could influence our observable CMB in case of existence of a tiny photon--mirror photon kinetic mixing,\cite{Holdom:1985ag,Foot:1991kb,Foot:2000vy} but its detection would not be possible with present experiments, given the very low estimates for the cross section of this interaction.
Nevertheless, the study of mirror CMB power spectrum is not only speculative, since it is a way to better understand the cosmology with mirror matter and our observable CMB.

\begin{figure}[pb]
\centerline{\psfig{file={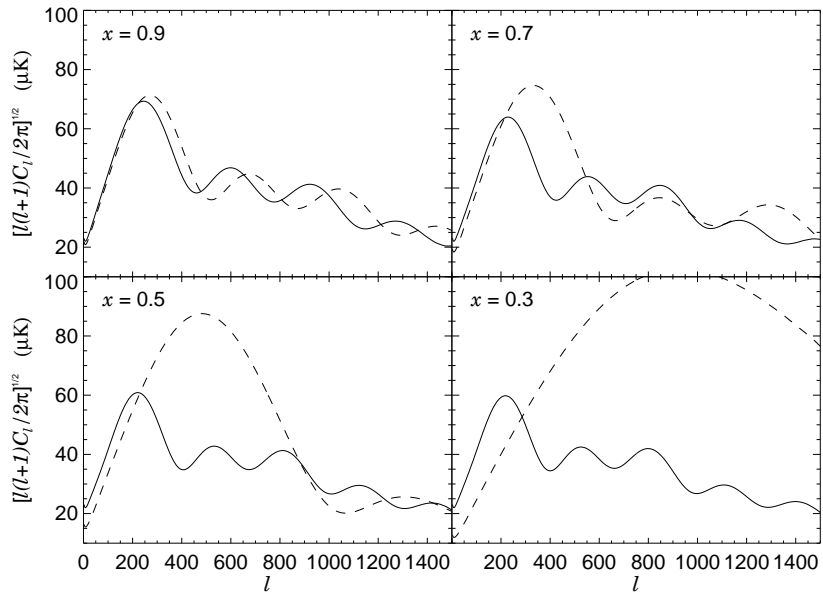},width=10.8cm}}
\vspace*{4pt}
\caption{Angular power spectra for ordinary (solid line) and mirror (dashed line) CMB photons. The models have $ \Omega_0 = 1 $, $ \Omega_m = 0.3 $, $ \omega_b = \omega_b' = 0.02 $, $ h = 0.7 $, $ n = 1.0 $, and $ x = 0.9 $ (top-left panel), $ x = 0.7 $ (top-right panel), $ x = 0.5 $ (bottom-left panel)$, x = 0.3 $ (bottom-right panel). \label{cmblssfig2}}
\end{figure}

We computed four models of mirror CMB, in order to have enough elements to compare with the corresponding observable CMBs. 
The chosen parameter values are those usually taken, and the amount of mirror baryons is the same as the ordinary ones, while $ x $ is taken as 0.9, 0.7, 0.5 or 0.3 in order to explore different scenarios. 
Thus the parameters of the models are: $ \Omega_0 = 1 $, $ \Omega_m = 0.3 $, $ \omega_b = \omega_b' = 0.02 $, 
$ x = 0.7$ or 0.5, $ h = 0.7 $, $ n = 1.0 $.
In Fig.~\ref{cmblssfig2} we plot the ordinary and mirror CMB spectra corresponding to the same model of Mirror Universe. 

The first evidence is that, being scaled by the factor $ x $ the temperatures in the two sectors, also their temperature fluctuations will be scaled by the same amount, as evident if we look at the lowest $ \ell $ values (the 
fluctuations seeds are the same for both sectors). 
Starting from the top-left panel of Fig.~\ref{cmblssfig2}, we see that the first mirror CMB peak is higher and shifted to higher multipoles than the ordinary one, while other peaks are both lower and at higher $ \ell $ values, with a shift growing with the order of the peak. 

Observing all the panels, we understand the effect of a change of the parameter $ x $ on the mirror CMB: 
(i) for lower $ x $ values the first peak is higher (for $ x = 0.5 $ it is nearly 1.5 the ordinary one); 
(ii) the position shifts to much higher multipoles (so that with the same horizontal scale we can no more see some peaks). 
The reason is that a change of $ x $ corresponds to a change of the mirror decoupling time. 
The mirror photons, which decouple before the ordinary ones, see a smaller sound horizon, scaled approximately by the factor $ x $; since the first peak occurs at a multipole $ \ell \propto ({\rm sound\;horizon}) ^{-1} $, we expect it to shift to higher values of $ \ell $ by a factor $ x^{-1} $, that is exactly what we observe in Fig.~\ref{cmblssfig2}.

It's possible to verify that increasing $ x $ the mirror CMB is more and more similar to the ordinary one, until for $ x = 1 $ (not shown in figure) the two power spectra are perfectly coincident (as expected, since in this case the two sectors have exactly the same temperatures, the same particle contents, and then their photon power spectra are necessarily the same). 

If we were able to detect both the ordinary and mirror CMB photons, we had two snapshots of the Universe at two different epochs, which were a powerful cosmological instrument, but unfortunately this is impossible with only gravitational interactions, because mirror photons are by definition completely invisible for us.


\subsection{The large scale structure for a Mirror Universe}
\label{lss_2}

Given the oscillatory behaviour of the mirror baryons (different from the smooth one of cold dark matter), we expect that mirror matter induces specific signatures also on the large scale structure power spectrum. 

\begin{figure}[pb]
\centerline{\psfig{file={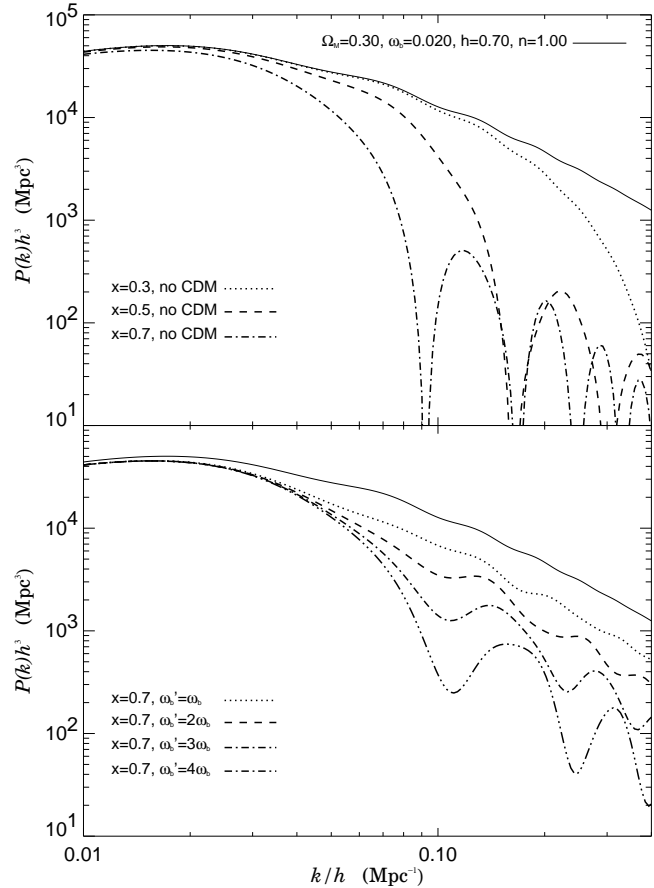},width=8.5cm}}
\vspace*{4pt}
\caption{LSS power spectrum in the linear regime for different values of $ x $ and $ \omega_b' = \Omega_b' h^2 $, compared with a standard model (solid line). In order to remove the dependences of units on the Hubble constant, we plot on the $ x $-axis the wave number in units of $ h $ and on the $ y $-axis the power spectrum in units of $ h^3 $. {\sl Top panel.} Mirror models with the same parameters as the ordinary one, and with $ x = 0.3, 0.5, 0.7 $ and $ \omega_b' = \Omega_m h^2 - \omega_b $ (no CDM) for all models. {\sl Bottom panel.} Mirror models with the same parameters as the ordinary one, and with $ x = 0.7 $ and $ \omega_b' = \omega_b, 2 \omega_b, 3 \omega_b, 4 \omega_b $. \label{cmblssfig3}}
\end{figure}

In order to evaluate this effect, we computed LSS power spectra using the same reference and mirror models used in Sec.~\ref{cmb_2} for the CMB analysis. 
The two panels of Fig.~\ref{cmblssfig3} show the LSS for the same models as in Fig.~\ref{cmblssfig1}. 
In order to remove the dependences of units on the Hubble constant, we plot, as usual, on the $ x $-axis the wave number in units of $ h $ and on the $ y $-axis the power spectrum in units of $ h^3 $. 
The minimum scale (the maximum $ k $) plotted is placed around the limit of the linear regime.

\begin{figure}[pb]
\centerline{\psfig{file={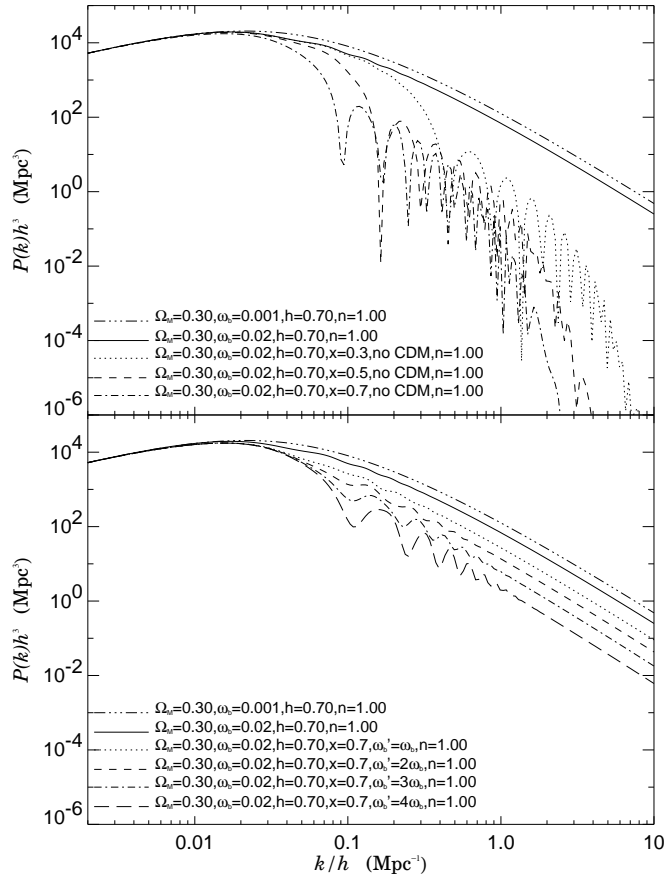},width=9.4cm}}
\vspace*{20pt}
\caption{LSS power spectrum beyond the linear regime for different values of $ x $ and $ \omega_b' = \Omega_b' h^2 $, compared with a standard model (solid line). The models have the same parameters as in Fig.~\ref{cmblssfig3}. For comparison we also show a standard CDM model with a negligible amount of baryons. 
\label{cmblssfig4}}
\end{figure}

\begin{figure}[pb]
\centerline{\psfig{file={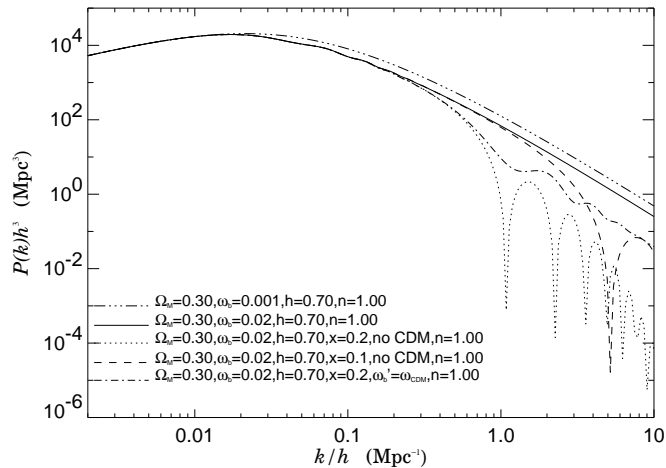},width=9.4cm}}
\vspace*{-142pt}
\caption{LSS power spectrum beyond the linear regime for two low values of $ x $ (0.1 and 0.2) and different amounts of mirror baryons ($ \Omega_b' = \Omega_m - \Omega_b $ or $ \Omega_b' = \Omega_{CDM} $), compared with a standard model (solid line). The other parameters are the same as in Fig.~\ref{cmblssfig3}. For comparison we show also a standard CDM model with a negligible amount of baryons.
\label{cmblssfig5}}
\end{figure}

In top panel of the figure we show the dependence on $ x $ for different mirror models without CDM; in this case, where all the dark matter is made of mirror baryons, the oscillatory effect is obviously maximum. 
The first evidence is the strong dependence on $ x $ of the beginning of oscillations: it goes to higher scales for higher $ x $, and below $ x \simeq 0.3 $ the power spectrum for a Mirror Universe approaches more and more the CDM one. 
This behaviour is a consequence of the $ x $ dependence of the mirror Silk scale, that increases for growing $ x $ (for details see Sec.~\ref{disseff-colldamp} and Refs.~\refcite{Berezhiani:2000gw,Ignatiev:2003js}--\refcite{Ciarcelluti:2004ij}): 
this dissipative scale induces a cutoff in the power spectrum, which is damped with an oscillatory behaviour (it will be more evident in Figs.~\ref{cmblssfig4} and \ref{cmblssfig5}, where we extend our models to smaller scales within the non linear region). 
Oscillations begin at the same time of the damping, and they are so deep (because there are many mirror baryons) to go outside the coordinate box. 
In any case the mirror spectra are always below the ordinary one for every value of $ x $.

The dependence on the amount of mirror baryons is instead shown in the bottom panel of the figure, where only a fraction of the dark matter is made of mirror baryons, while the rest is CDM. 
Contrary to the CMB case, the matter power spectrum strongly depends on $ \omega_b' $. 
The oscillations are deeper for increasing mirror baryon densities and the spectrum goes more and more away from the pure CDM one. 
We note also that the damping begins always at the same scale, and thus it depends only on $ x $ and not on $ \omega_b' $, as we know from the expression (\ref{ms_m}) of the mirror Silk scale obtained in Sec.~\ref{disseff-colldamp}.

The same considerations are valid for the oscillation minima, which become much deeper for higher mirror baryon densities, but shift very slightly to lower scales, so that their positions remain practically constant.


Let us now extend the behaviour of the matter power spectrum to lower scales, which already became non linear. 
Obviously, since our treatment is based on the linear theory, it is no longer valid in non linear regime. 
Nevertheless, even if it cannot be used for a comparison with observations, the extension of our models to these scales is very useful to understand the behaviour of the power spectrum in a mirror baryonic dark matter scenario, in particular concerning the position of the cutoff (we recall that its presence could help in avoiding the problem of the CDM scenario with the excessive number of small structures).

Therefore, in Figs.~\ref{cmblssfig4} and \ref{cmblssfig5} we extend the power spectra up to $ k/h = 10 $ Mpc$^{-1}$ (corresponding to galactic scales), well beyond the limit of the linear regime, given approximately by $ k/h < 0.4 $ Mpc$^{-1}$.

In Fig.~\ref{cmblssfig4} we plot in both panels the same models as in Fig.~\ref{cmblssfig3}.
For comparison we show also a standard model characterized by a matter density made almost completely of CDM, with only a small contamination of baryons ($ \Omega_b \simeq 0.2\% $ instead of $ \simeq 4 \% $ of other models). 
In top panel, the $ x $ dependence of the mirror power spectra is considered: the vertical scale extends to much lower values compared to Fig.~\ref{cmblssfig3}, and we can clearly see the deep oscillations, but in particular it is evident the presence of the previously cited cutoff. 
For larger values of $ x $ oscillations begin earlier and cutoff moves to higher scales. 
Moreover, note that the model with almost all CDM has more power than the same standard model with baryons, which in turn has more power than all mirror models for any $ x $ and for all the scales. 
In bottom panel we show the dependence on the mirror baryon content. 
It is remarkable that all mirror models stop to oscillate at some low scale and then continue with a smooth CDM-like trend. 
This means that, after the cutoff due to mirror baryons, the dominant behaviour is the one characteristic of cold dark matter models (due to the lack of a cutoff for CDM). 
Clearly, for higher mirror baryon densities the oscillations continue down to smaller scales, but, contrary to the previous case, where all the dark matter was mirror baryonic, there will always be a scale below which the spectrum 
is CDM-like.

An interesting point of the mirror baryonic scenario is his capability to mimic a CDM scenario under certain circumstances and for certain measurements. 
To explain this point, in Fig.~\ref{cmblssfig5} we show models with low $ x $ values (0.2 or 0.1) and all dark matter made of mirror baryons; we see that for $ x = 0.2 $ the standard and mirror power spectra are already practically 
coincident in the linear region. 
If we go down to $ x = 0.1 $ the coincidence is extended up to $ k/h \sim 1 $ Mpc$ ^{-1} $. 
In principle, we could still decrease $ x $ and lengthen this region of equivalence between the different CDM and mirror models, but we have to remember that we are dealing with linear models extended to non linear scales, then neglecting all the non linear phenomena (such as merging or stellar feedback), that are very different for the CDM and the mirror scenarios. 
In the same plot we also considered a model with $ x = 0.2 $ and dark matter composed equally by mirror baryons and CDM. 
This model shows that in principle it's possible a tuning of the cutoff effect reducing the amount of mirror matter, in order to better reproduce the cutoff needed to explain, for example, the low number of small satellites in 
galaxies.

These power spectra provide the linear transfer functions, which constitute the principal ingredient for the computation of the power spectrum at non linear scales.


\subsection{Comparison with observations}
\label{comp_obs}

So far we have studied the behaviour of the photon and matter power spectra varying the two mirror parameters, i.e. the ratio of the temperatures of two sectors $ x $ and the amount of mirror baryons $ \omega_b' $.

Here we compare these models with some experimental data, in order to estimate the compatibility of the mirror scenario with observations and possibly reduce the parameter ranges.

In the last decades the anisotropies observed in the CMB temperature became the most important source of information on the cosmological parameters: a lot of experiments (ground-based, balloon and satellite) were dedicated to its measurement. 
At the same time, many authors proved that its joint analysis with the fluctuations in the matter distribution (they have both the same primordial origin) are a powerful instrument to determine the parameters of the Universe. 
As in Secs.~\ref{cmb_2} and \ref{lss_2}, we analyze separately the variation of $ x $ and $ \omega_b' $ in the mirror models, using now both the CMB and LSS informations at the same time.

\begin{figure}[pb]
\centerline{\psfig{file={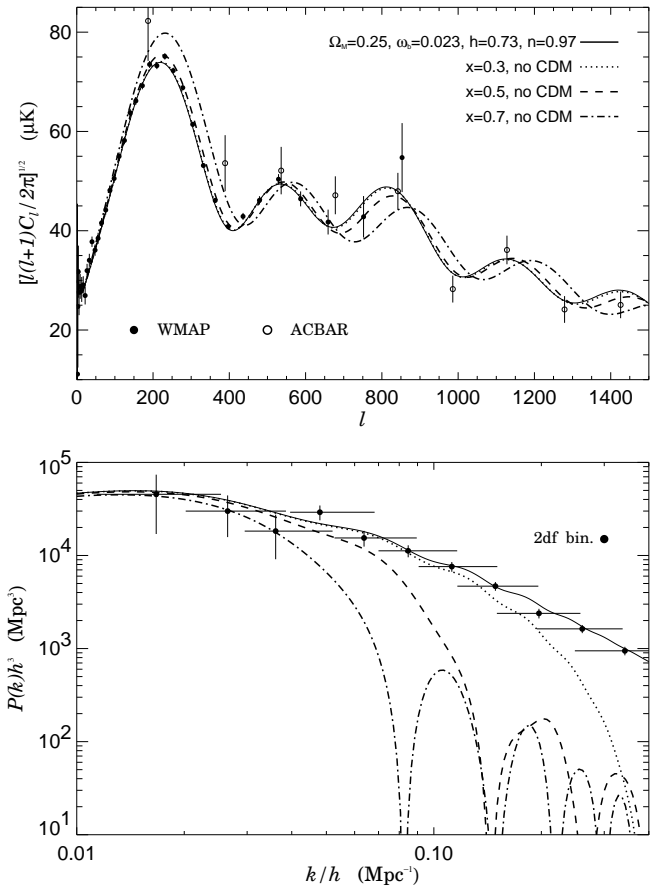},width=9.5cm}}
\vspace*{22pt}
\caption{CMB and LSS power spectra for various mirror models with different values of $ x $, compared with observations and with a standard reference model (solid line) of parameters $ \Omega_0 = 1 $, $ \Omega_m = 0.25 $, $ \Omega_\Lambda = 0.75 $, $ \omega_b = \Omega_b h^2 = 0.023 $, $ h = 0.73 $, $ n = 0.97 $. The mirror models have the same parameters as the standard one, but with $ x = 0.3, 0.5, 0.7 $ and $ \omega_b' = \Omega_m h^2 - \omega_b $ (no CDM) for all models. {\sl Top panel.} Comparison of the photon power spectrum with the WMAP and ACBAR data. {\sl Bottom panel.} Comparison of the matter power spectrum with the 2dF binned data. \label{cmblssfig18}}
\end{figure}

In order to compare our predictions with observations, we report the only available analysis of this kind, described in Refs.~\refcite{Berezhiani:2003wj,Ciarcelluti:2004ip}, where the authors used
for the CMB the WMAP\cite{Hinshaw:2003ex} and ACBAR\cite{Kuo:2002ua} data, and for the LSS the 2dF survey (in particular the binned power spectrum obtained by Tegmark {\it et al.}\cite{Tegmark:2001jh}).
The results obtained are still valid, even if better data are now available.
In order to compare with the standard CDM results, we use a reference cosmological model with scalar adiabatic perturbations and no massive neutrinos with the following set of parameters\cite{Spergel:2003cb}:
$ \Omega_m = 0.25, ~~ \omega_b = 0.023, ~~ \Omega_\Lambda = 0.75, ~~ h = 0.73, ~~ n = 0.97 $.
As usually, we include in this model the mirror sector; for the sake of comparison, in all calculations the total amount of matter $ \Omega_m = \Omega_{CDM} + \Omega_b + \Omega'_b$ is maintained constant. 
Mirror baryons contribution is thus always increased at the expenses of diminishing the CDM contribution.

\begin{figure}[pb]
\centerline{\psfig{file={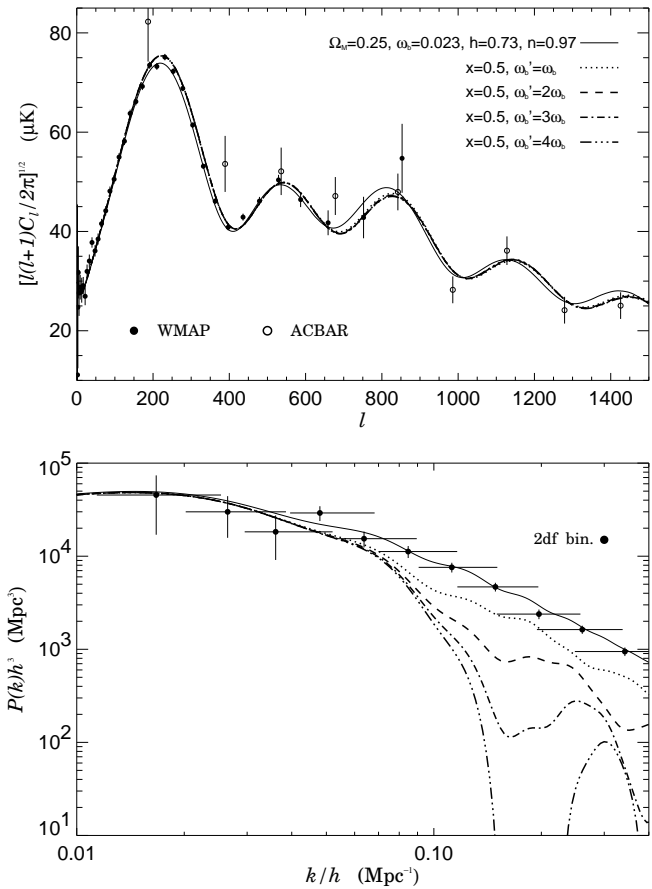},width=9.5cm}}
\vspace*{22pt}
\caption{CMB and LSS power spectra for various mirror models with different values of mirror baryon density, compared with observations and with a standard reference model (solid line) of parameters $ \Omega_0 = 1 $, $ \Omega_m = 0.25 $, $ \Omega_\Lambda = 0.75 $, $ \omega_b = \Omega_b h^2 = 0.023 $, $ h = 0.73 $, $ n = 0.97 $. The mirror models have the same parameters as the standard one, but with $ x = 0.5 $ and for $ \omega_b' = \omega_b, 2 \omega_b, 3 \omega_b, 4 \omega_b $. {\sl Top panel.} Comparison of the photon power spectrum with the WMAP and ACBAR data. {\sl Bottom panel.} Comparison of the matter power spectrum with the 2dF binned data. \label{cmblssfig19}}
\end{figure}

We start from Fig.~\ref{cmblssfig18}, where we assume that the dark matter is entirely due to mirror baryons and we consider variations of the $ x $ parameter (as in the upper panels of Figs.~\ref{cmblssfig1} and \ref{cmblssfig3}). 
In top panel, we see that with the accuracy of the anisotropy measurements the CMB power spectra for mirror models are perfectly compatible with data, except for the one with highest $ x $. 
Indeed, the deviations from the standard model are weak for $ x \lesssim 0.5 $, even in a Universe full of mirror baryons (see Sec.~\ref{cmb_2}). 
In lower panel, instead, the situation is very different: oscillations due to mirror baryons are too deep to be in agreement with data, and only models with low values of $ x $ (namely $ x \lesssim 0.3 $) are acceptable. 
Thus, we find the first strong constraint on the mirror parameter space: {\em models with high mirror sector temperatures and all the dark matter made of mirror baryons have to be excluded}.

\begin{figure}[pb]
\centerline{\psfig{file={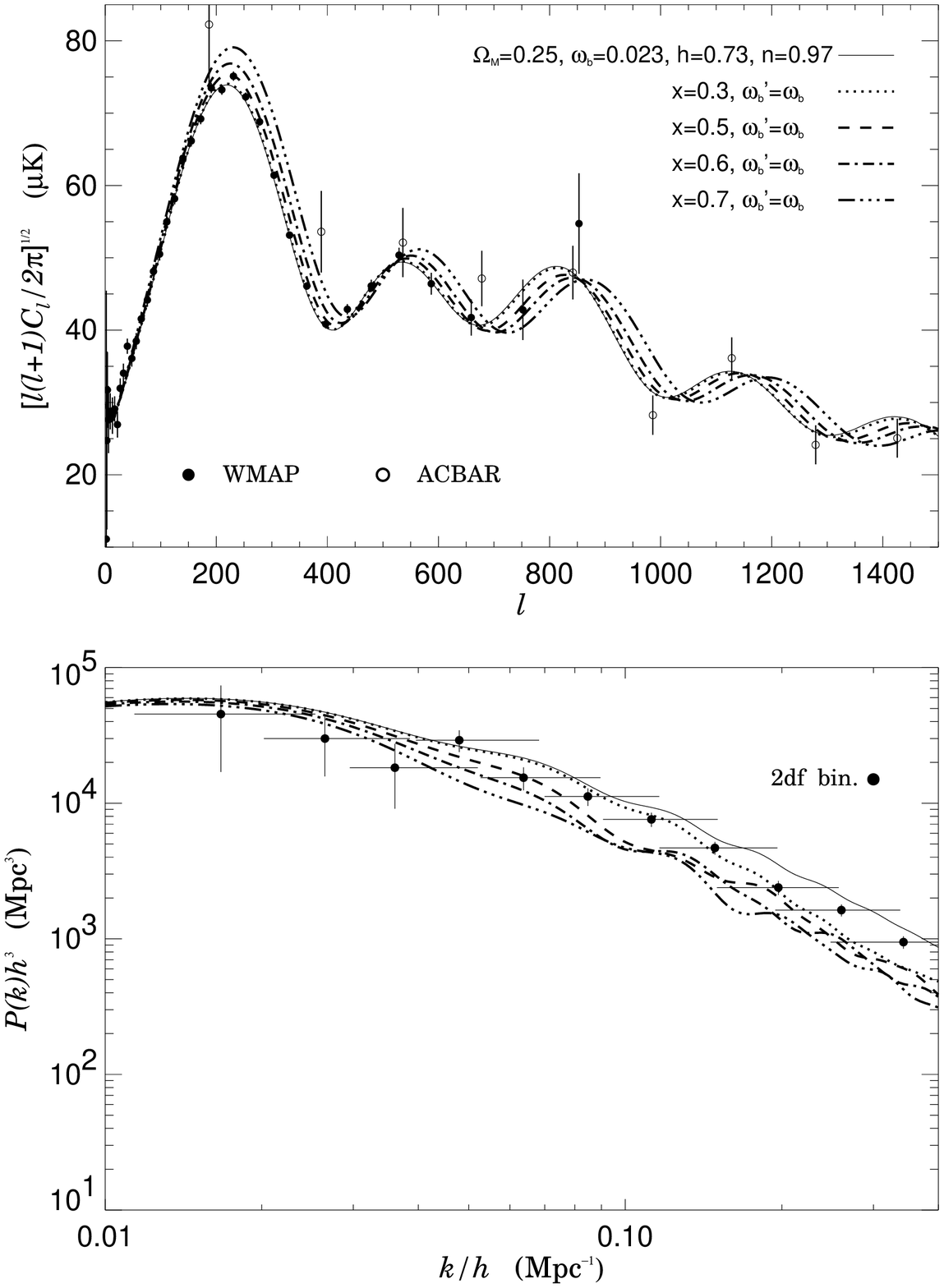},width=9.5cm}}
\vspace*{22pt}
\caption{CMB and LSS power spectra for various mirror models with different values of $ x $ and equal amounts of ordinary and mirror baryons, compared with observations and with a standard reference model (solid line) of parameters $ \Omega_0 = 1 $, $ \Omega_m = 0.25 $, $ \Omega_\Lambda = 0.75 $, $ \omega_b = \Omega_b h^2 = 0.023 $, $ h = 0.73 $, $ n = 0.97 $. The mirror models have the same parameters as the standard one, but with $ \omega_b' = \omega_b $ and $ x = 0.3, 0.5, 0.6, 0.7 $. {\sl Top panel.} Comparison of the photon power spectrum with the WMAP and ACBAR data. {\sl Bottom panel.} Comparison of the matter power spectrum with the 2dF binned data. \label{cmblssfig20}}
\end{figure}

In Fig.~\ref{cmblssfig19} we compare with observations models with the same $ x $, but different mirror baryon contents (as in the bottom panels of Figs.~\ref{cmblssfig1} and \ref{cmblssfig3}). 
The above mentioned low sensitivity of the CMB power spectra on $ \omega_b' $ doesn't give us indications for this parameter (even for high values of $ x $), but the LSS power spectrum helps us again, confirming a sensitivity to the mirror parameters larger than the CMB one. 
This is an example of the great advantage of a joint analysis of CMB and LSS power spectra, being the following conclusion impossible looking at the CMB only. 
This plot tells us that also {\em high values of $ x $ can be compatible with observations if we decrease the amount of mirror baryons in the Universe}. 
It is a second useful indication: in case of high mirror sector temperatures we have to change the mirror baryon density in order to reproduce the oscillations present in the LSS data.

\begin{figure}[pb]
\centerline{\psfig{file={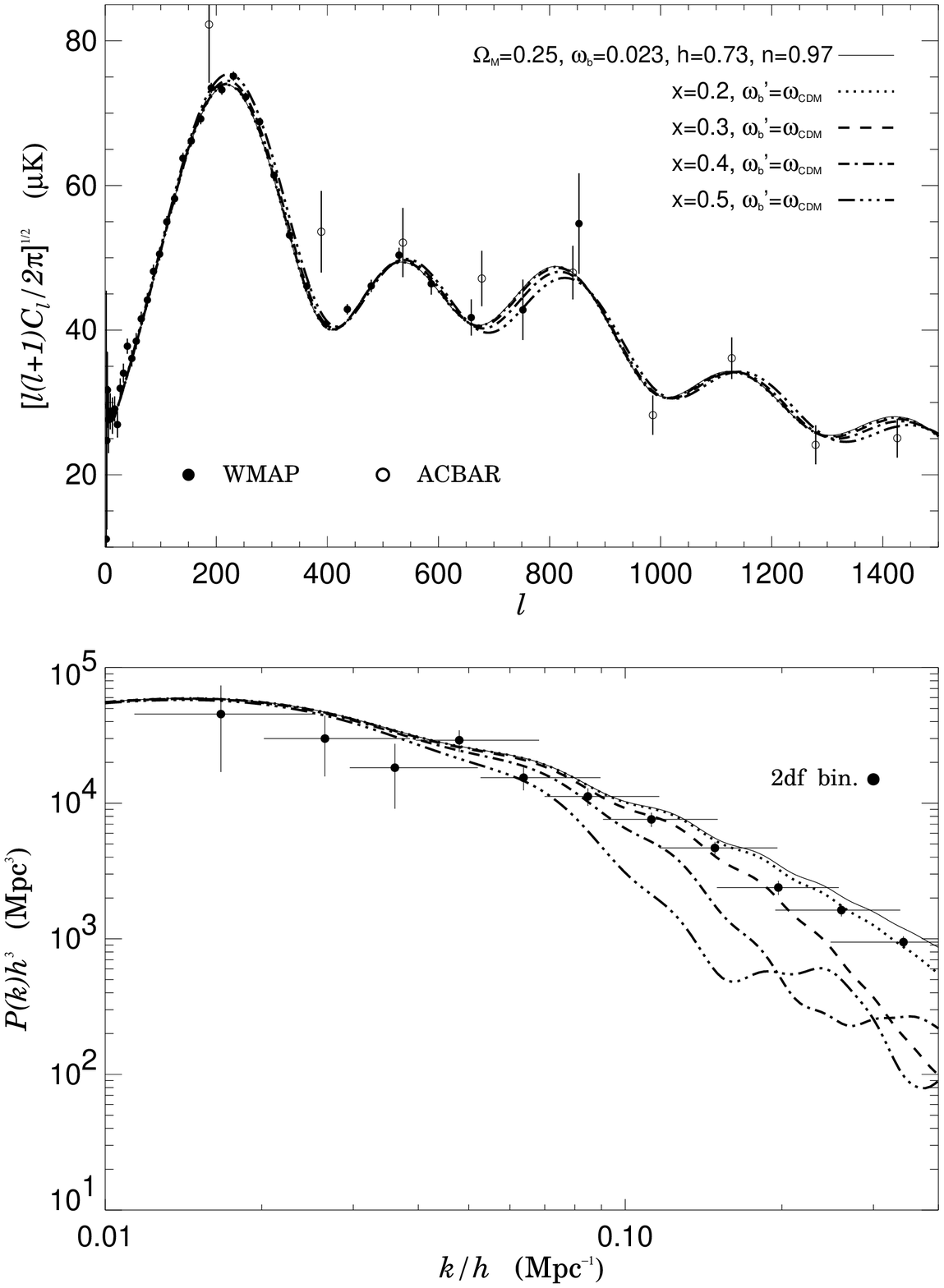},width=9.5cm}}
\vspace*{22pt}
\caption{CMB and LSS power spectra for various mirror models with different values of $ x $ and equal amounts of CDM and mirror baryons, compared with observations and with a standard reference model (solid line) of parameters $ \Omega_0 = 1 $, $ \Omega_m = 0.25 $, $ \Omega_\Lambda = 0.75 $, $ \omega_b = \Omega_b h^2 = 0.023 $, $ h = 0.73 $, $ n = 0.97 $. The mirror models have the same parameters as the standard one, but with $ \omega_b' = \omega_{CDM} $ and $ x = 0.2, 0.3, 0.4, 0.5 $. {\sl Top panel.} Comparison of the photon power spectrum with the WMAP and ACBAR data. {\sl Bottom panel.} Comparison of the matter power spectrum with the 2dF binned data. \label{cmblssfig21}}
\end{figure}

Therefore, after the comparison with experimental data, we are left with three possibilities for the Mirror Universe parameters:
\begin{itemize}
\item high $ x \lesssim 0.5 $ and low $ \omega_b' $ (differences from the CDM in the CMB, and oscillations in the LSS with a depth modulated by the baryon density);
\item low $ x \lesssim 0.3 $ and high $ \omega_b' $ (completely equivalent to the CDM for the CMB, and few differences for the LSS in the linear region);
\item low $ x $ and low $ \omega_b' $ (completely equivalent to the CDM for the CMB, and nearly equivalent for the LSS in the linear region and beyond, according to the mirror baryon density).
\end{itemize}
Thus, with the current experimental accuracy, we can exclude only models with high $ x $ and high $ \omega_b' $.
Our next step will be to consider some interesting mirror models and compute their power spectra. 

In Fig.~\ref{cmblssfig20} we plot models with equal amounts of ordinary and mirror baryons and a large range of temperatures. 
This is an interesting situation, because the case $ \Omega_b' = \Omega_b $ could be favoured in some baryogenesis 
scenario with some mechanism that naturally lead to equal baryon number densities in both visible and hidden sectors. 
These models are even more interesting when we consider both their CMB and LSS power spectra. 
In top panel of figure we see that the temperature anisotropy spectra are fully compatible with observations until $ x \simeq 0.5 $, without large deviations from the standard case. 
In bottom panel, we have a similar situation for the matter power spectra, with some oscillations and a slightly bigger slope, that could be useful to better fit the oscillations present in the data and to solve the discussed 
problem of the desired cutoff at low scales. 
Let us observe that we are deliberately neglecting the biasing problem, given that an indication on its value can come only from a fit of the parameters; then, we have indeed a small freedom to vertically shift the curves in order to better fit the experimental data.

Models of a Mirror Universe where the dark matter is composed in equal parts by CDM and mirror baryons are plotted in Fig.~\ref{cmblssfig21}. 
Here we concentrate on values of $ x $ lower than Fig.~\ref{cmblssfig20}, because the larger mirror baryonic density would generate too many oscillations in the linear region of the matter power spectrum. 
In top panel we show that, apart from little deviations for the model with higher $ x $, all other models are practically the same. 
In bottom panel, instead, deviations are big, and we can still use LSS as a test for models. 
Indeed, models with $ x \gtrsim 0.4 $ are probably excluded by comparison with observations,
and this conclusion is valid even if we consider the effect of a possible bias that affects the data.
Models with lower $ x $ are all consistent with observations.

Observing the figures, even if it is surely premature (given the lack of a detailed statistical analysis of mirror models), we are tempted to guess that mirror baryons could hopefully better reproduce the oscillations present in the LSS power spectra. 
Thus, we are waiting for a more accurate investigation (in particular for the large scale structure) in order to obtain indications for a standard or a Mirror Universe.


\section{Summary}

As in a puzzle representing the whole Universe, researchers working on mirror matter are slowly putting together all the pieces, verifying that their positions are the right ones, and going on...
Proceeding in this way, so far it's emerging a consistent picture of the Universe in presence of mirror dark matter, but not all the pieces are found yet.
The cosmology of mirror matter is more complex than for other usual dark matter candidates, since it is self interacting, shows many different physical phenomena,  and can form a large variety of structures at all scales, similarly to what our baryons do.

\begin{figure}[pt]
\centerline{\psfig{file={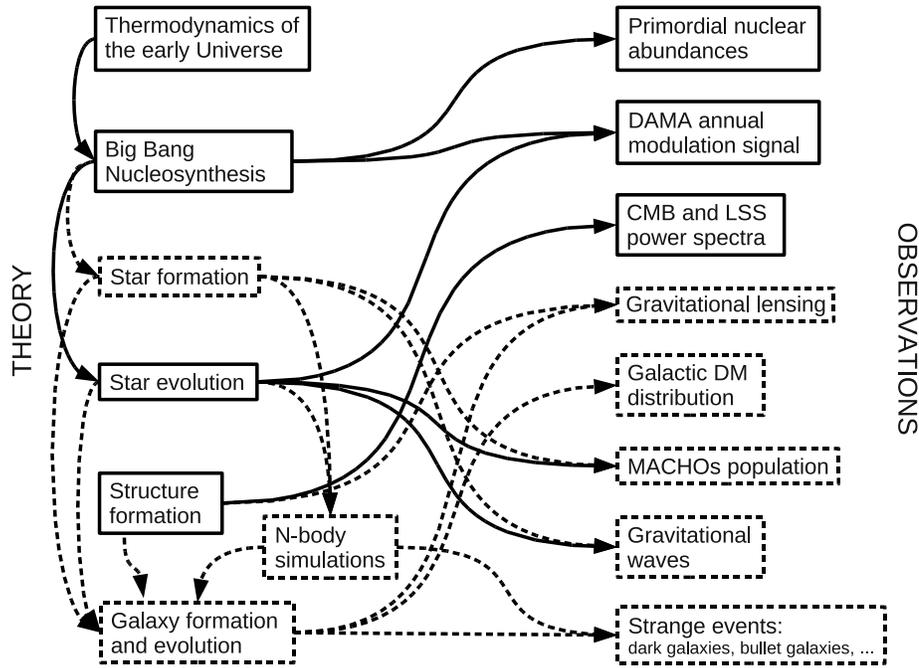},width=13cm}}
\caption{Current status of the astrophysical research with mirror dark matter: solid lines mark what is already done, while dashed ones mark what is still to be done. For an extensive explanation see the text. \label{resume}}
\end{figure}

The current situation of the astrophysical research in presence of mirror dark matter is shown in Fig.~\ref{resume}, with emphasis on the connections between the different theoretical studies and the predicted observable signatures.
Solid lines mark what is already done (and is compatible with current observational and experimental constraints), while dashed ones mark what is still to do.
For the last point a special importance is covered by star formation and N-body simulation studies, towards which people interested in mirror dark matter should address their efforts in the near future.

At present, we have already investigated the early Universe (thermodynamics and primordial nucleosynthesis), and the process of structure formation in linear regime, that permit to obtain predictions, respectively, on the abundances of primordial elements, and on the observed cosmic microwave background (CMB) and large scale structure (LSS) power spectra.
In addition, we have studied the evolution of mirror dark stars, which, together with the mirror star formation, are necessary ingredients for the study and the numerical simulations of non linear structure formation, and of the formation and evolution of galaxies. 
Furthermore, in future studies they should provide predictions on the observed abundances of MACHOs and on the background of gravitational waves.
Using the results of mirror Big Bang nucleosynthesis (BBN) and stellar evolution, we have verified that the interpretation, in terms of photon--mirror photon kinetic mixing, of the DAMA annual modulation signal is compatible with the inferred consequences on the physics of the early Universe.
Ultimately, we will be able to obtain theoretical estimates of gravitational lensing, galactic dark matter distribution, and strange astrophysical events still unexplained (as for example dark galaxies, bullet galaxy, ...), that can be compared with observations.

I summarize here just the topics described in this review, the reader interested in other aspects can refer to the references listed in bibliography.

A hidden mirror sector with unbroken parity symmetry has no free parameters at microscopic level, while macroscopically its presence is parametrized by two additional cosmological quantities, the ratio of photon temperatures in the two sectors, $x=T'/T$, and the ratio of baryonic densities, $\beta=\Omega_b'/\Omega_b$. 
The parameter $\beta$ has a lower bound $\beta \gtrsim 1$ in order to be cosmologically relevant, while $x \lesssim 0.7$ is required by the current limits on extra neutrino families computed at BBN epoch.

In presence of a mirror sector that interacts only gravitationally with ordinary particles, 
the evolution of thermodynamical quantities in the early Universe has been studied in detail for temperatures below $\sim 100$ ~MeV.
The equations were numerically solved, obtaining the interesting prediction that the effective number of extra-neutrino families raises for decreasing temperatures before and after primordial nucleosynthesis; this could help solving the apparent discrepancy in this number computed at BBN and CMB formation epochs.

This study is required for primordial nucleosynthesis analyses and numerical simulations, suggesting that, if we respect the bound on $x$, the presence of the mirror particles has a negligible influence on the ordinary BBN.
Instead, ordinary particles have a big influence on primordial nucleosynthesis in the mirror sector, producing much more helium He$'$ (up to $\sim 80\%$, depending on the value of $x$) and heavier elements (which nevertheless still have a negligible abundance\footnote{
The effects of higher primordial mass fractions of heavy elements on the opacity of matter, and consequently on the mirror star formation efficiency, still need to be carefully inspected.}).
Such a large value of the primordial He$'$ mass fraction will have important consequences for the processes of mirror star formation and evolution. 

Considering the interesting possibility that, besides gravity, ordinary matter interacts with mirror dark matter via photon--mirror photon kinetic mixing of strength $\epsilon \sim 10^{-9}$ (this value is compatible with all experimental and astrophysical bound applicable to mirror matter), the mirror scenario is able to explain the results of current dark matter direct detection experiments.
A photon--mirror photon mixing of this magnitude is consistent with constraints from ordinary primordial nucleosynthesis, as well as the more stringent constraints from CMB and LSS considerations.
In this case the primordial mirror helium He$'$, emerging from BBN in the mirror sector of particles, can reach a very large mass fraction, $Y' \approx 90\%$, with important implications for the mirror dark matter interpretation of the direct detection experiments.

Star formation represents a key ingredient for the galaxy formation, evolution and dark matter distribution, but still need to be studied.
Stellar evolution, instead, has been studied and it was found that stellar lifetimes are 1-2 orders of magnitude smaller than ordinary stars with the same masses.
This is qualitatively expected, since mirror stars have less hydrogen, that is the principal fuel of nuclear fusion during most stellar life, while at the same time more helium makes the stars more luminous and then more fuel-consuming.
A fast stellar evolution means also that in the mirror sector the enrichment of heavy elements is faster due to a higher supernova rate, and this helps the interpretation in terms of mirror matter of the results of direct detection experiments.

The difference in temperature of the ordinary and mirror sectors plays an important role for the cosmological evolution of the mirror world, since it shifts to earlier times the key epochs for structure formation, which proceeds in the mirror sector under different conditions.
Considering adiabatic scalar primordial perturbations as the input, the trends of all the relevant scales for structure formation (Jeans length and mass, Silk scale, horizon scale) for both ordinary and mirror sectors, were analyzed and compared with the cold dark matter (CDM) case. 
These scales are functions of the fundamental macroscopic parameters $x$ and $\beta$, and in particular they are influenced by the differences between the cosmological key epochs in the two sectors.
The analysis of the temporal evolution of perturbations for different values of mirror temperature and baryonic density shows that for $ x $ less than a typical value $ x_{\rm eq} \simeq 0.3 $, for which the mirror baryon-photon decoupling happens before the matter-radiation equality, mirror baryons are equivalent to the CDM for the linear structure formation process.
Indeed, the smaller the value of $x$, the closer mirror dark matter resembles standard CDM during the linear regime.
This study shows some interesting results:
1) the mirror Jeans mass $ M'_{\rm J} $ is always smaller than the ordinary one, thus making easier the growth of perturbations; 
2) there exists a dissipative mirror Silk scale $ M'_{\rm S} $ (analogous to the Silk scale for ordinary baryons), that for $ x \sim x_{\rm eq} $ has the value of a typical galaxy mass; 
3) for $ x < x_{\rm eq} $ we obtain $ M'_{\rm J} \sim M'_{\rm S} $, so that all the primordial perturbations with masses larger than the mirror Silk masscomel can grow uninterruptedly.

Differences in the evolution of matter and radiation components translate into different signatures in the CMB and LSS power spectra computed for scalar adiabatic perturbations in a flat Universe.
Their analysis demonstrated that the LSS spectrum is particularly sensitive to the mirror parameters, due to the presence of both the oscillatory features of mirror baryons and the collisional mirror Silk damping. 
For $ x < x_{\rm eq} $ the mirror baryon-photon decoupling happens before the matter-radiation equality, so that CMB and LSS power spectra in linear regime are equivalent for mirror and CDM cases.
For higher values of $x$ the LSS spectra strongly depend on the amount of mirror baryons. 
Qualitatively comparing with the CMB and LSS data, it was found that for $ x \lesssim 0.3 $ the entire dark matter could be made of mirror baryons, while in the case $ x > 0.3 $ the pattern of the LSS power spectrum excludes the possibility of dark matter consisting entirely of mirror baryons, but they could be present as a mixture (up to $ \sim 50\% $) with the conventional CDM.
The results are the following:
1) the present LSS data are not compatible with a scenario where all the dark matter is made of mirror baryons, unless we consider enough small values $ x \lesssim 0.3 \simeq x_{\rm eq} $;
2) high values of $ x $, $ x > 0.5 $, can be excluded even for a relatively small amount of mirror baryons, since we observe relevant effects on LSS power spectra down to values of mirror baryon density of the order $ \Omega'_b \sim \Omega_b $;
3) intermediate values of $ x $, $ 0.3 < x < 0.5 $, can be allowed if the mirror matter is a subdominant component of dark matter, $ \Omega'_b \lesssim \Omega_b \lesssim \Omega_{CDM} $;
4) for small values of $ x $, $ x < 0.3 $, the mirror and the CDM scenarios are indistinguishable as far as the CMB and the linear LSS power spectra are concerned.
In this case, in fact, the mirror Jeans and Silk lengths, which mark region of the spectrum where the effects of mirror baryons are visible, decrease to very low values, which undergo non linear growth from relatively large redshift.
This is an interesting opportunity for mirror matter, since for low mirror temperatures we could obtain models completely equivalent to the CDM scenario at larger scales (when it works well), but with less power at smaller scales (when it shows possible problems).
Thus, the present situation can exclude only models with high $ x $ and high $ \Omega_b' $.

The results of the linear evolution of perturbations are the inputs for the study of non-linear structure formation down to galactic and subgalactic scales, hopefully with the aim of N-body simulations, that has still to be done.

Estimates of the present population of galactic MACHOs and the gravitational waves from the mirror sector, that could be observable with the next generation of detectors, can be obtained after the study of mirror stellar formation, used together with the already existing predistions of mirror stellar evolution.

At stellar scales, there is also the possibility that a non-negligible amount of mirror matter is trapped inside stars, and can have observable consequences.
They have been studied for the peculiar case of neutron stars, for which it's predicted that the mass-radius relation is not unique, but depends on the history of each star.

Concluding, the astrophysical tests so far used show that the mirror scenario is viable and consistent at all investigated scales, making mirror particles a promising candidate for dark matter.
However, we still need to obtain a complete picture of the Mirror Universe.


\section*{Acknowledgments}

I acknowledge colleagues working on mirror matter, and in particular those who started to study it at early epochs, since their pioneering works made possible the advancement of the knowledge that we have reached today in this topic.


\end{document}